# Assessment of Isoprene as a Possible Biosignature Gas in Exoplanets with Anoxic Atmospheres

Z. Zhan[1], S. Seager[1,2,3], J. J. Petkowski[1], C. Sousa-Silva[1], S. Ranjan[1], J. Huang[4], W. Bains[1,5]

[1] Dept. of Earth, Atmospheric, and Planetary Sciences, MIT,
77 Massachusetts Ave, Cambridge, MA 02139, USA
[2] Dept. of Physics, MIT,
77 Massachusetts Ave, Cambridge, MA 02139, USA
[3] Dept. of Aeronautics and Astronautics, MIT,
77 Massachusetts Ave, Cambridge, MA 02139, USA
[4] Dept. of Chemistry, MIT,
77 Massachusetts Ave, Cambridge, MA 02139, USA
[5] Rufus Scientific, 37 The Moor, Melbourn, Royston, Herts SG8 6ED, UK.

## Abstract

Research for possible biosignature gases on habitable exoplanet atmosphere is accelerating, although actual observations are years away. This work adds isoprene, $C_5H_8$, to the roster of biosignature gases. We found that formation of isoprene geochemical formation is highly thermodynamically disfavored and has no known abiotic false positives. The isoprene production rate on Earth rivals that of methane (~500 Tg $yr^{-1}$). Unlike methane, on Earth isoprene is rapidly destroyed by oxygen-containing radicals. Although isoprene is predominantly produced by deciduous trees, isoprene production is ubiquitous to a diverse array of evolutionary distant organisms, from bacteria to plants and animals—few, if any at all, volatile secondary metabolite has a larger evolutionary reach. While non-photochemical sinks of isoprene may exist, such as degradation of isoprene by life or other high deposition rates, destruction of isoprene in an anoxic atmosphere is mainly driven by photochemistry. Motivated by the concept that isoprene might accumulate in anoxic environments, we model the photochemistry and spectroscopic detection of isoprene in habitable temperature, rocky exoplanet anoxic atmospheres with a variety of atmosphere compositions under different host star UV fluxes.

Limited by an assumed 10 ppm instrument noise floor, habitable atmosphere characterization using JWST is only achievable with transit signal similar or larger than that for a super-Earth sized exoplanet transiting an M dwarf star with a $H_2$-dominated atmosphere. Unfortunately isoprene cannot accumulate to detectable abundance without entering a run-away phase, which occurs at a very high production rate, ~100 times Earth's production rate. In this run-away scenario isoprene will accumulate to > 100 ppm and its spectral features are detectable with ~20 JWST transits. One caveat is that some spectral features are be hard to be distinguish from that of methane. Despite these challenges, isoprene is worth adding to the menu of potential biosignature gases.

# 1. Introduction

For ninety years researchers have considered oxygen as a biosignature gas[1] worth searching for on planets and moons (Jeans 1930). The upcoming 2021 launch of the James Webb Space Telescope (*JWST*; (Gardner *et al.* 2006)), which will be capable of observing the atmospheres of a handful of prime small rocky exoplanets transiting M dwarf stars, has stimulated the study of potential biosignatures gases that could be detected by this and other future space telescope missions.

Beyond JWST, the large ground-based telescopes now under construction (GMT (Johns *et al.* 2012), ELT (Skidmore *et al.* 2015), and TMT (Tamai and Spyromilio 2014)) are expected to come online in the coming decade, and with the right instrumentation are expected to be able to study rocky planets around M dwarf stars by direct imaging. ESA's Atmospheric Remote-sensing Infrared Exoplanet Large-survey (ARIEL) (Gardner *et al.* 2006; Pascale *et al.* 2018) is planned for launch in 2028 and may be able to reach down to observe transiting super-Earth-sized exoplanets around the smallest M dwarf stars. These facilities will provide an excellent opportunity to detect biosignature gases.

Yet oxygen alone does not tell the full tale as life on Earth produces thousands of gases other than oxygen. Some volatiles produced by life on Earth, such as methane ($CH_4$), and nitrous oxide ($N_2O$) are prominent in Earth's atmosphere and therefore have been studied in the context of exoplanet atmosphere biosignature gases. Other gases produced by life are present only as trace gases (< 1 ppbv) in Earth's atmosphere. The possibility that life elsewhere may generate gases different than gases produced by life on Earth and in larger quantities has motivated studies of gases such as dimethyl sulfide (DMS), dimethyldisulfide (DMDS), methyl chloride ($CH_3Cl$), and phosphine ($PH_3$) (Domagal-Goldman *et al.* 2011; Pilcher 2003; Segura *et al.* 2005; Sousa-Silva *et al.* 2020). For a review of exoplanet atmosphere biosignature gases, see (Grenfell 2018; Kiang *et al.* 2018; Schwieterman *et al.* 2018; Seager *et al.* 2016).

[1]Gases produced by life that accumulate in a planetary atmosphere and are remotely detectable are called "biosignature gases".

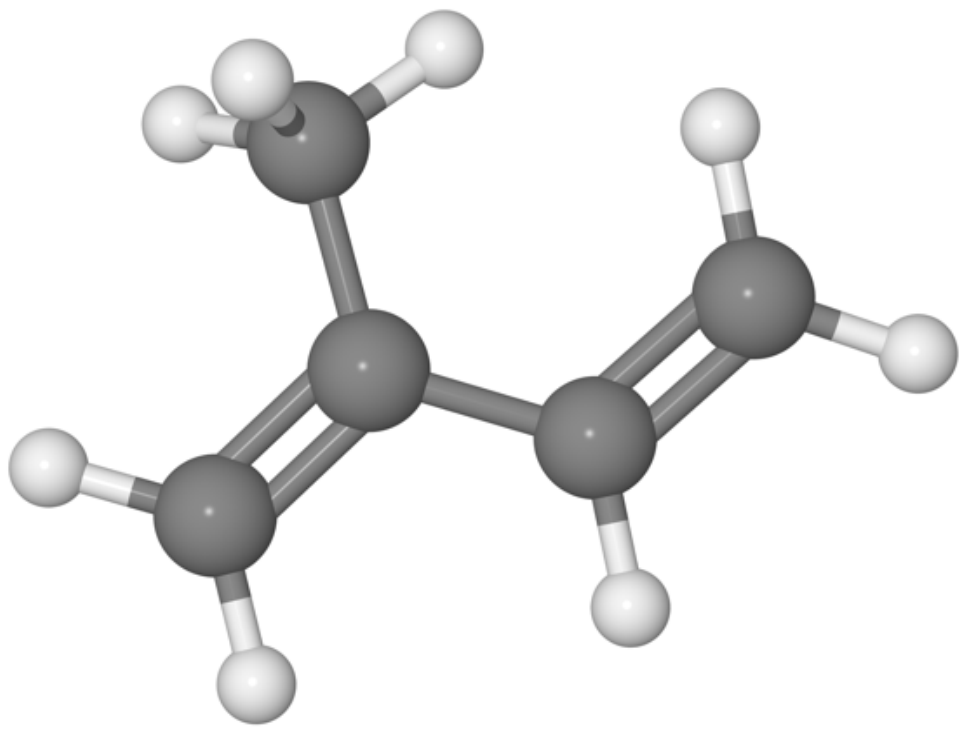

Figure 1. The chemical structure of isoprene. Carbon atoms are shown in dark grey and hydrogen atoms in white. Isoprene ($C_5H_8$ or 2-methyl-1,3 butadiene) is a conjugated-diene with a methyl group attached to the second position. Conjugated dienes are two double bonds separated by one single bond.

In this work, we add isoprene ($C_5H_8$, Figure 1) to the list of biosignature gases to be considered for detection in future missions. Isoprene is a hydrocarbon containing two carbon-carbon double-bonds connected by one carbon-carbon single-bond, or a “conjugated diene”.

Isoprene on Earth is predominately produced by deciduous trees and land plants. The production rate of isoprene is about 500 Tg $yr^{-1}$ (Sharkey *et al.* 2008), which is comparable to the production rate of methane, also 500 Tg $yr^{-1}$ (e.g., (Dlugokencky *et al.* 2011)). For a more detailed decomposition of isoprene sources and sinks, see (Figure 2). To our knowledge isoprene has not yet been evaluated in detail as an exoplanet biosignature gas (although it has been briefly mentioned by (Grenfell 2017; Seager *et al.* 2012)).

On first consideration, one might disregard isoprene as a potential biosignature gas because of its short lifetime (< 3 hr) in Earth’s atmosphere. The short lifetime results in a very low isoprene atmospheric abundance, ranging from 1 to 5 part per billion (ppbv) only in localized regions above cities and forests (Sharkey *et al.* 2008) to no detection above deserts. In Earth's atmosphere isoprene is primarily regarded as a precursor to secondary organic aerosols. This is because once isoprene is released into the atmosphere it is rapidly destroyed by reactions with ·OH, and subsequent reactions with by $O_2$, to form diverse and reactive products. The intermediate products subsequently react with a wide variety of atmospheric components including trace gases, and $NO_3^-$ and $Cl^-$ radicals), eventually forming aerosols[2] (Fan and Zhang 2004; Teng *et al.* 2017) (Figure 3).

However, the lack of $OH^-$ isoprene under anoxic conditions motivate its assessment as a biosignature gas. Earth's atmosphere had no oxygen during its initial 2.4 Gyr and isoprene could, in principle, accumulate in anoxic atmospheres to detectable levels (Holland 2006).

[2]For example, the blue haze characteristic of forest-covered mountains (e.g., the Blue Ridge Mountains, a physiographic province of the Appalachian Mountain range) is the end product of isoprene radical chemistry (Claeys *et al.* 2004).

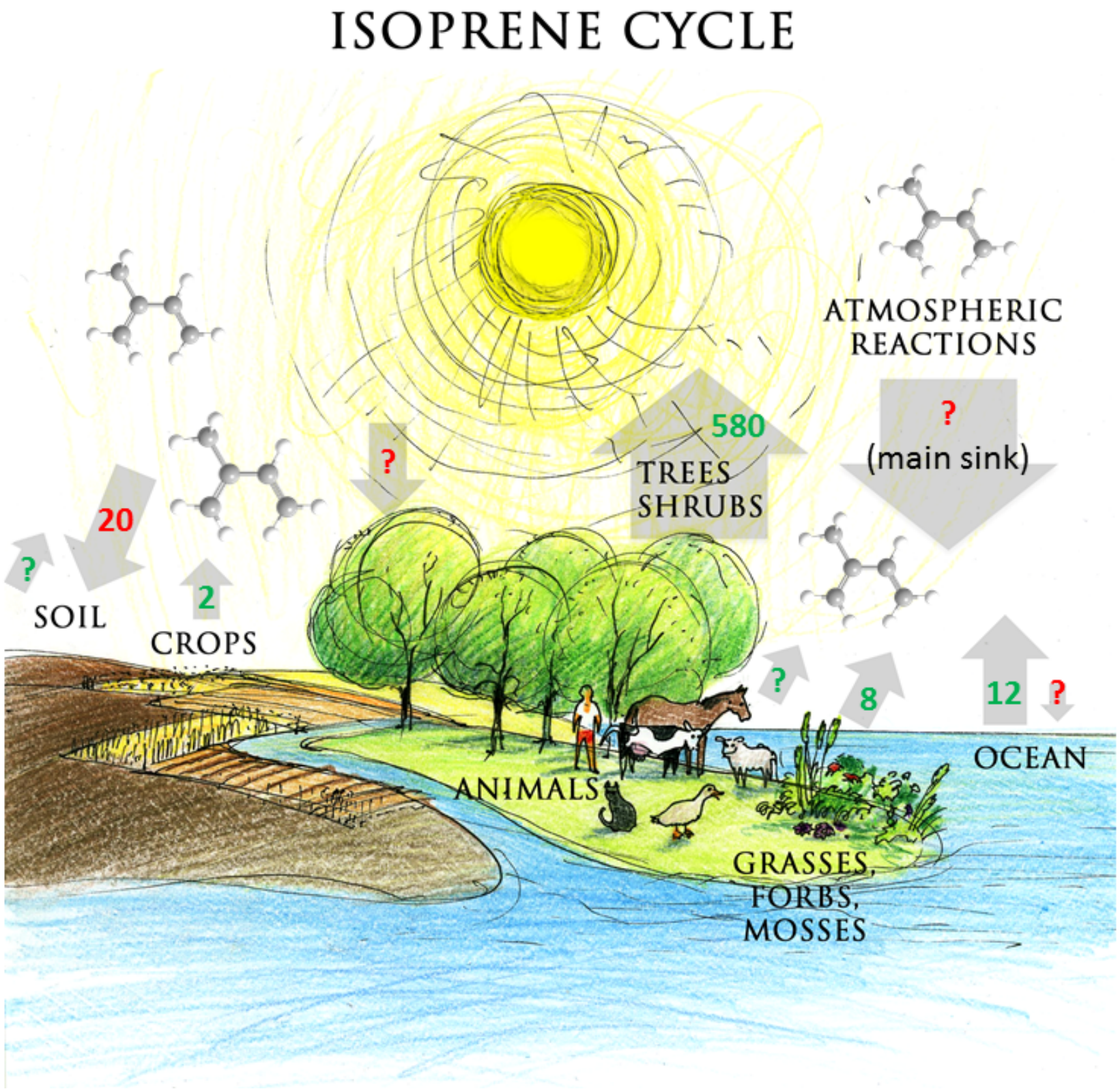

**Figure 2.** Schematic of the major sources and sinks of isoprene in the Earth's atmosphere. The isoprene sources (up arrows and green numbers) and sinks (down arrows and red values) are shown. The thickness of arrows provides a relative estimation of the contribution of various sources and sinks of isoprene (McGenity TJ et al. 2018).

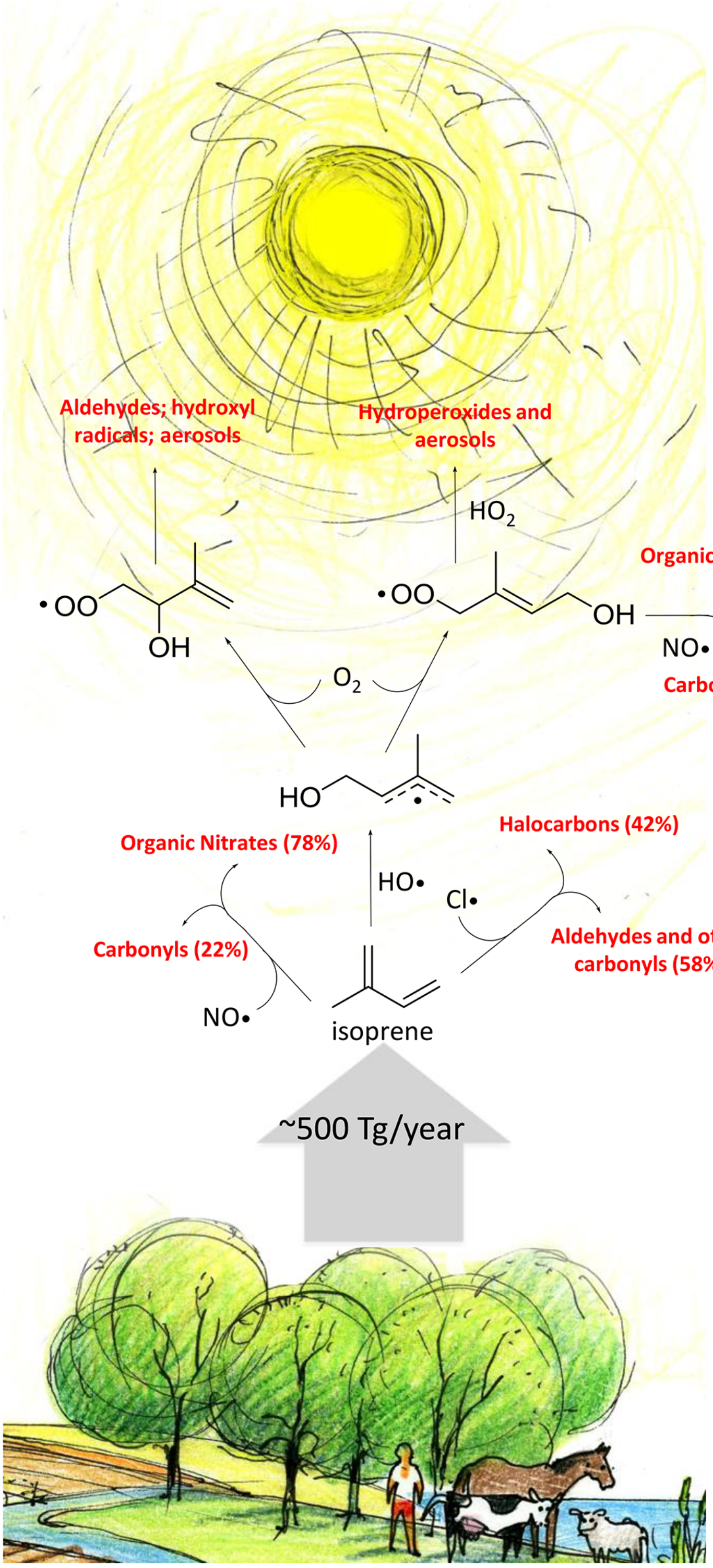

**Figure 3.** Schematic of the fate of the isoprene in the Earth's atmosphere. The oxidation of isoprene by OH radicals is the main pathway for the destruction of isoprene in Earth's atmosphere (Teng *et al.* 2017). Less predominant destruction pathways include the reaction with NO· and Cl· radicals (Fan and Zhang 2004). We note that the photochemically-driven reactions with minor radical species (e.g., NO· species) can also proceed with downstream isoprene radical products (Iso[$O_2$]), see e.g. (Fan and Zhang 2004) for detailed pathways of photochemically-driven atmospheric oxidation of isoprene species.

In this paper, we evaluate isoprene as a biosignature gas. We first summarize isoprene's sources and sinks (Section 2), including isoprene's overall production on Earth (Section 2.1), with details on isoprene's biological production from diverse organisms, both aerobic and anaerobic (Section 2.2), followed by a review of the known destruction mechanisms for isoprene (Section 2.3). Next, we outline our inputs and methods to assess the detectability of isoprene for a diverse set of anoxic atmosphere scenarios (Section 3). We discuss our main findings in (Section 4): first we present the production rates required for isoprene to accumulate to a detectable level in a given atmosphere scenario (Section 4.1); next we assess whether or not isoprene can be detected using JWST with a reasonable number of transit observations (Section 4.2); then we show that isoprene is not produced thermodynamically in the atmosphere and therefore that isoprene as a biosignature gas has no false positives in habitable exoplanet atmospheres (Section 4.3). Finally, we conclude the paper with a discussion of our results, limitations, and caveats (Section 5).

# 2. Isoprene Sources and Sinks

Before we study the detection of isoprene in an exoplanet atmosphere, we first explore how isoprene is created and destroyed. On Earth, isoprene production is biological (Sections 2.1 and 2.2). We explore the destruction pathways of isoprene, which is mainly by direct photolysis and with OH radicals and $O_2$ (Figure 3).

## 2.1 Isoprene Productions on Earth

Globally, life on Earth produces 400 - 600 Tg yr$^{-1}$ of isoprene (Arneth *et al.* 2008; Guenther *et al.* 2006; Guenther *et al.* 2012). The biological production rate of isoprene on Earth is roughly equal to global emission of methane from all sources (525 Tg yr$^{-1}$) (Guenther *et al.* 2006; Guenther *et al.* 2012; Seinfeld and Pandis 2016) and significantly exceeds production rates of other volatile organic molecules made by life on Earth such as DMS (38.4 Tg yr$^{-1}$), nitrous oxide ($N_2O$) (20 Tg yr$^{-1}$)[3], and methyl chloride ($CH_3Cl$) (3.5 Tg yr$^{-1}$) (Figure 4) (Guenther *et al.* 2006; Korhonen *et al.* 2008; Tian *et al.* 2015; Yokouchi *et al.* 2015). Isoprene is the most abundantly produced biological volatile organic compound on Earth and constitutes more than 1/3 (by mass) of the total amount of all natural volatile organic compounds released into Earth's atmosphere (Guenther *et al.* 2006; Sharkey *et al.* 2008). For some plants, isoprene can comprise up to 20% of the carbon release rate by the plants (Sharkey and Loreto 1993). We note that Earth's isoprene production rate pales in comparison to the production rate of Earth's most obvious biosignature gas: $O_2$ (300,000 Tg yr$^{-1}$), with one caveat that most of the $O_2$ are respired and only 0.1% contribute to net $O_2$ emission (Knoll *et al.* 2012).

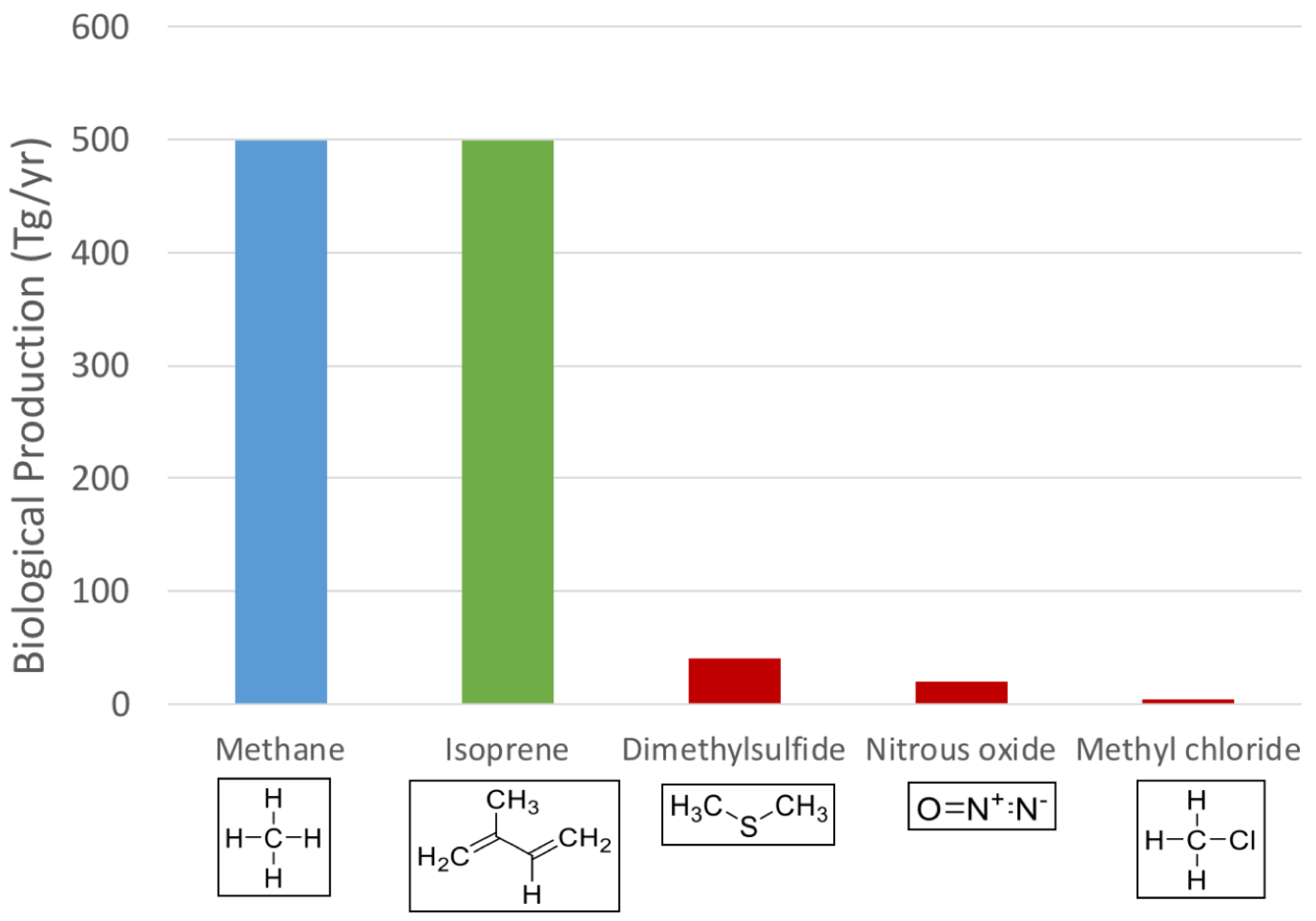

**Figure 4.** Estimated biological production of five different gases in Tg yr$^{-1}$. Isoprene (green bar) has a production rate in the same range of that of methane (blue bar). Other gases considered as biosignature gases have much lower biological production rates. Data from the following sources: (Guenther *et al.* 2006), (Tian *et al.* 2015), (Korhonen *et al.* 2008), (Yokouchi *et al.* 2015).

Gases such as $CH_3Cl$ (Segura *et al.* 2005) and DMS (Arney *et al.* 2018; Domagal-Goldman *et al.* 2011; Pilcher 2003; Seager *et al.* 2012) were suggested before as potential biosignature gases due to their large production rate by marine life on Earth and low destruction rate, which lead to relative stability in some atmospheres. Global annual production rates of isoprene are much higher than those of $CH_3Cl$ and DMS (production rates of major volatile secondary metabolites by life on Earth are compared in Figure 4).

Isoprene has a short atmospheric lifetime of less than 3 hours in the modern terrestrial atmosphere (Section 2.3). The very high destruction rate of isoprene in $O_2$-dominated environments leads to a very low effective abundance of isoprene in Earth's atmosphere. Isoprene concentration in the atmosphere varies geographically and seasonally, ranging from 1 - 5 ppb above forests (Sharkey *et al.* 2008) to no detection above deserts. But even in the high-producing areas, above deciduous forests, isoprene concentration does not exceed 5 ppb (Sharkey *et al.* 2008). Low production rates in other areas means that the global average level of isoprene in Earth's modern troposphere is less than ppt levels. Such low atmospheric abundances make the remote detection of isoprene in Earth's atmosphere a challenging task; in fact, isoprene has not been detected in the transmission spectra of Earth's atmosphere (Schreier *et al.* 2018) as measured by the ACE-FTS Earth observation mission (Bernath 2017; Hughes *et al.* 2014).

## 2.2 Biological Production of Isoprene

In this section, we review the biological production of isoprene by life on Earth. We discuss the diversity of species that synthesize isoprene (Section 2.2.1), briefly review isoprene's biosynthesis, and explore production of isoprene by anaerobic lifeforms on Earth (Section 2.2.2) and summarize the variety of biological functions of isoprene (Section 2.2.3). We leave an in-depth discussion of the structural and phylogenetic diversity of isoprenoids (isoprene polymers or molecule with isoprene-like structure) and a detailed review of the known isoprenoid biosynthetic pathways for Appendix II.

---

[3] With anthropogenic sources the total production of $N_2O$ is 30 Tg/year.

### 2.2.1 The Extent of Formation of Isoprene by Life on Earth

Isoprene is produced by a very large number of evolutionarily diverse organisms including algae, animals, bacteria, fungi, plants and protists (Bäck *et al.* 2010; Exton *et al.* 2013; Fall and Copley 2000; Gelmont *et al.* 1981; King *et al.* 2010; Kuzma *et al.* 1995; Moore *et al.* 1994; Sharkey 1996). The majority (~90%) of the global production of isoprene is from terrestrial plants, mostly by tropical trees and shrubs (Sharkey *et al.* 2008) (see Figure 2 for an overview of the isoprene cycle in the atmosphere). Animals are responsible for the release of a significant fraction of the remaining 10% of isoprene's yearly global emissions. Production of isoprene was extensively studied in many animal species, but the majority of research was done on isoprene production in rodents and humans (Sharkey 1996). For example, nursing mice and rats emit substantial amounts of isoprene (Sharkey 1996). Isoprene is also the most abundant hydrocarbon in the exhaled breath of humans (Gelmont *et al.* 1981; King *et al.* 2010; Sharkey 1996).

In addition to plants and animals, many bacteria, both aerobic and anaerobic, produce isoprene. The true extent of isoprene synthesis in prokaryotes is still difficult to estimate as only a few phyla have been tested for isoprene production (e.g. *Proteobacteria*, *Actinobacteria,* and *Firmicutes*) (Alvarez *et al.* 2009; Fall and Copley 2000; Kuzma *et al.* 1995; Schöller *et al.* 1997; Schöller *et al.* 2002). Bacteria from the genus *Bacillus*, both terrestrial and marine, were shown to be the highest producers of isoprene among tested prokaryotes (Kuzma *et al.* 1995; McGenity *et al.* 2018). Some *Bacillus* species are also the only bacteria known so far to naturally produce isoprene completely anaerobically (see Section 2.2.2 below; (Fall *et al.* 1998)).

The endogenous production of isoprene in archaea has not been widely investigated and isoprene has not yet been found to be produced by archaea.

In summary, isoprene production on Earth is not only abundant but also is widespread and present in a large number of evolutionarily diverse organisms, from bacteria to mammals, and is made by at least two, evolutionarily distinct metabolic pathways (see Appendix II). No other volatile secondary metabolite has a larger evolutionary reach than isoprene.

### 2.2.2 Biosynthesis of Isoprene

Isoprene biosynthesis has only been extensively studied in plants. In plants isoprene synthase (IspS; PDB ID: 3n0g; EC 4.2.3.27) is responsible for the catalysis of the last step in the isoprene biosynthesis pathway—the elimination of pyrophosphate from the isoprene precursor dimethylallyl pyrophosphate (DMAPP) and release of isoprene (Figure 5) (Köksal *et al.* 2010).

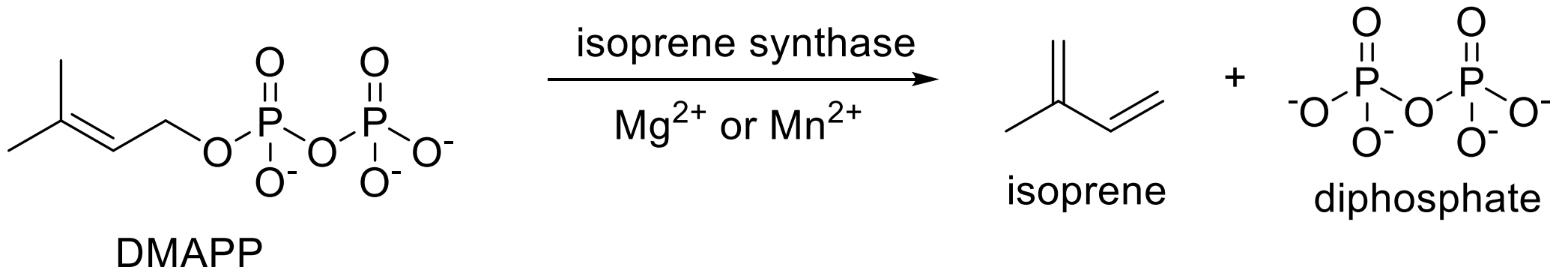

**Figure 5.** Biological production of isoprene. Isoprene synthase is a $Mg^{2+}$- or $Mn^{2+}$-dependent terpenoid synthase that catalyzes the cleavage of inorganic diphosphate from the isoprene precursor dimethylallyl diphosphate (DMAPP) to yield isoprene.

The mechanisms of biosynthesis of isoprene in non-plant species are largely unknown despite confirmed widespread isoprene production by a diverse host of organisms. Early studies suggested that mammals synthesize isoprene in the liver through a different pathway than plants (Deneris *et al.* 1985; Sharkey 1996). For more detail, see Appendix II.

Interestingly, despite plentiful evidence for bacterial production of isoprene, bacterial isoprene synthase has been only partially characterized and is thought to be evolutionarily unrelated to the plant isoprene synthase (McGenity *et al.* 2018; Sivy *et al.* 2002). Isoprene production has been detected in fungi (Berenguer *et al.* 1991) and animals even if they too, like bacteria, do not seem to have plant isoprene synthase homologs. We conducted a bioinformatic search of genomic databases for sequences similar to plant isoprene synthase sequences and confirmed that no homologues of plant isoprene synthase have been found in bacteria, archaea, fungi or animals. This confirms that isoprene synthetic pathways have evolved independently at least twice.

Impressively, all species belonging to the three domains of life (Bacteria, Archaea, and Eukarya) possess isoprenoid biosynthetic pathways. This means that all species are capable of creating complicated natural compounds containing the isoprene "motif", even though not all species release isoprene as an isolated molecule (Table IIB) (Firn 2010). For details on isoprenoid biosynthetic pathways see Appendix II.

While on Earth life that produces isoprene is aerobic ($O_2$-dependent) or facultatively anaerobic (e.g. *Escherichia coli* or *Bacillus subtilis*), the biosynthesis of isoprene does not require molecular oxygen (unlike for e.g. the biosynthesis of sterols). This means that isoprene could be in principle made by strictly anaerobic organisms, in anoxic atmospheres. The synthesis of isoprene by recombinant anaerobic bacteria and archaea is known (Beck *et al.* 2014; Murphy *et al.* 2017). For example, the methanogenic and anaerobic archaea *Methanosarcina acetivorans* is capable of efficient isoprene production upon heterologous expression of isoprene synthase from plants (Murphy *et al.* 2017). There are also a small number of studies of native anaerobic isoprene production in natural environments. A few anaerobic bacteria, such as *Bacillus cereus* 6A1 and *Bacillus lichenformis* 5A24, have been shown to naturally produce isoprene anaerobically, and in substantial quantities, with production rates of 40 - 60 nmol $g^{-1}$ $hr^{-1}$ (Fall *et al.* 1998), comparable to that of terrestrial plants as we demonstrate in Section 4.1.2 where we discuss in detail the global production rate achievable for an Archean anoxic biosphere comprised purely of isoprene-producing prokaryotes.

Indeed, the capability for isoprene biosynthesis appears to be independent of aerobic metabolism. Such few laboratory studies on anaerobic production of isoprene establish the precedent that alien life could in principle discover an anaerobic biosynthetic pathway to produce isoprene, even on planets that have atmospheres very different than Earth's (e.g. $H_2$-dominated). We note that $H_2$-dominated atmospheres are not detrimental for life and that life can survive and actively reproduce in an $H_2$-dominated environment (Seager *et al.* 2020). There is however the question of sufficient evolutionary incentive for the production of huge amounts of isoprene by an anaerobic biosphere. We discuss this problem next.

The reasons why Earth's aerobic biosphere makes isoprene in such impressively large amounts is not known, and it is not yet known what are the evolutionary pressures that govern isoprene production by life on Earth (Sharkey and Monson 2017). It is however likely that the functions of isoprene for life on Earth are many and are not limited to one single dominant role (see Section 2.2.3 below for discussion of various biological functions of isoprene). The biological functions of isoprene may be related to UV shielding and reactive-UV-radical protection (as evidenced by plants' response to UV, heat etc.), isoprene might also be used as a signaling molecule (Harvey and Sharkey 2016; Zuo *et al.* 2019). It is impossible to predict what biological functions a specialized secondary metabolite like isoprene could have in an anaerobic setting. One could speculate that the protective role of isoprene against UV radiation and/or other stressors could be universal to all life, even anaerobic one, and therefore could justify its abundant production on an anoxic world, especially as the anoxic world would have no ozone layer to protect against UV

It is likely that more endogenous isoprene production among archaea and other anaerobic organisms awaits discovery. We hope that this paper stimulates further research into this understudied aspect of isoprene biology.

### 2.2.3 Biological Functions of Isoprene

The biological roles of isoprene have mostly been studied in plants, as plants are responsible for more than 90% of isoprene production on Earth. The consensus is that isoprene protects the photosynthetic apparatus of tree leaves from heat stress, especially the sudden changes in temperature caused by varying exposure to sunlight (Sharkey *et al.* 2008), although other functions have been proposed (Jones *et al.* 2016; Laothawornkitkul *et al.* 2008; Sharkey and Monson 2017; Velikova *et al.* 2012; Vickers *et al.* 2009). A range of observations supports this thermal protection role for isoprene (Logan *et al.* 2000; Peñuelas *et al.* 2005; Taylor *et al.* 2019), although its mechanism of action is not known.

The function of isoprene in bacteria, fungi or animals is far less studied than its function in plants. In a facultative anaerobe bacterium *Bacillus subtilis*, isoprene synthesis is elevated as a response to hydrogen peroxide treatment (Hess *et al.* 2013; Xue and Ahring 2011) or in response to non-optimal growth conditions (e.g. elevated temperature and salinity) (Xue and Ahring 2011). It has also been suggested that isoprene might play a role as a signaling molecule in the regulation of spore development of *Bacillus subtilis* (Fall and Copley 2000; Sivy *et al.* 2002; Wagner *et al.* 1999). The role of isoprene as an interspecies signaling molecule was also postulated. Isoprene could also act as a repellant for microbe-grazing springtails (hexapods) (Fall and Copley 2000; Gershenzon 2008; Michelozzi *et al.* 1997). Despite the fact that animals produce significant amounts of isoprene, our knowledge of its biological function in animals is still limited.

## 2.3 Isoprene Atmospheric Chemistry

Here we list the dominant known pathways for isoprene loss in the atmosphere. Isoprene is not known to reform from any of its reaction products by any known atmospheric processes, unlike other atmospheric gases such as $H_2O$ or $O_2$. Isoprene has very low water solubility, so isoprene itself is not likely to be absorbed into an aerosol or rained out, though we do model this process. We therefore consider the source of isoprene to be solely biological production, and the main sink of isoprene to be photochemistry.

In Earth's atmosphere, isoprene's main destruction pathways are 1.) direct photolysis and 2.) reaction with ·OH radicals and subsequently followed by reaction with $O_2$. Isoprene also reacts with other radicals but their abundance is too low compared to ·OH radicals to make a significant impact (Fan and Zhang 2004). However, there might be as yet unknown isoprene destruction pathways in anoxic atmospheres that might affect the overall destruction rate of isoprene significantly. Directed experimental studies (e.g., similar to (He *et al.* 2019)) on the chemistry of isoprene in different atmospheric scenarios (especially anoxic ones) are needed to fully understand the scope of isoprene's possible reactions in diverse exoplanet atmospheres.

The reaction rate constants used in the following subsections are constants in the Arrhenius rate equation: $k = Ae^{-E/RT}$, where $k$ is the reaction rate constant ($cm^3\ s^{-1}$ for a second order reaction), $A$ is a constant ($cm^3\ s^{-1}$), $E$ is the activation energy ($J\ mol^{-1}$), $R$ is the gas constant ($J\ mol^{-1}\ K^{-1}$), and $T$ is temperature (K).

### 2.3.1 Destruction by ·OH Radicals and $O_2$

Isoprene's reaction with ·OH is the first step in a series of reactions that end with aerosol formation (Figure 6; e.g., (Zhang *et al.* 2000)). The rate coefficient for the destruction of isoprene by reaction with ·OH is $k = 10.0 \pm 1.2 \times 10^{-11}$ $cm^3$ $molecule^{-1}$ $s^{-1}$ at 294 K (Zhang *et al.* 2000). The ·OH radical can attack different position of the isoprene molecule to create intermediate sets of radicals with the general formula,

$$C_5H_8 + \cdot OH \rightarrow \cdot C_5H_8[OH] \quad \text{Eq. 1}$$

In an anoxic atmosphere, $\cdot OH$ will still be present (from $H_2O$ photodissociation) but at much lower levels than in an oxygenic atmosphere (Hu et al. 2012). Therefore, the fate of $\cdot C_5H_8[OH]$ radicals, in absence of oxygen, will depend on the trace constituents of a given atmosphere. To our knowledge, the reaction network for ·C5H8[OH] is not known for anoxic conditions; therefore, as with the other products of isoprene destruction we neglect its chemistry to focus on the prospects for isoprene buildup, with the understanding that this approximation may lead to underestimates of the isoprene concentration at a given isoprene surface flux since we neglect the possibility of isoprene recombination and/or UV shielding from isoprene photochemical products.

In Earth's atmosphere, the intermediate $\cdot C_5H_8[OH]$ radicals then react with $O_2$. The product oxidized isoprene radicals ($\cdot C_5H_8[OH][O_2]$) subsequently react with $NO_x\cdot$ species in the atmosphere and other reactive trace gases, contributing to the overall destruction rate of isoprene (Fan and Zhang 2004; Zhang *et al.* 2000). We ignore the subsequent reactions of isoprene radicals for anoxic atmospheres where oxygen is not present.

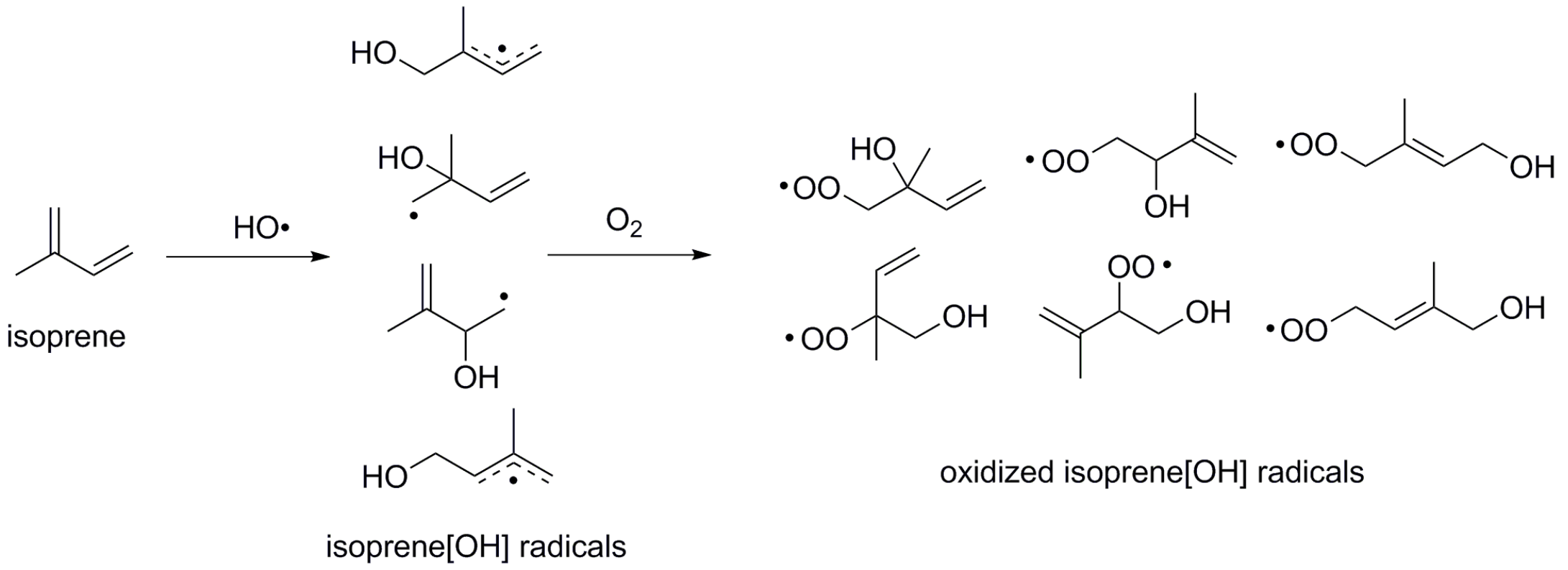

**Figure 6.** A mechanistic diagram for the reactions of ·OH with isoprene and the subsequent ·C5H8[OH] radical reactions with $O_2$. The dots indicate the location of the radicals. The dashed lines indicate the delocalized electrons. The four isoprene intermediates (left side) immediately react with $O_2$ resulting in six different radicals (right side). The six radicals react with other trace atmospheric components to form aerosols. Figure adapted from (Zhang *et al.* 2000).

## 2.3.2 Destruction by $O_3$

Isoprene can react directly with $O_3$ with the rate coefficient $k = 9.6 \pm 0.7 \times 10^{-18}$ $cm^3$ $molecule^{-1}$ $s^{-1}$ at 286 K (Karl *et al.* 2004). We include the destruction rate of isoprene by $O_3$ for completeness; the reaction rate is several orders of magnitude smaller than the dominant pathways (reaction with ·OH, O·).

### 2.3.3 Destruction by O· Radicals

Isoprene can react directly with O· with the rate coefficient $k = 3.5 \pm 0.6 \times 10^{-11}$ $cm^3$ $molecule^{-1}$ $s^{-1}$ at 298 K (Paulson *et al.* 1992).

### 2.3.4 Destruction by H· Radicals and $H_2$ molecules

To our knowledge, there is no data published on the reactivity of isoprene with hydrogen (H·) radicals. While there are plenty of documented reactions of H· radicals with ethene, propene, butene (Linstrom and Mallard 2001), it is beyond the scope of this work to extrapolate from these reactions to isoprene. Further experimental work is needed to confirm or rule out the possibility of efficient conversion of isoprene with H· radicals in $H_2$-dominated conditions. It is also not known if any of the $\cdot C_5H_8[OH]$ radicals, formed upon reacting with ·OH, can efficiently react with H· radicals in an $H_2$-dominated environment to revert back to isoprene and water.

Hydrogenation of isoprene (or isoprene units) using molecular hydrogen, resulting in saturation of a double bond, is a standard reaction utilized in human industry. However, such reactions requires higher than habitable temperatures (> 400 K), catalysts and meticulous environments (Abdelrahman *et al.* 2017). It is also unknown if lightning might be a catalyst for such reactions to occur in the atmosphere.

### 2.3.5 Destruction Through UV Radiation

The general chemical formula for photodissociation of isoprene is

$$C_5H_8 + h\nu\,(\lambda_{peak} \cong 218\text{ nm}) \rightarrow \cdot C_5H_7^- + H^+, \qquad \text{Eq. 2}$$

where $h\nu$ is the energy of a photon. The quantum yield of isoprene photolysis is also not well known, to our best knowledge. We take a conservative approach and assume the quantum yield of 1, which means that any high energy UV photon that is absorbed by an isoprene molecule will dissociate it[4].

### 2.3.6 Destruction by Life

Apart from the atmospheric sinks of ·OH, $O_2$, and UV photolysis, approximately 4% (20 Tg $yr^{-1}$) of the yearly production of isoprene is directly consumed as a carbon source by a variety of soil microorganisms (Cleveland and Yavitt 1998; Shennan 2005). While on Earth the biological sink of isoprene is small, it may be reasonable to assume that any atmosphere that is enriched with isoprene will have a sizable population of living organisms utilizing isoprene as a carbon source, further contributing to its removal from the atmosphere. The efficiency of biologically-driven removal of isoprene will be largely dependent on the unique biochemistry and ecology of life inhabiting the planet. The impact of the potential biological destruction of isoprene has therefore not been included in our model.

[4] We note that the conjugated diene structure of isoprene stabilizes the radical formed by photolysis so that lower energy photons can cleave the C-H bond. The conjugated diene also has a high cross section for the absorption of UV photons. These two features mean that conjugated dienes such as isoprene are photolyzed with higher efficiency compared to other hydrocarbons.

#### 2.3.7 Aerosol and Haze Formation

On Earth, isoprene radical-induced aerosols are a major source of secondary organic aerosols. The production of haze is also the primary pathway for isoprene destruction on Earth (e.g., (Seinfeld and Pandis 2016)). For example, the blue haze of some forest-covered mountains (Claeys *et al.* 2004) is a product of isoprene radical-induced aerosols.

In the context of anoxic atmospheres, hazes and aerosols would be different from those found in Earth's atmosphere, with haze composition depending on the molecules and radicals available to react with isoprene and isoprene's destruction products.

## 3. Inputs and Methods for the Assessment of Detectability

The goal of this section is to provide a framework to assess whether or not a biosignature gas may be detectable given a proposed exoplanet atmospheric context and the reality of telescope observations. Our ability to detect a biosignature gas depends on the dominant molecular composition of the exoplanet atmosphere, observatory capabilities, and instrumental effects. Whether or not a biosignature gas is detectable seldom has a simple, fixed answer.

We start with addressing isoprene molecular absorption inputs, including UV cross sections used in the photochemistry calculations, Infrared (IR) cross sections used in calculating molecular absorption, and haze extinction cross sections (Section 3.1). Next, we describe the photochemistry code used to compute the mixing ratio profile used for each atmosphere archetype (Section 3.2) and additional parameters to compute atmospheric simulations (Section 3.3). Then we describe the simulation of transmission spectroscopy and secondary eclipse thermal emission spectroscopy (Section 3.4). Finally, we discuss observation strategies (Section 3.5) and describe the framework to assess the detection of isoprene (Section 3.6).

### 3.1 Isoprene Molecular Inputs

Molecular absorption cross sections of isoprene are required for calculating the photochemistry rate in the ultraviolet-visible (UV-Vis) regime and simulating its absorption spectral features in an exoplanet atmosphere in the infrared (IR) regime.

### 3.1.1 UV-Vis Cross Section

The isoprene UV-Vis cross section is shown in (Figure 7). The data has been taken from (Dillon *et al.* 2017) and is used for calculating the UV photolysis rate (see Section 3.3). The isoprene UV-Vis absorption peaks at 218 nm with $\sigma_{peak}$ = 7.93 ± 0.02 × $10^{-17}$ $cm^{-2}$ $molecule^{-1}$ and covers a wavelength range of 118 nm to 278 nm.

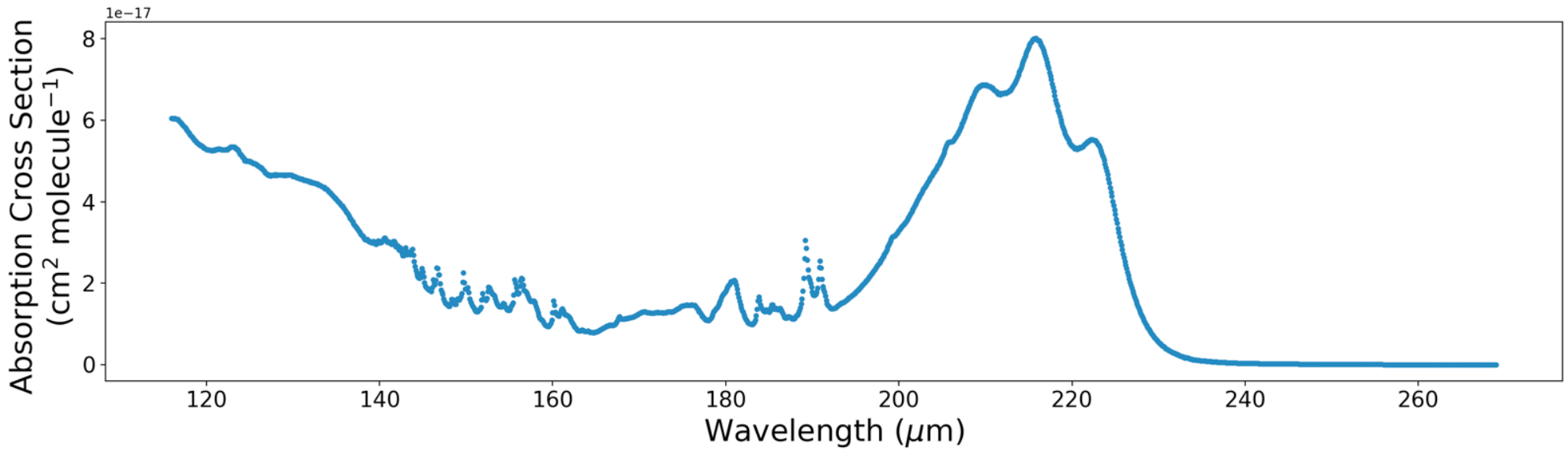

**Figure 7.** Isoprene UV-Vis absorption cross section taken from (Dillon *et al.* 2017). The axes shows absorption cross section [$cm^2$ $molecule^{-1}$] vs. wavelength [μm]. The peak absorption of isoprene UV-Vis is at 218 nm.

### 3.1.2 IR Cross Sections and Uncertainty Estimates

The isoprene IR absorption cross sections are shown in (Figure 8). The data are measured by (Brauer *et al.* 2014) and are collected and calibrated by the "HITRAN online Absorption Cross Sections Database" (Gordon et al. 2017).

The isoprene cross section datasets are measured at standard pressure for 278 K, 298 K, and 323 K with a 1/8 $cm^{-1}$ resolution. In this study, we opt to only use the 298 K data to assess the detectability of isoprene because it has the least uncertainties. More specifically, measurements at standard pressure and temperature do not require heating/cooling of the experimental setup, and will, therefore, have the least variation between the source and background reference spectra. As a side note, the methodology for the consideration of the noise floor differs between the 298 K data and the 278 K / 323 K data. As a result of this difference in treatment, it is not possible to reliably extrapolate the measured cross sections to temperatures beyond those measured. Including all three measurements with different noise floor treatments may introduce additional uncertainties to our models.

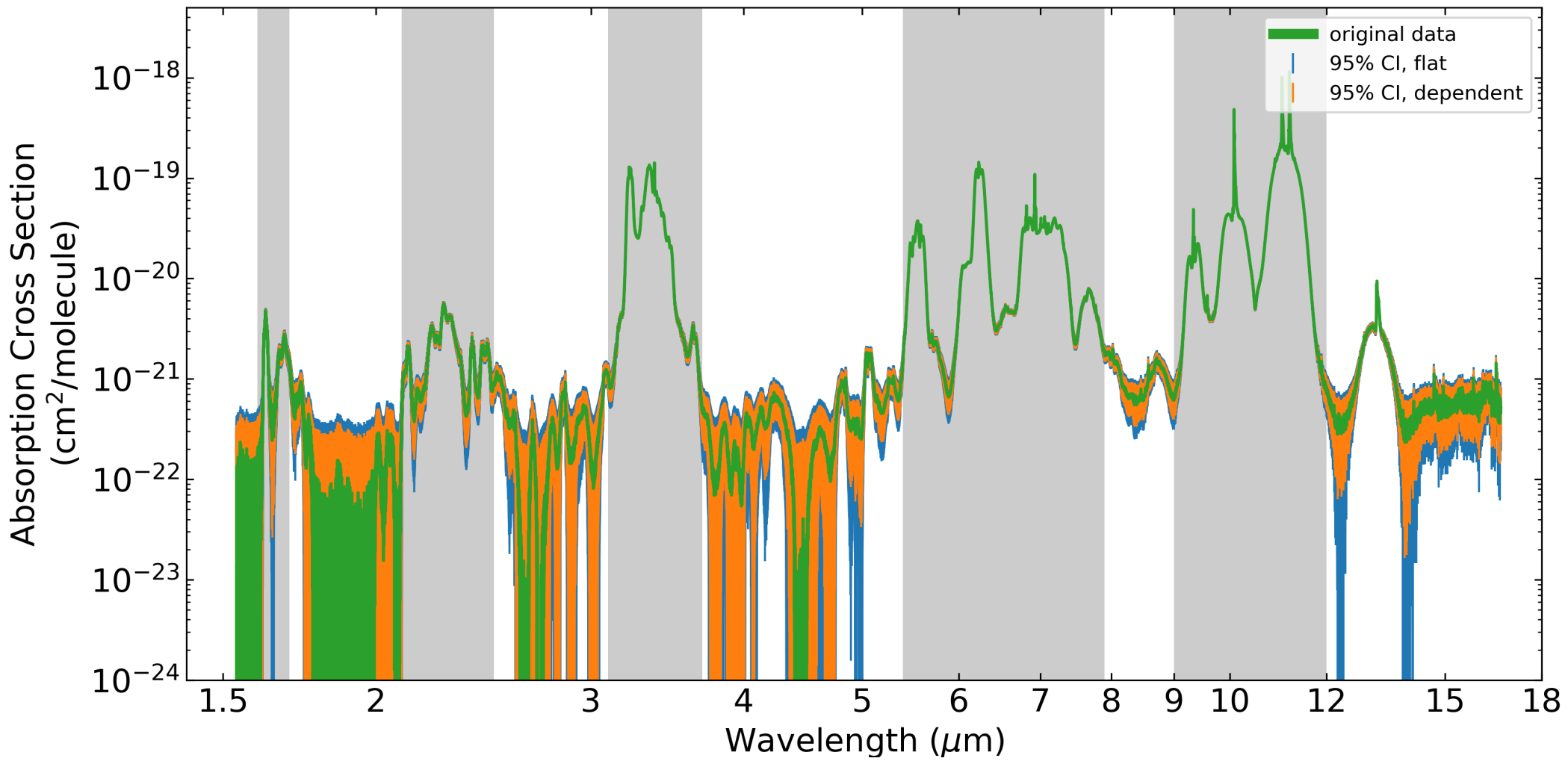

**Figure 8.** Isoprene high-resolution IR cross sections at standard pressure and temperature in log-log scale from 1.3 to 18 μm. Shown in green are isoprene cross sections, as collected by (Brauer *et al.* 2014) and calibrated by HITRAN (Gordon *et al.* 2017). Shown in blue are the isoprene cross sections with an uniform uncertainty of 3 x $10^{-22}$ $cm^2$ $molecule^{-1}$. Shown in orange is the estimated wavelength dependent uncertainty based on methods described in (Chu et al. 1999). The uncertainties represent the 95% confidence interval of the data. Marked in grey are the five regions of isoprene spectral features that we consider to evaluate the detectability of isoprene (1.6 - 1.7 μm, 2.1 - 2.5 μm, 3.1 - 3.7 μm, 5.4 - 7.9 μm and 9 - 12 μm). We omit assessment of the spectral features in other regions due to high uncertainties and omit features longer than 12 μm due to high instrumental noise from the JWST MIRI LRS Instrument (Batalha et al. 2017).

Unlike absorption cross sections calculated from molecular line lists, the uncertainties of absorption cross sections calculated from lab-measured transmission spectra cannot be ignored, especially for data points with opacities that approach the instrument noise-floor.

We estimated the wavelength-dependent uncertainties (95% confidence interval of the measured data) as described in equation 4 in (Chu *et al.* 1999) as follows:

$$U \approx 2 \cdot (Ba^2 + Ca + D)^{\frac{1}{2}} \qquad \text{Eq. 3}$$

where *U* is the expanded uncertainty, *a* is the absorption cross section, and *B*, *C*, *D* are coefficients unique for each molecule. The uncertainty U is the 95% confidence interval. We compare the wavelength-dependent uncertainties to a uniform 3 x $10^{-22}$ $cm^2$ $molecule^{-1}$ noise-floor as described in (Brauer *et al.* 2014) and approximately validate this method by finding the same averaged value. Therefore both methods are sufficient to identify which data points we can trust and which may be no different than noise, but in general, the uncertainty of the wavelength-dependent method scales with the cross section values (uncertainties for the largest peaks are larger than the uniformed uncertainty and uncertainties for the small peaks near noise floor are smaller than the uniformed uncertainty).

We note that the specific values of the *B*, *C*, and *D* coefficients for isoprene are not measured by (Chu *et al.* 1999) not provided in the original work ((Brauer *et al.* 2014) and are also missing in the NIST spectral database (Linstrom and Mallard 2001). We therefore adopt the values of $B = 1.6 \times 10^{-4}$, $C = 1.1 \times 10^{-9}$, $D = 2.7 \times 10^{-14}$ from $C_4H_6$ (1-3-butadiene). Although we expect this substitution to introduce some additional uncertainty, it is the most appropriate approximation; $C_4H_6$ is structurally similar to isoprene (2-Methyl-1,3-butadiene), though it has one less methyl- group.

### 3.1.3 Isoprene Spectral Features

Isoprene has 33 fundamental IR-active vibrational modes, associated with several functional groups containing carbon-carbon and carbon-hydrogen bonds. The fundamental vibrational modes of isoprene have previously been assigned from both measured and theoretically calculated spectra (Brauer *et al.* 2014; Panchenko and De Maré 2008).

We assess the detectability of isoprene in the context of JWST's observation capabilities (see Section 3.4 and Section 4.2). We have divided the isoprene spectral features into five wavelength regions: 1.6 - 1.7 μm, 2.1 - 2.5 μm, 3.1 - 3.7 μm, 5.4 - 7.9 μm and 9 - 12 μm. We omitted spectral features in other regions due to the high measurement error-bars and omitted spectral features above 12 μm due the high instrumental noise of JWST MIRI LRS beyond 12 μm (Batalha *et al.* 2017).

Isoprene cross section features in the 1.6 - 1.7 μm and 2.1 - 2.5 μm regions are formed by rovibrational overtones of the 3.1 - 3.7 μm region bands. Features in these two regions lack reliable experimental measurements (Brauer *et al.* 2014) and detectability of these two spectral features should be taken with some caution. This paper motivates future measurements and theoretical simulations of isoprene spectra in the visible and near-infrared as it would expand the assessment of isoprene detection using more readily available instruments that cover these spectral regions.

Isoprene spectral features in the 3.1 - 3.7 μm region are primarily comprised of the following two features: 1.) The narrow bands around 3.2 μm, which are composed of the $\nu_1$ and $\nu_2$ asymmetric stretching modes. 2) The broader bands (3.3 - 3.7 μm), which are composed of the $\nu_3$ (symmetric stretch), $\nu_4$ (asymmetric stretch) and $\nu_{24}$ (deformation) modes (Brauer *et al.* 2014). These two features arise from the stretching modes of X=C-H ($sp^2$ hybridized) and X-C-H ($sp^3$ hybridized), where X denotes another atom (other than H).

Isoprene spectral features in the 5.4 - 7.9 μm region are comprised of the following two features: 1) The narrow features around 6.5 μm, which are composed of the symmetric and asymmetric stretching modes ($\nu_8$ and $\nu_9$) associated with the C=C double bond; 2) The features in the 6.6 - 7.9 μm region that composed of the $\nu_{10-13}$ and $\nu_{24}$ modes associated with deformation and scissoring rovibrational motions (Brauer *et al.* 2014).

Isoprene spectral features in the 9 - 12 μm region are comprised of several bands associated with the wagging modes of the carbon-hydrogen functional groups ($\nu_{26}$, $\nu_{27}$, and $\nu_{28}$), and one associated with the rocking motion of the X=C-H functional group ($\nu_{17}$ mode) (Brauer *et al.* 2014).

### 3.1.4 Haze Extinction Cross section

We anticipate that abundance of isoprene, a hydrocarbon, in an atmosphere could lead to the presence of a haze layer in the atmosphere similar to the haze layer induced by organic molecules described in (Arney *et al.* 2018). The presence of a haze layer may hinder detection of isoprene spectral features and should be quantified.

Studying the effects of isoprene-induced haze requires wavelength-dependent refractive indices and haze particle size distribution, but neither are available for isoprene as studies regarding reactions of the isoprene-induced radicals (or products of isoprene reactions described in Section 2.3) are limited. Isoprene-induced haze on Earth does not have measurements in IR thus far, and in any case, the Earth's isoprene-induced haze is an oxygenated product not likely to be the same as isoprene-induced haze on an anoxic exoplanet. Therefore we estimate the isoprene-induced haze extinction cross section using wavelength-dependent refractive index measurements and haze particle size distributions of other hydrocarbons in an reducing environment.

We adopt the wavelength-dependent refractive indices of Titan's methane-induced haze (measured by (Khare *et al.* 1984), or tholins, as a proxy for that of isoprene-induced hazes. In our solar system, Titan (Khare *et al.* 1984), Pluto (Zhang *et al.* 2017) Venus have extensive haze layers in their atmosphere. The composition of Pluto's haze is yet to be confirmed. The Venusian haze is not known, but is likely to contain high concentrations of sulfur-containing molecules derived from $SO_2$ and $H_2SO_4$ (Takagi *et al.* 2019). Therefore Titan's atmospheric haze is the closest analog to isoprene haze on habitable exoplanets, both are hydrocarbon-induced haze. To show that our results are not specific to the case of Titan tholins, we compare our results with those using refractive indices of HCN (Khare *et al.* 1994), $C_2H_2$ (Dalzell and Sarofim 1969) and octane (Anderson 2000).

We approximated the mean size distribution as a Gaussian distribution with a mean particle size equal adopted from (He *et al.* 2018; Hörst *et al.* 2018)), who measured the diameters of haze particles for different temperatures and metallicities. For $CO_2$-dominated and $N_2$-dominated atmospheres, we used a mean size of 89 nm and a standard deviation of 25 nm, which is approximated from the 300 K, 1000x metallicity case from He *et al.* (2018). For $H_2$-dominated atmospheres, we used a mean size of 53.8 nm and a standard deviation of 16 nm, which is approximated from the 300 K, 10000x metallicity case from He *et al.* (2018). We choose to use a Gaussian distribution approximation rather than using the original measurement in order to avoid overfitting the data for isoprene.

Finally, we used miepython[5] to calculate isoprene-induced haze's extinction cross section with the assumed wavelength-dependent refractive indices and haze particle size distribution. The cross section is averaged from 1000 radii sampled from the Gaussian distribution. For simplicity, we assumed the haze particle to be spherical and we assume the mean size and size distribution is constant as a function of height.

## 3.2 Photochemistry Model

We use the photochemical model from (Hu *et al.* 2012) to calculate the concentration of isoprene in exoplanet atmospheres as a function of surface production flux. The code includes photolysis, reactions with radicals and molecules, dry deposition to the surface, and rainout as sinks of atmospheric gases. The code has been validated by computing the atmospheric composition of current Earth and Mars, matching observations of major trace gases in both atmospheres. The photochemical model by (Hu *et al.* 2012) have been used in a variety of papers (e.g., (Sousa-Silva *et al.* 2020)); we provide a brief summary description of our photochemical model here.

The (Hu *et al.* 2012) photochemical model computes the steady-state chemical composition of a planetary atmosphere scenario. We have adapted the model to include isoprene, including photolysis, rainout, and reactions with O·, $O_3$, and ·OH as sinks on isoprene. Due to the lack of reaction network studies for isoprene-induced radicals in anoxic environments, we assumed fractions of the photochemical products of isoprene result in haze formation while the rest is not tracked; we note that this assumption formally underestimates isoprene concentrations as haze will shield isoprene from UV photolysis. By contrast, our lack of a detailed reaction network means that we may omit chemical cycles by which destruction of one molecule of isoprene may lead to additional destructions; this may lead to overestimates of isoprene concentrations. More detailed characterization of the reactions of isoprene and its products are required to resolve this challenge. In section 4.3 we discuss how different assumed haze-to-isoprene mass fractions (from 1 ppm to 10%) will impact the effect of haze on the transmission spectra. We assume the mass fraction to be constant as a function of height.

The model handles over 800 chemical reactions (and photochemical reactions), formation and deposition for aerosols (including elemental sulfur and sulfuric acid); our exoplanet scenarios (Section 3.3) employ a subset of ~450 of the reactions, excluding primarily nitrogen chemistry and high-temperature reactions; see Hu et al. 2012 for the rationale for these choices. The model also treats dry and wet deposition, thermal escape, and surface emission. The model is flexible to simulate both oxidized and reduced conditions. The model uses delta-Eddington two-stream method to compute the ultraviolet and visible radiation in the atmosphere. The optical depth used

[5] Miepython (https://github.com/scottprahl/miepython) is a python implementation of fast extinction cross section calculation describe in (Wiscombe 1979). We note that the imagery part of the refractive index is taken as a negative number in miepython.

is calculated with molecular absorption, Rayleigh scattering, and aerosol Mie scattering.

The stellar UV spectral flux data is an input for the photochemistry code. We take the input stellar fluxes from (Seager *et al.* 2013). For the UV flux from a solar-type star, (Seager *et al.* 2013) use the Air Mass Zero (AM0) reference spectrum produced by the American Society for Testing and Materials (http://rredc.nrel.gov/solar/spectra/am0/) with the average quiet-Sun emission spectrum from (Curdt *et al.* 2004), in total covering the wavelength range from 110 nm – 10 μm. For the M dwarf star, we used the measured spectrum for GJ 1214 b from 115 - 300 nm (France *et al.* 2013), and the 3000 K NextGen model spectrum at > 300 nm (Allard *et al.* 1997).

## 3.3. Simulation Scenarios

We now describe our exoplanet benchmark scenarios. Our exoplanet benchmark scenarios are based on those in (Hu *et al.* 2012) and (Seager *et al.* 2013) (they are also used in (Sousa-Silva *et al.* 2020)). We discuss in detail below the 6 simulation scenarios considered here, representing $H_2$-dominated, $N_2$-dominated, and $CO_2$-dominated atmosphere, each exposed to a Sun-like star and an M dwarf star.

### 3.3.1 Astronomical Scenarios

We consider stellar irradiation environments corresponding to the modern Sun and an M5V star (“M dwarf star”) with a visual magnitude of 10.

The semi-major axes of the planets are taken to be 1.6 AU, 1.0 AU, and 1.3 AU if orbiting a Sun-like star, and 0.042 AU, 0.026 AU, 0.034 AU if orbiting an M dwarf star for $H_2$-dominated, $N_2$-dominated, and $CO_2$-dominated atmosphere, respectively. The semi-major axis are chosen to maintain surface temperature of 288 K (Hu *et al.* 2012).

We calculate photochemical models for an Earth analog planet (1 Earth-mass, 1 Earth-radius). We follow (Hu *et al.* 2012) in projecting these photochemical models to the scenario of a massive super-Earth with 10 Earth-mass and 1.75 Earth-radius, by assuming the molecular mixing ratio to be a function of pressure, which is invariant of surface gravity. The preference for the larger planet is beneficial for observation with near-future space telescopes for mass measurement with radial velocity and atmosphere characterization with both transmission (e.g. more likely to retain a reducing, $H_2$-dominated atmosphere) and secondary eclipse spectroscopy (e.g. higher planet/star flux ratio).

### 3.3.2 Atmospheric Scenarios

We consider three different atmosphere scenarios according to their redox state: a reducing $H_2$-dominated atmosphere, an intermediate redox state $N_2$-dominated atmosphere, and a weakly oxidizing $CO_2$-dominated atmosphere. We only consider anoxic atmospheres because it is already well known from Earth's atmosphere that isoprene cannot accumulate in an $O_2$-dominated atmosphere. The exact composition and vertical mixing ratio profile for the starting atmosphere scenarios which we use as seeding conditions for calculating vertical mixing ratio profiles with varying surface flux are shown in Appendix I, (and originally come from (Sousa-Silva *et al.* 2020)) for details on the $H_2$-dominated and $CO_2$-dominated atmosphere scenarios. The physical concept behind the $H_2$-dominated atmosphere scenario is that the planet outgassed $H_2$ during planet formation and managed to either maintain its $H_2$ atmosphere or has interior reservoirs with planetary processes to replenish it.

| Atmosphere Archetypes | Stellar Type | Main Composition |
|---|---|---|
| $H_2$-dominated | Ma | H, **$H_2$**, O, $CH_4$, **$H_2O$**, **$N_2$**, **CO**, $O_2$, **$CO_2$** |
| | Sun | H, **$H_2$**, O, $CH_4$, **$H_2O$, $N_2$**, **CO**, $O_2$, **$CO_2$**, C, $CH_3$, $C_2H_2$, $C_2H_6$, |
| $N_2$-dominated | Ma | H, **$H_2$**, **O**, $CH_4$, **$H_2O$, $N_2$**, **CO**, $O_2$, **$CO_2$** |
| | Sun | H, **$H_2$**, **O**, $CH_4$, **$H_2O$, $N_2$**, CO, $O_2$, **$CO_2$**, C |
| $CO_2$-dominated | Ma | H, **$H_2$**, **O**, $CH_4$, **$H_2O$, $N_2$**, **CO**, **$O_2$**, **$CO_2$**, $H_2O_2$, $O_3$ |
| | Sun | H, **$H_2$, O**, $CH_4$, **$H_2O$**, **$N_2$**, **CO**, **$O_2$**, **$CO_2$**, |

**Table 1**: Atmosphere composition adopted from photochemistry output of (Hu et al. 2012) model for the 6 simulation scenarios considered. For spectroscopy calculation and detection assessment, we only consider molecules that have reached a local mixing ratio of at least 100 ppb at any height in the atmosphere. Molecules that fail to meet this mixing ratio criteria are unlikely to contribute sufficient opacity to the simulated transmission and emission spectra and are therefore not included. Molecules more than 100 ppm at any height are marked in bold. For detailed mixing ratio profile, see (Appendix 1, Figure I-1). Addition of isoprene may drastically change the mixing

### 3.3.3 Atmosphere Temperature, Pressure and Abundances

Using the six starting simulation scenarios (three atmosphere scenarios for the Sun-like star and three for the M dwarf star) as seeds, we calculate the mixing ratio profile of isoprene with varying surface flux using the photochemistry code from (Hu *et al.* 2012) for each scenario. The vertical mixing strength (Eddy diffusion profile) of an atmosphere archetype is linearly scaled from the ratio of Earth atmosphere's mean molecular mass and the atmosphere archetype. We adopt the eddy diffusion coefficients scaling factors of 6.3, 1.0, and 0.68 for the $H_2$, $N_2$ and $CO_2$ dominated atmospheres from (Hu *et al.* 2012), which is based on the scale height of the atmosphere archetypes.

The surface pressure of the planet is set to 1 bar, and the surface temperature is set to 300 K. In the troposphere, the temperature decrease with altitude based on a dry adiabatic lapse rate. The tropopause is set to 160 K for the $H_2$-dominated atmospheres, 180 K for the $CO_2$-dominated atmospheres, and 200 K for the $N_2$-dominated atmospheres. The different tropopause temperatures used for each atmosphere reflect the different efficiencies of gases as coolants in the upper atmosphere. $H_2$ is a more effective coolant than $N_2$, so the stratosphere of $H_2$-dominated atmospheres will be colder than $N_2$-dominated atmospheres.

Temperatures above the troposphere are set to be constant, i.e. no temperature inversion (See Appendix I for the temperature-pressure profiles for the three-atmosphere archetypes). This assumption may be violated for hazy atmospheres, where absorption of UV photons by high-altitude haze may drive an inversion, analogous to ozone in modern Earth's atmosphere. We expect our results to be relatively insensitive to the stratospheric temperature, since the ultimate limit on accumulation of isoprene comes from UV photolysis. Nevertheless, we performed a sensitivity test of the resulting transmission spectra by replacing the temperature-pressure profile with that of Earth's temperature-pressure profile. We find that there is negligible difference for isoprene as a trace gas (Section 4.1.1) because isoprene only accumulate in the lower atmosphere and isoprene in a run-away phase (Section 4.1.3) because isoprene is close-to uniformly distributed in the atmosphere.

We exclude trace gases that do not exceed 100 ppb at any altitude from our spectral model because their absorption does not contribute enough variation to the atmosphere spectra and is unlikely to be detectable without significant instrumentation advancements capable of reaching a < 1 ppm noise floor.

## 3.4. Atmosphere Spectra Simulation

To assess the detection of trace gases and biosignature gases in exoplanet atmosphere scenarios, we constructed the "Simulated Exoplanet Atmosphere Spectra" (SEAS) model to simulate transmission and secondary eclipse thermal emission spectroscopy following the principles described in (Seager *et al.* 2013). The methods for calculating spectra are similar to those described in both (Kempton *et al.* 2017; Miller-Ricci *et al.* 2009). SEAS accepts user-input temperature-pressure profiles and mixing ratio profiles; the mixing ratio profiles are especially important for the study of super-Earth atmospheres as they are severely impacted by photochemistry and atmosphere chemistry.

The molecular cross sections used by SEAS are interpolated from a pre-generated grid of cross sections for a grid of pressure ($10^5$ Pa to $10^{-2}$ Pa in multiples of 10) and temperature (150 K to 400 K in steps of 25 K). For molecules that have line lists from (Gordon *et al.* 2017), the cross sections are calculated using the HAPI package (Kochanov *et al.* 2016). For molecules that have line lists in ExoMol (Gordon *et al.* 2017; Tennyson *et al.* 2016), cross sections are calculated using the ExoCross package (Yurchenko *et al.* 2018).

We validated SEAS by generating Earth spectra and comparing transmission spectra results with data from the Atmospheric Chemistry Experiment data set (Bernath *et al.* 2005) and comparing emission spectra with results from MODTRAN spectrum simulation code (Berk *et al.* 1998).

### 3.4.1 Transmission Spectra

The SEAS transmission spectrum code calculates the radiative transfer of stellar radiation passing through each layer of the transiting planet atmosphere. Below we detailed the exact step to calculate the effective height of the atmosphere and the transit depth for each wavelength.

1. We defined each layer of the atmosphere with a height of 1 scale height starting from the surface to the top of the atmosphere and assume local thermodynamic equilibrium within each layer of the atmosphere.
2. Since the atmosphere is taken to be homogenous within each layer, we approximate the 3D spherical shell as a 2D ring.
3. Since each layer is curved and stellar radiation is radial, the stellar radiation along the limb path of each layer will penetrate sections of the current layer and sections of layers above the current layer. Therefore the optical depth of each layer is the sum of the optical depth through each section as follow:

$$A(\lambda)_i = \sum_{i}^{z} \sum_{j=0}^{m} n_{i,j}(P,T) \cdot \sigma_{i,j}(P,T,\lambda) \cdot l_i \qquad \text{Eq. 4}$$

   where $A(\lambda)_i$ is the total wavelength-dependent absorption for the *i* th layer, *j* denotes each molecule, *n* is the number density, $\sigma(\lambda)$ is the wavelength-dependent absorption cross section, *P* is the pressure, *T* is the temperature, $l$ is the path length, *z* denotes the total number of layers above the *i* th layer which the stellar radiation passes through and *m* denotes the total number of molecules.
4. The transmission by each layer is calculated using Beer-Lambert's Law: $T = e^{-A}$, and absorbance is *1-T*.
5. The effective height of the atmosphere is calculated by summing the effective height of each layer, which is calculated by multiplying the absorbance by the scale height of each layer.
6. The transit depth of the atmosphere is calculated by summing the flux attenuated by each layer of the atmosphere, which is calculated by multiplying the absorbance by the cross sectional area of each layer.

We note that the cross sections of haze particles have units of [ $cm^2$ $particle^{-1}$] and therefore the particle density is calculated from the gas number density and has a unit of [particle $cm^{-3}$]. The particle density of isoprene-induced haze at a given layer is calculated by dividing the particle vapor density with the average particle mass. The particle vapor density is the product of the air density, the mixing ratio of isoprene and an assumed haze-to-isoprene mass fraction (See Section 3.2).

To consider detection of isoprene via transmission spectra, we first assessed model scenarios in which the atmospheres have no clouds or haze. This result represents an upper bound on the detectability of the isoprene features in transmission (Section 4.2). Next we study how detection will be hindered if haze is included (Section 4.3).

### 3.4.2 Secondary Eclipse Thermal Emission Spectroscopy

The SEAS thermal emission code is similar to those described in (Seager 2010) and Sousa-Silva et al., (2020) and uses the same temperature-pressure profiles, mixing ratio profiles and molecular cross sections as in the transmission code.

The emission code integrates the blackbody radiation for each wavelength from the surface and up through each layer of the atmosphere, as the radiation is absorbed and reemitted by gases in each layer. The surface is set to be a pure black body and the top layer is set to be transparent. We do not consider scattering in the current emission code. The final spectrum is calculated by integrating the emerging flux by the cross sectional area of the planet. To add the presence of clouds, we consider 50% cloud coverage by averaging between a cloudy and cloud-free spectrum.

## 3.5 Simulated Exoplanet Observation

We simulate observations of the six simulation scenarios as described in (Section 3.3.2) with varying amounts of isoprene as computed by our photochemistry model in (Section 3.2). We used the astronomical parameters defined in (Section 3.3.1) and a 10 $M_{Earth}$,1.75 $R_{Earth}$-planet transiting a star with an K-band apparent magnitude of 10 (JWST observes in near-mid IR). The star can either be 1) a Sun-like star 2) a 3000 K, 0.26-$R_{sun}$ M dwarf star.

Since Isoprene has many broad spectral features spanning a wide range of wavelength as discussed in (Section 3.1.2), we opt to assess detection of isoprene using JWST's NIRSpec (G140M, G235M, G395M) and MIRI (LRS) observation modes.

For transmission spectroscopy, we combined the simulated spectra from SEAS and observational noise simulated using Pandexo (Batalha *et al.* 2017), the community JWST exposure time calculator and noise simulator. To account for potential unknowns, we added a 10 ppm noise floor as suggested by (Batalha *et al.* 2017).

For secondary eclipse thermal emission spectroscopy, we approximate our telescope specifications based on JWST and the use of the NIRSpec and MIRI instruments (Bagnasco *et al.* 2007; Wright *et al.* 2010), or a 6.5 m space telescope with a quantum efficiency of 25%. Since stellar flux is the source of the noise, we do not model instrumental noise and instead used a 50% photon noise multiplier. Finally, we compute the SNR for each bin using the equation below:

$$SNR = \frac{|F_{out} - F_{in}|}{\sqrt{\sigma^2_{F_{out}} + \sigma^2_{F_{in}}}} \qquad \text{Eq. 5}$$

where $F_{in}$ is the flux density within the absorption feature, $F_{out}$ is the flux density of the surrounding continuum of the feature, and $\sigma F_{out}$ and $\sigma F_{in}$ is the respective uncertainty.

While we have estimated instrumental noise, our estimates on stellar noise remain rudimentary. We raise the caveat that additional sources of astrophysical noise may make detecting molecular features challenging. Other astrophysical phenomenon, such as non-homogeneity of stellar disks, could introduce false spectral features from 0.3 - 5.5 μm with intensities around 70 ppm for M5V stars (Rackham *et al.* 2018), potentially hinder detection of biosignature gases.

## 3.6 Biosignature Gas Detectability Assessment

Determining the detectability of a biosignature gas in our atmosphere scenarios is expanded based on the methods defined in (Seager *et al.* 2013; Sousa-Silva *et al.* 2020; Tessenyi *et al.* 2013), where we assess if the spectral features of the target biosignature gas can be identified within a reasonable number of transit (secondary eclipse) observations.

First we assess whether or not the atmosphere (and the biosignature gas) can be detected by applying a null-hypothesis test. More specifically, we assess whether or not the simulated wavelength-dependent transit depth data can be explained with a straight line (transmission) or with a blackbody radiation curve (emission). If so, then the simulated observation cannot pass the null-hypothesis test and we deem the atmosphere scenario to be not-detectable.

Next, if the simulated observation for the atmosphere scenario passed the null hypothesis, we then compare the goodness-of-fit of a model atmosphere that contains the biosignature gas and a model atmosphere without the biosignature gas. The goodness-of-fit is computed using the reduced chi-square statistic using the following equation:

$$x_v^2 = \frac{x^2}{v} = \frac{1}{v} \cdot \left[ \sum_i \frac{(O_i - C_i)^2}{\sigma_i^2} \right] \qquad \text{Eq. 6}$$

where $\chi_v^2$ is the reduced chi square, $v$ is the degree of freedom (or the number of wavelength bins), $\chi^2$ is the chi squared, $O_i$ is simulated observational data, $C_i$ is the simulated model, $\sigma_i$ is the variance (or error as calculated from Pandexo noise simulator for a specific instrument), and finally $i$ denotes each wavelength bin. We note that binning the spectra reduces the variance at the expense of reducing the degree of freedom.

The simulated observation data has moderate spectral resolution (R > 1000 for NIRSpec and R = 160 for MIRI) but also has large error bars for each data point. Since we are interested in detecting the broad spectral features of isoprene and not the narrow, detailed individual features, we can trade resolution for increased SNR to improve our model's goodness-of-fit. R = 10 - 20 is where isoprene becomes indistinguishable with methane for features at shorter wavelengths ( < 4 μm) and this concept will be further explored in Section 4.3.

Finally, we repeat calculation of the above detection metric by binning down our simulated spectra to spectral resolutions of R = 10, 20, 50, 100 to find most optimal choice and iterate from 1 to 100 transits until detection is reached. If no detection is found with any spectral resolution and 100 transits (theoretical upper limit for a planet in the habitable zone of an M dwarf star given JWST's expected cryogenic lifetime of 5 years), we deem the spectral features not detectable.

## 4. Results

Despite its promising potential, isoprene does not satisfy all criteria to be a good biosignature gas. Namely, isoprene is unable to accumulate in the upper atmosphere at even 10 times Earth's production level, and in fact isoprene must enter a "run-away phase" to accumulate to detectable abundances. In addition, isoprene can be spectrally indistinguishable from methane at short wavelengths (< 4 μm) with JWST's spectral resolution. Regardless, isoprene should still be added into the roster of biosignature gas to consider in future studies because isoprene has no abiotic false-positives and has the potential to be produced in large quantities as it is by life on Earth.

We demonstrate that with Earth's production rate, isoprene molecules are predominantly concentrated near the surface (< 10 km) (Section 4.1.1) making detection impossible with near future technologies (Section 4.2.1). An additional key point is that with a production rate 100 to 1000 times higher than Earth's isoprene production rate—which we assert is challenging but within reasonable estimates (Section 4.1.2) —isoprene can enter a run-away phase, enabling isoprene molecules to populate the upper atmosphere at significant concentration (> 100 ppm) (Section 4.1.3) and become detectable (Section 4.2.2).

Unfortunately, (within the context of observing the atmosphere of a super-Earth with JWST) despite its detectability in a run-away phase, we show that isoprene's spectral features can be confused with that of methane and other hydrocarbons More specifically, isoprene spectral features at 3.1 - 3.7 μm overlap with that of methane. Moreover, isoprene's spectral features at 9 - 12 μm lie in a wavelength region populated by hundreds of other hydrocarbon gases (Section 4.2.3). On Earth any isoprene is immediately converted into haze, so we also discuss the impact of haze on hindering detection of isoprene (Section 4.2.4).

Far-future telescopes that can achieve detection at higher spectral resolution than JWST may make detection of isoprene possible. Therefore, given the abundant chemical reactions involving isoprene in known biochemistry and the fact that it does not have any abiotic false positives (Section 4.3), it would be hasty to discard isoprene as a potential biosignature gas.

## 4.1 Isoprene Accumulation

We calculated the column-averaged mixing ratio profile of isoprene and other gases as a function of surface isoprene flux, using the photochemistry code described in (Section 3.3). We list the column-averaged mixing ratio for isoprene given the surface production rate for each simulation scenario in Table 2.

To assess isoprene's ability to accumulate in an atmosphere, we first examine the distribution of isoprene in an exoplanet atmosphere given a production rate similar to that on Earth, ~ 3 × $10^{10}$ molecules $cm^{-2}$ $s^{-1}$ (Section 4.1.1). Next, to go beyond Earth's conditions, we calculate isoprene mixing ratio profile for a range of isoprene surface production rates. We vary the production rate from $10^3$ molecules $cm^{-2}$ $s^{-1}$ to $10^{15}$ molecules $cm^{-2}$ $s^{-1}$ in steps of 10. We chose the maximum of $10^{15}$ molecules $cm^{-2}$ $s^{-1}$ to represent the highest isoprene production rate found in a niche environment on Earth (Section 4.1.2)

| Atmospheric Scenario | $H_2$-dominated | | $N_2$-dominated | | $CO_2$-dominated | |
|---|---|---|---|---|---|---|
| $P_{iso}$ / Stellar Type | M dwarf | Sun like | M dwarf | Sun like | M dwarf | Sun like |
| $10^{10}$ | 0.0006 | / | 0.0002 | / | 0.0016 | / |
| $10^{11}$ | 0.050 | / | 0.0079 | / | 6.22 | / |
| $10^{12}$ | 3.30 | / | 1.98 | / | 635.81 | / |
| $10^{13}$ | 865.41 | 0.001 | 4426.8 | 0.0001 | 7341.80 | 0.075 |
| $10^{14}$ | / | 0.040 | / | / | / | 1.84 |
| $10^{15}$ | / | 24.95 | / | / | / | 46.78 |

**Table 2**: Isoprene column-averaged mixing ratios (in units of ppm) corresponding to various isoprene surface fluxes (i.e. production rates) for our six atmosphere archetypes. For reference, biological production of isoprene on Earth is approximately 500 Tg $yr^{-1}$, or 2.7 × $10^{10}$ molecule $cm^{-2}$ $s^{-1}$ and biological production of $CH_4$ on Earth is also about 500 Tg $yr^{-1}$, or 1.2 × $10^{11}$ molecule $cm^{-2}$ $s^{-1}$. Our photochemistry model simulated equilibrium atmospheric abundances for a range of surface fluxes from $10^3$ to $10^{15}$ molecules $cm^{-2}$ $s^{-1}$. Fluxes < $10^{10}$ and < $10^{13}$ molecules $cm^{-2}$ $s^{-1}$ are omitted for M dwarf star and Sun-like stars respectively because the resulting column-averaged mixing ratios are negligible (< 1 ppb). Fluxes > $10^{13}$ molecules $cm^{-2}$ $s^{-1}$ are omitted for M dwarf stars because isoprene reaches a run-away phase and to exceed 1% of the atmosphere by volume, a likely unrealistic value. Omitted entries are denoted by the "/" symbol. The column-averaged mixing ratio quickly transitions from < 1 ppm to > 100 ppm around $10^{12}$ - $10^{13}$ molecules $cm^{-2}$ $s^{-1}$ surface flux values.

### 4.1.1 Isoprene Remains a Trace Gas at Earth's Production Rate

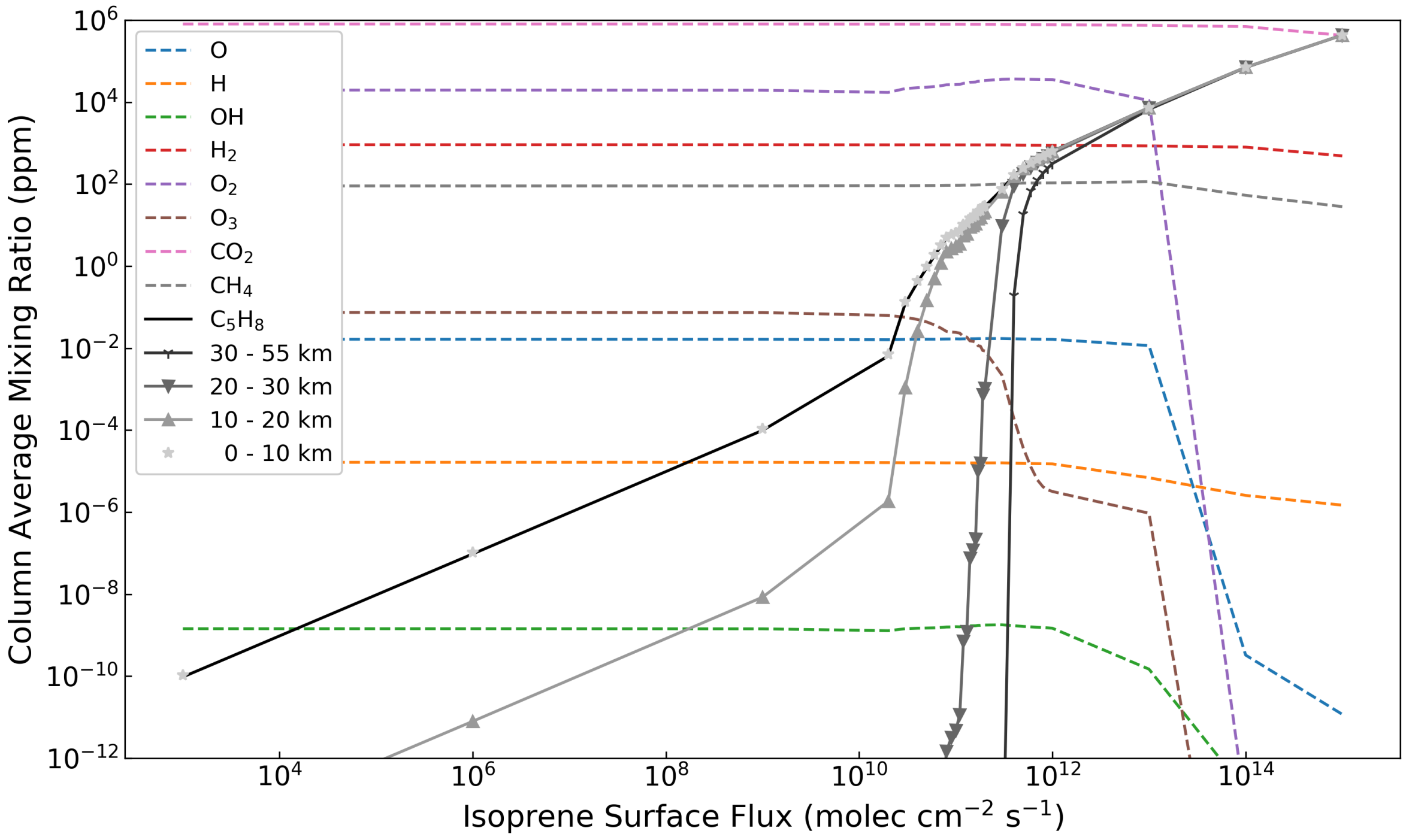

**Figure 9.** Column-averaged mixing ratios of isoprene and other major atmospheric gases in a $CO_2$-dominant atmosphere orbiting an M dwarf star as a function of isoprene surface flux. Dominant atmosphere species and isoprene-reacting radicals are plotted in various colored dash lines. The isoprene column-averaged mixing ratio (in units of ppm) for different isoprene surface fluxes (i.e. biological production rate) (in units of molecules cm$^{-2}$ s$^{-1}$) are shown by a solid black curve. The abundance of isoprene at 0 - 10 km from the surface overlap with the solid black curve and is additionally indicated by the light grey stars. The abundances of isoprene at 10 - 20 km, 20 - 30 km, and > 30 km from the surface are shown in different shades of grey and are additionally shown by different types of triangles. For low surface fluxes, isoprene remains a trace gas throughout the atmosphere (< ppm levels) with abundances increasing linearly with surface fluxes. For surface fluxes above $3 \times 10^{10}$ molecules cm$^{-2}$ s$^{-1}$, isoprene abundance rapidly increases. For isoprene surface fluxes above $3 \times 10^{11}$ molecules cm$^{-2}$ s$^{-1}$, isoprene abundance at the upper atmosphere (where most transmission spectral features originate) reach the same level as surface abundance.

At Earth's isoprene production rate of $3 \times 10^{10}$ molecules cm$^{-2}$ s$^{-1}$, isoprene remains a trace gas in all of three exoplanet atmosphere scenarios, at less than 1 ppb (column-averaged mixing ratio) (Table 2). At surface fluxes lower than $1 \times 10^{11}$ molecules cm$^{-2}$ s$^{-1}$, isoprene is concentrated near the surface where it is created. Any isoprene that diffuses to the upper atmosphere is readily destroyed. To illustrate this finding, the column-averaged mixing ratio of isoprene at four different altitudes for a $CO_2$-dominated atmosphere of an exoplanet transiting a M dwarf star is shown in Figure 9.

Isoprene may remain a trace gas even at higher production rates than on Earth, because large isoprene sinks may exist on planets different from Earth. The sinks could realistically include: life that has evolved to consume the abundant isoprene; photochemical destruction pathways as yet unknown in anoxic atmospheres; and/or higher deposition rates than those exist on Earth.

As a trace gas, isoprene is not detectable via transmission spectroscopy even with far-future space telescopes. In (Section 4.2.1), we explore the potential to detect isoprene as a trace gas via thermal emission spectroscopy using JWST.

### 4.1.2 Maximum Isoprene Production Estimate

For isoprene to accumulate to higher levels than a trace gas, it must be produced by life at rates hundreds if not thousand times the global production rate on Earth (Figure 9). In this section we establish that high isoprene production niche environments exist on Earth, up to one million times Earth's globally-averaged isoprene production rate (Section 2.2).

One such niche is a modern tropical environment, where isoprene production is optimal for trees, the main producer of isoprene on Earth due to their widespread abundance. In the Amazon rainforest, African rainforest, and Southeast Asia, the isoprene production rate averages > 1 mg $m^{-2}$ $s^{-1}$ with core areas averaging > 10 mg $m^{-2}$ $s^{-1}$ (McFiggans et al 2019). In comparison, the global averaged isoprene production rate on Earth is ~ $3 \times 10^{-5}$ mg $m^{-2}$ $s^{-1}$ (converted from 500 Tg $yr^{-1}$, or ~ $3 \times 10^{10}$ molecules $cm^{-2}$ $s^{-1}$). For 12% of Earth's habitable history (e.g., during Phanerozoic eon (The last 541 million years), Earth had a pole-to-pole tropical climate (e.g., during Carnian Pluvial Event (Dal Corso *et al.* 2012; Royer *et al.* 2004)). Therefore, in a hypothetical scenario where Earth's total land mass (~ 30 % of total surface area) is completely filled with isoprene producers and attains the 1 mg $m^{-2}$ production rate, it is possible for global average to reach 0.3 mg $m^{-2}$ $s^{-1}$, or $3 \times 10^{14}$ molecules $cm^{-2}$ $s^{-1}$. However, while trees are the main producer of isoprene on Earth, we cannot ignore the fact that trees are also the main producer of oxygen and the presumed condition for isoprene to accumulate is an anoxic atmosphere.

For a purely anoxic environment, we assume an Archean biosphere that is comprised of anaerobic, isoprene-producing prokaryotes such as those found in lab studies (Fall *et al.* 1998), see Section 2.2 for details), which is capable of naturally producing isoprene in high quantities with average production rates of 50 nmol $g^{-1}$ $hr^{-1}$. Using an average bacteria density of 1 g $cm^{-3}$ (Loferer-Krößbacher *et al.* 1998) and assuming a global biomass layer of 1 cm thick covering all of Earth's total land mass (~ 30 % total surface area), it is possible for global average to reach $2.5 \times 10^{13}$ molecules $cm^{-2}$ $s^{-1}$.

With these assumptions, we show that the high isoprene production rates required for isoprene to accumulate in the upper atmosphere can be supported by species on Earth and the theoretical upper limit to isoprene production rate is $10^4$ that of Earth's current production rate. Therefore, it is plausible to explore detection of isoprene under these conditions.

### 4.1.3 Isoprene Run-away

Very high surface fluxes of isoprene will send isoprene accumulation into a run-away state (see Figure 9). In a run-away phase isoprene rapidly accumulates in the upper atmosphere to high levels, up to hundreds of ppm. The run-away is a result of "photochemical self-shielding" whereby the isoprene production flux saturates its UV-driven sinks, resulting in a dramatic increase in lifetime and hence accumulation. This run-away phenomena is has been discussed for abiotic CO (Kasting 2014; Kasting *et al.* 2014; Kasting *et al.* 1984; Kasting *et al.* 1983; Zahnle 1986), has been alluded to for CH4 (Segura et al. 2005), and has been observed by us for other biosignature gases (Huang et al., in prep, (Sousa-Silva *et al.* 2020). We plan a detailed study in Ranjan et  al. (in prep). In this subsection, we consider the case where isoprene has entered a run-away phase, a scenario in which the abundance of isoprene in the upper atmosphere reaches a similar level to the abundance of isoprene in the lower atmosphere.

The run-away phase is important because it shows that cases exist in which isoprene can populate the upper exoplanet atmosphere such that it can be detected via transmission spectroscopy in simulations of the JWST. Recall that transmission spectra are only sensitive to the upper atmosphere. The lower atmosphere has extremely long path lengths for light to travel through it, making the atmosphere optically thick, which in turn results in featureless spectra.

The most favorable atmosphere scenario for isoprene accumulation is a $CO_2$-dominated atmosphere, because $CO_2$ shields isoprene more strongly from UV irradiation than $N_2$-dominated atmosphere or $H_2$-dominated atmosphere do.

The run-away effect is highly dependent on the quantity of UV flux from the host star. UV fluxes from Sun-like stars are significantly (more than 1000 times) higher than UV fluxes from M dwarf stars. Isoprene is unlikely to enter the run-away phase for planets orbiting Sun-like stars. There, the production rate required is around $3 \times 10^{14}$ molecules $cm^{-2}$ $s^{-1}$ for a $CO_2$-dominated atmosphere, approaching the maximum isoprene production rate even in niche environments on Earth. In contrast, for planets orbiting M-dwarf stars, isoprene's transition to the run-away phase occurs around $3 \times 10^{11}$ molecules $cm^{-2}$ $s^{-1}$ for a $CO_2$-dominated atmosphere and $1 \times 10^{12}$ molecules $cm^{-2}$ $s^{-1}$ for a $H_2$-dominated or $N_2$-dominated atmosphere. The surface flux required to enter the run-away phase is within 1 - 2 orders of magnitude of Earth's globally-averaged surface isoprene flux of $2.7 \times 10^{10}$ molecules $cm^{-2}$ $s^{-1}$. With isoprene surface fluxes above $3 \times 10^{11}$ molecules $cm^{-2}$ $s^{-1}$, but below $3 \times 10^{12}$ molecules $cm^{-2}$ $s^{-1}$, the corresponding atmosphere volume mixing ratio is 100 ppm or greater, and isoprene can accumulate in the upper atmosphere. At isoprene surface fluxes above $3 \times 10^{12}$ molecules $cm^{-2}$ $s^{-1}$, the corresponding atmosphere column-averaged volume mixing ratio is 1000 ppm (0.1 %) or greater. There are sufficient isoprene molecules in the atmosphere to balance photochemical destruction, thus allowing isoprene molecules to diffuse to the upper atmosphere; the resulting mixing ratio profile is well mixed such that isoprene has a constant mixing ratio up to 50 km above the surface.

Therefore, if life produces enough isoprene to enter a run-away phase, isoprene can accumulate to become a major atmospheric gas. In this case, life will have re-engineered the atmosphere, reminiscent of cyanobacteria's oxygenation of Earth's atmosphere. We note that the run-away hypothesis ignores potential unknown chemical or surface sinks in anoxic atmospheres that would limit the accumulation of isoprene in the atmosphere. Therefore realistic situations might require further investigation (see e.g., Ranjan *et al.*, in prep.).

## 4.2. Isoprene Detectability in Exoplanet Atmospheres

Isoprene detectability can be separated into two categories. The first category is where isoprene does not enter a run-away phase and remains a trace gas (i.e., does not accumulate above a column-averaged mixing ratio of 1 ppm). The second category is where isoprene is a major atmospheric gas, resulting from its production by life at high enough levels that isoprene enters a run-away phase (i.e., isoprene accumulates above a column-averaged mixing ratio of 100 ppm). The transition from a column averaged mixing ratio of 1 to 100 ppm occurs rapidly as a function of surface flux (Figure 9) so we omit discussion of this transition phase.

### 4.2.1 Detecting Isoprene as a Trace Gas is Challenging

As a trace gas, isoprene molecules are concentrated near the surface. We did not find any transmission spectra scenario in which isoprene can accumulate above the troposphere for a surface flux below $1 \times 10^{12}$ molecules $cm^{-2}$ $s^{-1}$. Our spectra simulations confirmed that the isoprene spectral features are less than 10 ppm in transit depth, smaller than JWST's assumed noise floor. Therefore it is not possible to detect isoprene as a trace gas via transmission spectroscopy.

We additionally explore whether isoprene can be detected via secondary eclipse thermal emission spectroscopy (emission spectroscopy) for planets transiting a M dwarf star. In emission spectroscopy, spectral features scale with the temperature gradient and in general detection might be more promising than for transmission spectroscopy.

For the terrestrial exoplanet atmosphere scenarios we considered in this study, the largest change in temperature occurs in the lower atmosphere layers, from the planet surface to "tropopause". Therefore, in scenarios where isoprene is a trace gas and concentrate near the surface, it is worth investigating detection via emission spectroscopy.

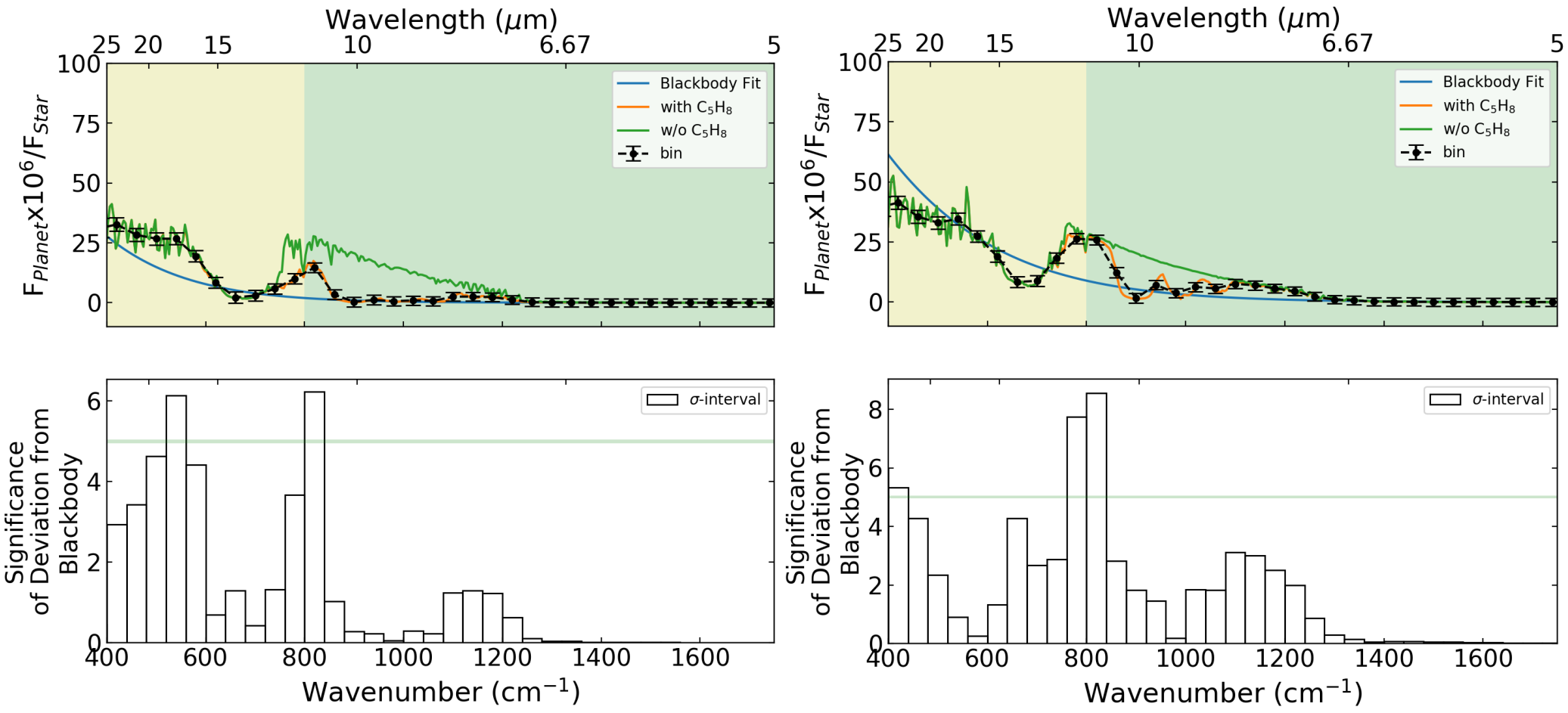

**Figure 10:** Simulated secondary eclipse thermal emission spectra for a $H_2$-dominated (left) and $N_2$-dominated (right) atmosphere of an exoplanet transiting an M dwarf star with an isoprene surface flux of 1 × $10^{12}$ molecule $cm^{-2}$ $s^{-1}$. The simulated atmosphere uses input and parameters as listed in (Section 3). We show the planet-to-star flux ratio vs wavelength in μm (top) and the statistical significance of a modeled atmosphere vs wavenumber in $cm^{-1}$ (bottom). The horizontal axes are applied to both the top and bottom panel. In the top panel, we show the simulated atmospheres with and without isoprene as represented by the green and orange curves, respectively. The best fit blackbody curves are shown in blue. Simulated observations of atmospheres with isoprene are represented by the black error bars. **In the bottom panel**, we show the statistical significance of a simulated atmosphere with isoprene as compared to the black body fit, in units of σ-interval. The green line represents the 5-σ statistical significance threshold.

Since isoprene accumulates best in $CO_2$-dominated atmospheres, we modelled this case. We found that in a $CO_2$-dominated atmospheres, secondary eclipse detection for a planet transiting an M dwarf star is possible given a surface flux of 1 × $10^{11}$ molecules $cm^{-2}$ $s^{-1}$, which is three times that of Earth's isoprene surface flux (Figure 10). Detection of isoprene in $H_2$- and $N_2$-dominated atmospheres is possible given a surface flux of 1 × $10^{12}$ molecules $cm^{-2}$ $s^{-1}$. For a planet with a habitable surface temperature of ~ 300 K, the peak thermal emission is between 10 - 15 μm. There is only one very broad spectral feature of isoprene that lies in this spectral region, between 9 - 12 μm.

Unfortunately, given a spectral resolution of R ~ 10 - 20, although we could detect isoprene, it would be hard to distinguish isoprene from other molecules that could be absorbing in this spectral region. Future 30-meter diameter aperture ground-based telescopes with dedicated instruments that focus on the N-band and using a spectral resolution of R > 100 will be able to identify the individual, narrow spectral features that made up the broad 9 - 12 μm spectral features if given generous observation time (100+ transits).

### 4.2.2 Detection of Isoprene as a Major Atmospheric Gas

We assess the detection of isoprene via transmission spectra for isoprene in the run-away phase via transmission spectra. For exoplanets with anoxic atmospheres orbiting M dwarf stars, the high isoprene accumulation scenario occurs given an isoprene production rate at least 1 × $10^{12}$ molecules $cm^{-2}$ $s^{-1}$ for any atmosphere scenario we studied. We found that at this high production rate, isoprene can accumulate to a column average mixing ratio of 100 to 1000 ppm and the isoprene spectral features would be prominent compared to those of other molecules, potentially allowing isoprene to be identified (Figure 11).

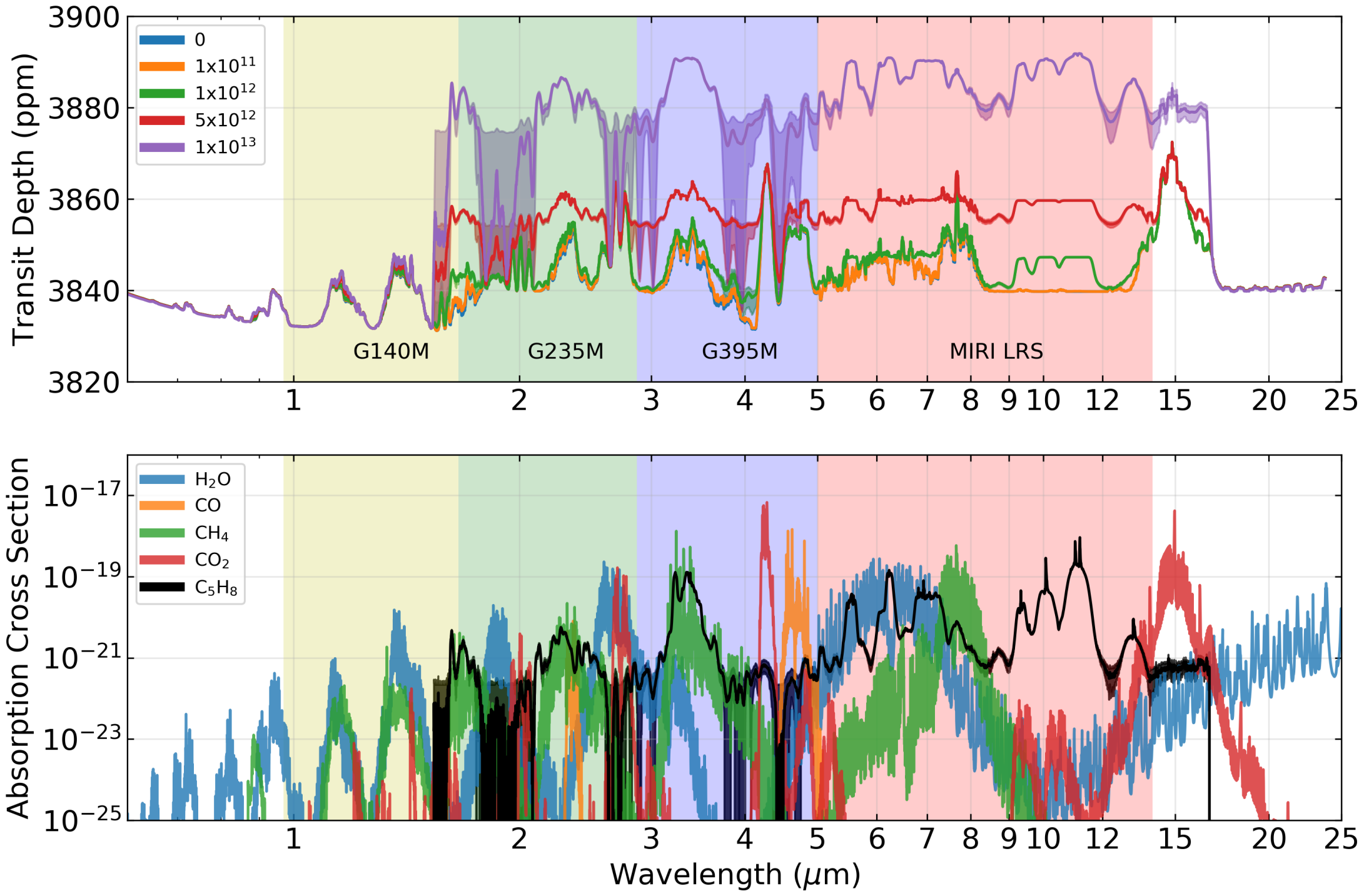

**Figure 11** Upper panel: Simulated spectra of exoplanets with $H_2$-dominated atmospheres transiting a M dwarf star for a range of isoprene surface fluxes from 0 to 1 × $10^{13}$ molecule $cm^{-2}$ $s^{-1}$. The y-axis shows transit depth (ppm) and the x-axis shows wavelength (μm). The spectra are simulated from 0.3 - 23 μm, covering the wavelength span of most of JWST's observation modes. The yellow, green and blue region shows the spectral coverage of NIRSpec and the red region shows that of MIRI LRS. At low surface mixing ratios, the isoprene spectral features are not prominent as isoprene is mostly concentrated near the surface and rapidly decays as a function of altitude. Isoprene features are not noticeable until the surface flux is above 1 × $10^{11}$ molecule $cm^{-2}$ $s^{-1}$. Above 1 × $10^{12}$ molecule $cm^{-2}$ $s^{-1}$, increasing the surface flux of isoprene rapidly increases the amount of isoprene in the atmosphere and significantly increases the strength of isoprene's transmission spectral features. Lower panel: Comparison of isoprene cross sections (Brauer *et al.* 2014) with cross sections of dominant molecules (except $H_2$, its absorption is mainly in the form of collision-induced absorption) in the atmosphere such as $H_2O$, CO, $CH_4$, and $CO_2$ (Gordon *et al.* 2017).

For terrestrial exoplanets transiting an 10th magnitude M5V dwarf star, only the $H_2$-dominated atmospheres (with surface pressure set to 1-bar and surface temperature set to 300 K) are detectable with JWST in transmission spectroscopy in near to mid-IR, as they have a large scale-height due to the low mean molecular weight. In this scenario, isoprene can be detected if it is produced at fluxes above 3 × $10^{12}$ molecules $cm^{-2}$ $s^{-1}$, or 100 times that of Earth's isoprene production rate. Isoprene has many spectral features, we find that detection can be achieved with 20 transits using any of the four modes of JWST (NIRSpec G140M, G235M, G 395M, MIRI LRS) we assessed (Figure 12).

For $N_2$-dominated and $CO_2$-dominated atmospheres, we did not find a scenario in which the atmosphere is detectable via transmission spectroscopy with JWST without investing 100 transits per observation mode. While it is theoretically possible to accumulate 100 transits for a planet orbiting a late M dwarf star over five years (the cryogenic lifetime of JWST), it is unrealistic given the competitive nature of the telescope observation time. We therefore conclude that detection of isoprene in a non-$H_2$ rich atmosphere should only be considered plausible for exoplanet dedicated future observatories with far better collecting power.

For exoplanets transiting a Sun-like star, we found no atmosphere scenarios are detectable via transmission spectroscopy because the ratio of planet to star radius, or $(R_{planet}/R_{star})^2$, was too small, resulting in too small a transit depth even for H2-dominated atmospheres.

Assessing the detectability of isoprene as a major atmospheric gas via secondary eclipse thermal emission spectroscopy is at present problematic due to our assumption of an isothermal atmosphere above the 0.01 bar ($10^3$ Pa) level (an assumption in our photochemistry model (Hu et al. 2012)). Secondary eclipse thermal emission spectra can in principle arise in the lower atmosphere, which in our model does have a temperature gradient. But when isoprene is a major atmosphere gas, above column-averaged mixing ratio of 100 ppm, there are enough isoprene molecules to saturate the atmosphere and render it opaque from the surface to near or above the 0.01 bar level. Since the atmosphere is isothermal above 0.01 bar, the result is a featureless spectrum: a blackbody curve of much lower temperature than the surface temperature (Appendix I, Figure I-2). To correct for this would require adapting the photochemistry model to use a self-consistent temperature-pressure profile, which is beyond the scope of this work.

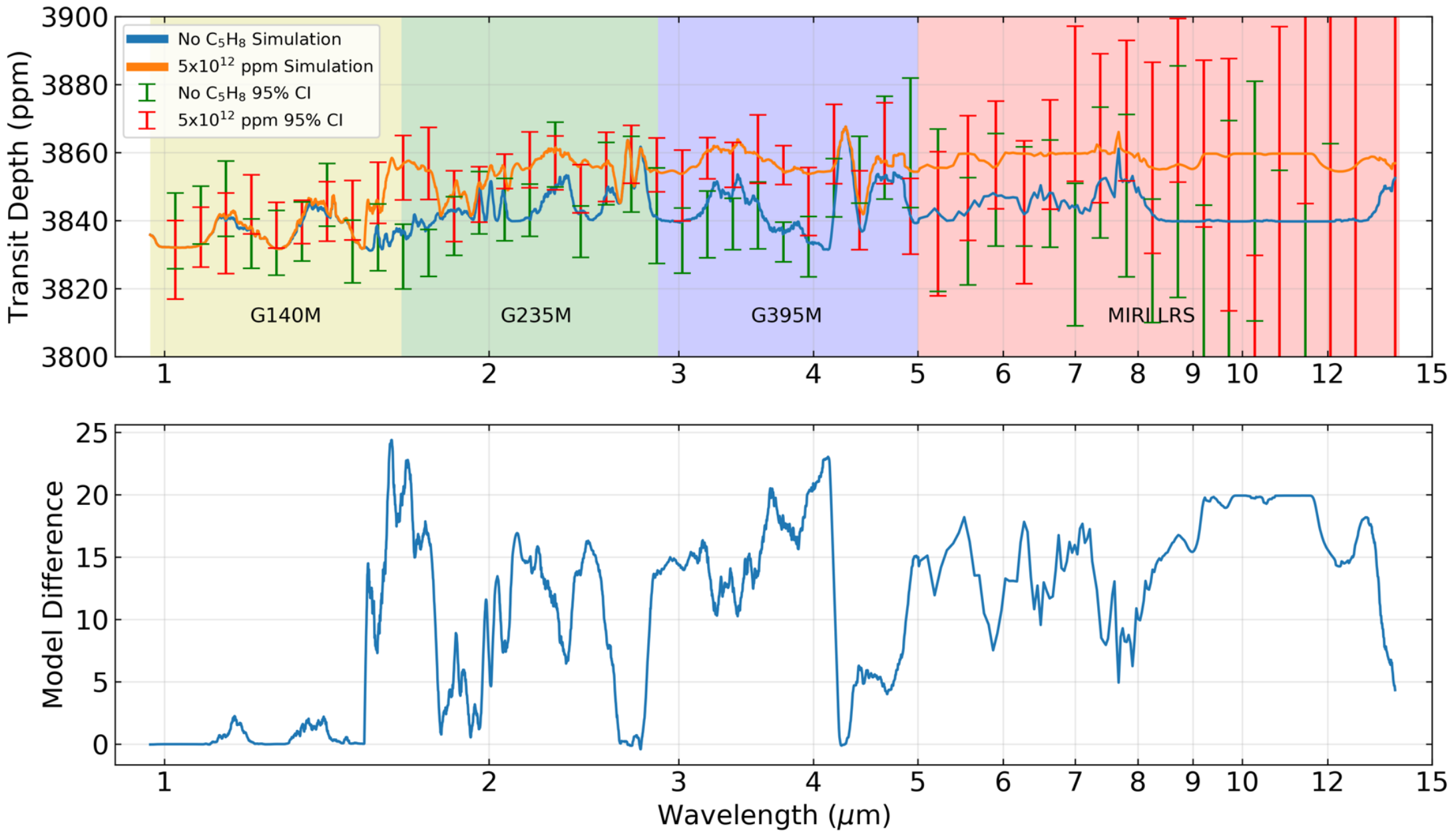

**Figure 12. Upper Panel**: Simulated $H_2$-dominated atmosphere observation using JWST for a 10 $M_{Earth}$, 1.75 $R_{Earth}$ super-Earth transiting an M dwarf star given 20 transit observations per instrument (80 transits in total). The y-axis shows transit depth (ppm) and x-axis shows wavelength (μm). The simulated observation spans the wavelength range of the NIRSpec and MIRI instruments. We compare a model with no isoprene surface flux (blue line, green error-bar) and a model with $5 \times 10^{12}$ molecule $cm^{-2}$ $s^{-1}$ (orange line, red error-bar). The error bars are 95% confidence intervals for each model uniformly binned to a spectral resolution of R = 10. For this comparison, we take the isoprene cross section as is and did not account for lab measurement error estimates. The error bar is attributed from observational noise only to more accurately reflect observation simulation. **Lower Panel**: Difference between the two simulated spectra showing the spectral features of isoprene peaks. We show that within each instrument, there are more than more than 20 ppm transit depth differences between the two models, therefore it is possible to achieve statistical significance and detect isoprene.

### 4.2.3 Isoprene Identification is Hindered by Methane

Detecting isoprene on an exoplanet atmosphere is challenging; requiring either very high isoprene production rate and/or long observation times. Another layer of challenge is that isoprene identification can be hindered by methane, unlike other biosignature gases we proposed: $PH_3$, which have distinguishable features around 4.5 μm ((Sousa-Silva *et al.* 2020) and $NH_3$, which have distinguishable features around 1.5 μm, 2 μm (Huang, et al. In prep). Therefore we report on isoprene spectral distinguishability because, as observing capabilities improve in the future, the ultimate limiting factor for isoprene detection will remain the spectral strength, location and distinguishability of its features, and these are immutable.

When compared to the expected major gases in habitable terrestrial planet atmospheres, isoprene broad band (R ~ 10 - 20) spectral features are distinguishable from molecules such as $H_2O$, $CO_2$, CO, $NH_3$, and $H_2S$ (see Appendix Figure I-3). However, distinguishing isoprene from $CH_4$ will be challenging and requires further discussion. The column averaged mixing ratio of methane in the six simulation scenarios are in the range between 1 - 100 ppm (See Appendix I Figure I-1 for detailed mixing ratio profile). For $H_2$-dominated atmospheres, $CH_4$ should be readily present (Seager *et al.* 2013). In $CO_2$-dominated atmospheres where $CH_4$ may not exist, isoprene contamination by $CH_4$ will not be an issue though other hydrocarbons might still confuse isoprene identification (see Section 5.2 for further discussion).

In an idealistic world with no observational constraints, the spectral features of all molecules are unique. Given the limitation of numbers of photons, however, instruments always have a finite spectral resolution. For detection of habitable exoplanet atmospheres, we must further trade spectral resolution in favor of increasing the (SNR) by binning the data after the data is obtained. We find that the data needs to be binned to a relatively low spectra resolution of R = 10 - 20 in order to reach statistical significance (as demonstrated in Figure 12).

At resolution lower than R = 20, distinguishing between isoprene spectral features and methane spectral features at shorter than 4 μm is not possible. Both methane and isoprene share the C-H stretch feature and its overtones, thus if methane is present in significant quantities, then the isoprene 1.6 - 1.7 μm, 2.1 - 2.5 μm, and 3.1 - 3.7 μm features will be masked by methane. In contrast, the 5.4 - 7.9 μm features and the 9 - 12 μm features do not overlap with methane spectral features (Figure 13). Therefore, to confidently detect isoprene in $H_2$-dominated atmospheres, the identification of both the 5.4 - 7.5 μm features and the 9 - 12 μm features is required, therefore requiring the use of both NIRSpec and MIRI, potentially more than doubling the number of transit observations required (since there are fewer photons at longer wavelengths).

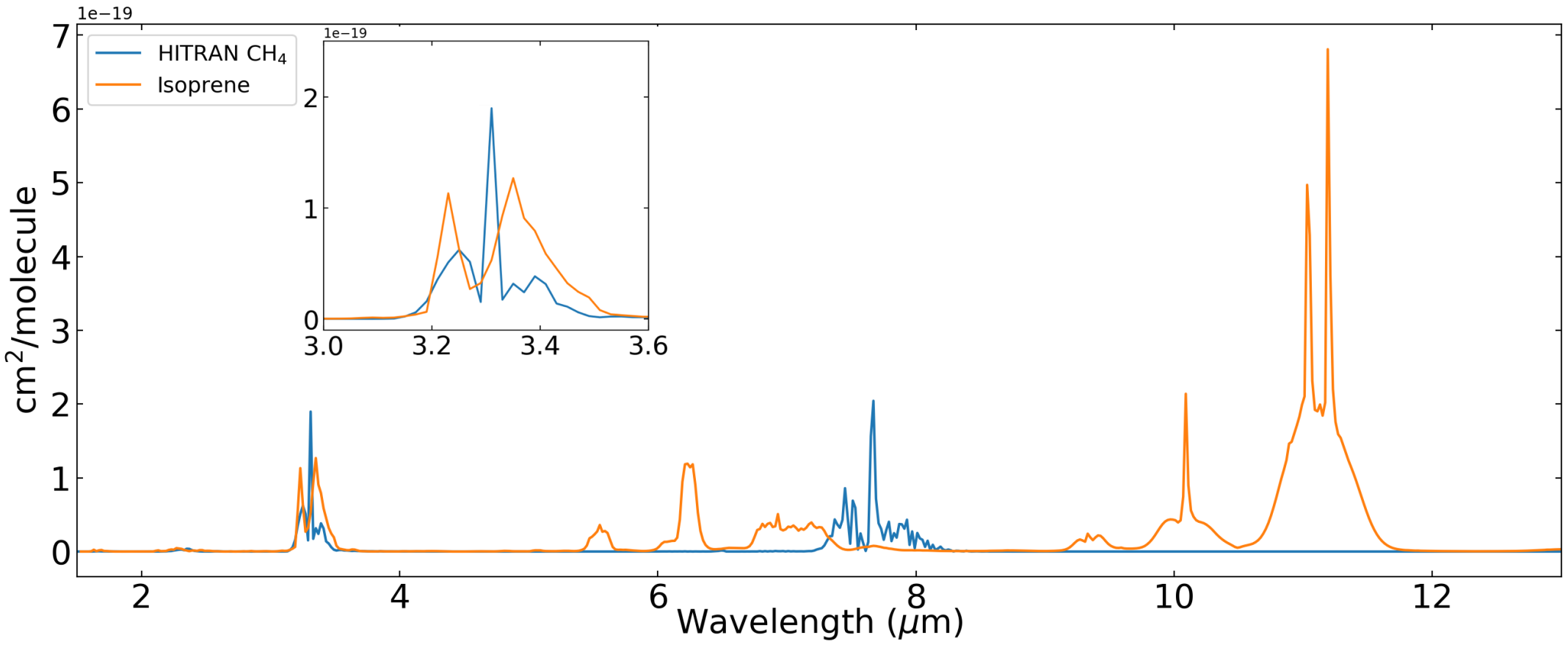

**Figure 13.** Comparison of isoprene (orange) and methane (blue) cross sections binned to R = 20. The axis shows cross section [$cm^{-2}$ $molecule^{-1}$] vs wavelength [μm]. Cross setions for isoprene is measured by Brauer *et al.* (2014). Cross section for methane is calculated using HITRAN line lists with standard pressure and temperature (Gordon *et al.* 2017). The small panel shows a zoomed-in version of the isoprene-methane-shared spectral feature at 3.1 - 3.7 μm. We show that distinguishing between isoprene and methane spectral features at 3.1 - 3.7 μm is are still possible at a spectral resolution of R = 20. Distinction between isoprene and methane at spectral resolution lower than this limit is not possible and therefore detection of isoprene's 3.1 - 3.7 μm feature require sufficient spectral resolution. We also show that isoprene's 5.4 - 7.5 μm features and the 9 - 12 μm features do not overlap with methane's spectral features, so detection of these two features can be done with a lower spectral resolution.

## 4.3. Impact of Haze on Isoprene Detection

We turn to an assessment of the detectability of isoprene in anoxic atmospheres with the presence of isoprene-induced haze. We find that haze will hinder but not completely mask detection of isoprene unless the ratio between the total mass of isoprene-induced haze vs the total mass of isoprene (haze-to-isoprene mass ratio) is above 10%. We consider a detection to be hindered if certain spectral features (in a confined wavelength range) are not detectable. We consider detection to be masked if no spectral features are detectable at any wavelength range. We simulate a wide range of scenarios to study how haze abundance, composition, size[6], and size distribution can hinder isoprene detection.

Our models show that the most important factor that affects the detection of isoprene spectral features is the abundance of haze, followed by the size of the haze particles. The abundance of haze governs the overall opacity of the atmosphere while the particle size plays a key role in determining the wavelength dependence of the opacity. Our simulations also suggest that the composition and the mean diameter of the haze particles matter less than the abundance and size distribution.

We quantify how different haze-to-isoprene mass ratios affect a simulated transmission spectrum (Figure 14). We used our $CO_2$-dominated atmosphere archetype with an isoprene surface flux of $1 \times 10^{13}$ molecules $cm^{-2}$ $s^{-1}$ and other parameters described in Section 3. A sub-micron[7] size particle's extinction cross section at wavelengths longer than 3 μm decrease as the wavelength increases. The 1.6 - 1.7 μm and 2.1 - 2.5 μm features are sensitive to haze and can be diminished in strength for an haze-to-isoprene mass ratio of 0.0001. The 3.1 - 3.7 μm isoprene spectral feature is not affected for a haze-to-isoprene mass ratio of 0.0001 or lower but is masked for a ratio > 0.001. The 5.4 - 7.5 μm spectral feature is not affected for a ratio of 0.001 or lower but is masked for a ratio > 0.01. Finally, the 9 - 12 μm region is not affected for a ratio of 0.01 or lower but is masked for a ratio > 0.1. Therefore, if the haze-to-isoprene mass ratio is 0.1 or less, detection of isoprene spectral features is still possible.

For context, the haze-to-methane mass ratio on Titan is approximately 0.1 at the thickest part of the haze layer at 400 km altitude. The number density of methane at the same altitude is $10^{12}$ molecule $cm^{-3}$ and the number density is around $10^8$ particles $cm^{-3}$ assuming the particles are 12.5 nm spheres (Fan *et al.* 2019) and using Titan tholin refractive indexes from (Khare *et al.* 1984). Although it is debatable whether or not habitable exoplanet atmospheres commonly attain the same abundance of haze as that of Titan, we adopt the haze-to-isoprene mass ratio of 0.1 as the upper bound.

---

[6] In this work, we use size to mean the radius of the particle.

[7] To reduce confusion, we use micron to denote particle radius and μm to denote spectral wavelength

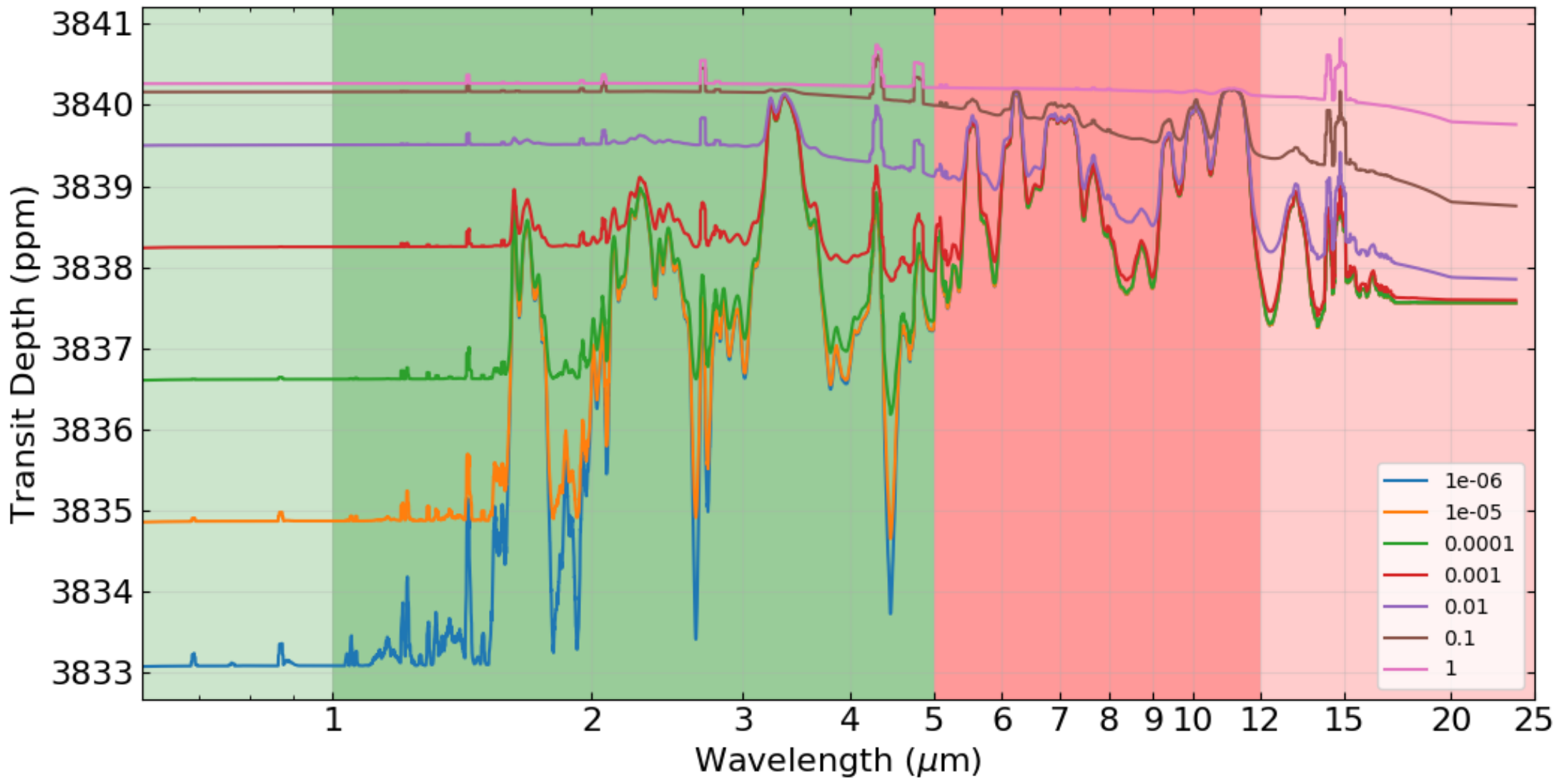

**Figure 14:** Haze abundance effects on simulated transmission spectra demonstrated using a $CO_2$-dominated atmosphere with an isoprene surface flux of $1 \times 10^{13}$ molecules $cm^{-2}$ $s^{-1}$ for a habitable super Earth transiting an M dwarf star. The axes are transit depth [ppm] vs. wavelength [μm]. The isoprene-induced haze size distribution is a Gaussian distribution with mean of 89 nm and standard deviation of 25 nm and is approximated from the 10,000x metallicity, 300 K scenario as described in Horst et al. (2018). The refractive index proxy for isoprene is $C_2H_2$ (Dalzell et al 1969). The different colored curves show the simulated transmission spectra as a function of haze/isoprene mass ratio from $1 \times 10^{-6}$ to 1.

Experiments and models both show that the mean radius of haze particles at standard temperature and pressure are 0.1 micron or smaller (He *et al.* 2018; Hörst *et al.* 2018), a particle size that affects shorter (i.e., visible to NIR) rather than longer (i.e., MIR) wavelengths. This effect is well known from Mie theory, and includes the point that once a particle radius is much smaller than the wavelength of light, scattering is in the Rayleigh regime and scales with wavelength as $\lambda^{-4}$. We study how different haze particle size affects simulated transmission spectra using the same inputs as for the haze-to-mass ratio study above and shown in Figure 14. We set the haze-to-isoprene mass ratio to 0.01 and vary the mean particle radius from 0.01 to 10 micron.

For completeness we explore the effect of mean particle size on the extinction cross section for a size of 0.01 - 10 micron. We found that the haze extinction cross section does not scale linearly with the size of the haze particles when assuming a fixed haze-to-isoprene mass ratio (total mass is constant). Taking Mie theory into account, the maximum strength for the extinction cross section is for a mean particle size of 0.1 microns (for shorter wavelength) to 1.0 microns (for longer wavelength). For haze particle size smaller than 0.1 micron, increasing the particle size increases the total extinction cross section. Overall, different mean particle size will hinder or mask detection of different part of the spectra and are shown in more detail in Figure 15.

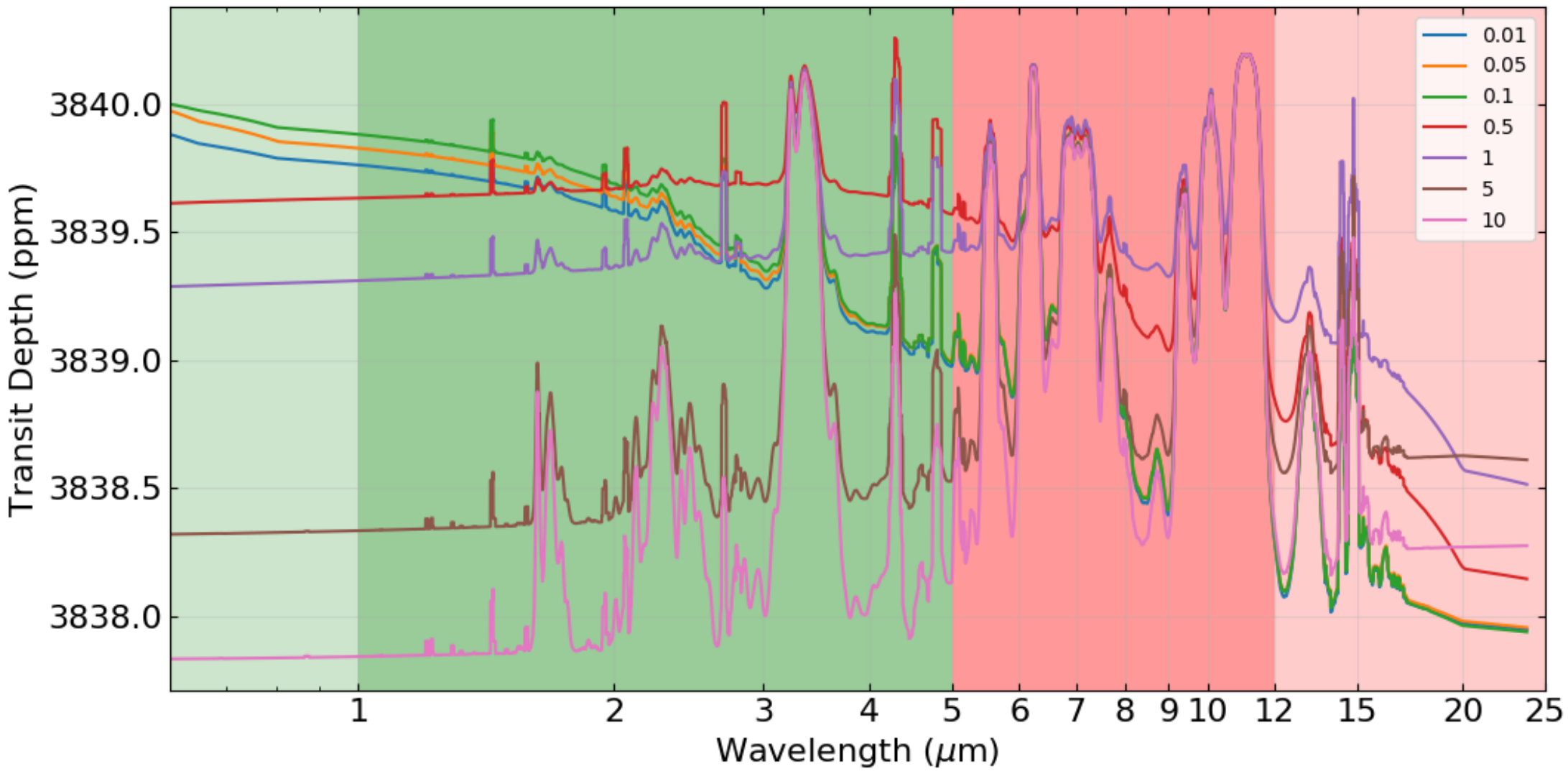

**Figure 15:** Haze abundance effects on simulated transmission spectra demonstrated using a $CO_2$-dominated atmosphere with an isoprene surface flux of 1 × $10^{13}$ molecules $cm^{-2}$ $s^{-1}$ for a habitable super Earth transiting an M dwarf star. The haze/isoprene mass ratio of this simulation is set to 0.01 and the refractive index proxy for isoprene is $C_2H_2$ (Dalzell et al 1969). The different colored curves show the simulated transmission spectra with varying haze mean particle size. Changes in the broad-band attenuation and spectral attenuation are not linear with increase in the mean particle radius.

Finally, we examine whether different types of refractive indices (as a proxy for haze particle composition) will impact detection. Based on the four types of materials we considered (Titan tholins (Khare *et al.* 1984), HCN (Khare *et al.* 1994), $C_2H_2$ (Dalzell and Sarofim 1969) and octane (Anderson 2000)), we found that while there are minor differences in wavelength-dependent opacities, none of these have unique absorption features that will hinder or mask isoprene spectral features.

## 4.4. Lack of Isoprene False Positives

On Earth, isoprene is exclusively produced by life (Section 2.2). Even so, one might suggest that isoprene could be formed geochemically, as a geological false positive, by reduction of carbon dioxide, by reduction of carbon monoxide, or by hydrogenation or dehydrogenation of hydrocarbons. However, the geochemical formation of isoprene on temperate, rocky planets is thermodynamically disfavored (Table 3). For any of the proposed isoprene formation reaction pathways, given the gas concentrations in terrestrial volcanoes, the calculated energy of formation of isoprene makes any geochemical formation scenario very unlikely (a positive Gibbs free energy (ΔG) means the reaction requires energy) (Table 3).

| Proposed geochemical isoprene formation pathways | *ΔG* of reaction (kJ/mol) |
|---|---|
| $5CO_2 + 14H_2 \rightarrow C_5H_8 + 10H_2O$ | 1670.1 |
| $5CO + 9H_2 \rightarrow C_5H_8 + 5H_2O$ | 1294.8 |
| $5CH_4 \rightarrow C_5H_8 + 6H_2$ | 477.7 |

**Table 3.** Proposed geochemical isoprene formation pathways and energy of formation of isoprene from plausible geochemical volatile concentrations. *ΔG* is the free energy of reaction at 298 K assuming typical geochemical concentrations of $H_2O$, CO, $CO_2$, $CH_4$ and $H_2$ and 1ppm isoprene. Average fractions of 'wet' gas from magmatic and hydrothermal volcanic systems from the ASM database (Bains *et al.* 2017): $H_2$=0.0028, $H_2O$=0.9223, CO=0.000615, $CO_2$=0.05332, $CH_4$=$4.206 \times 10^{-5}$. Non-biological formation of isoprene from any terrestrial geological source is very unlikely for any of the proposed reaction pathways (positive *ΔG*).

Molecules in a planetary atmosphere may potentially originate from several common abiotic sources including: the interstellar medium (ISM); as leftover from planetary formation; meteorites and comets; photochemical processes in upper planetary atmospheres; and geological processes on planetary surfaces. None of these processes or environments can efficiently abiotically produce isoprene, as detailed below.

Isoprene and other hydrocarbons containing conjugated double bonds are not known to be a product of ISM chemistry (Ehrenfreund and Cami 2010; McBride *et al.* 2013) and are not identified among known interstellar medium molecules (McElroy *et al.* 2013). We note however that many other hydrocarbons, including unsaturated ones, have been detected in interstellar medium (e.g. CH, $C_2H$, $CH_2$, $CH_3$, $CH_4$, $C_4H$, $C_5H$, $C_6H$, $C_6H_2$, $C_6H_6$, as reviewed by (McElroy *et al.* 2013). Similarly, isoprene and other hydrocarbons containing conjugated double bonds are not detected in meteorites, including in the Murchison meteorite (Levy *et al.* 1973; Pizzarello and Shock 2010; Sephton 2004), a meteorite famous for containing structurally diverse types of organic chemicals.

Photochemical processes in upper planetary atmospheres are also known to not produce isoprene[8]. For example, isoprene and other complex hydrocarbons containing conjugated double bonds are not known to be produced in the atmosphere of Titan. Moreover, the simplest conjugated-double bond hydrocarbon (1,3-butadiene) is also not likely to be formed abiotically. Limited theoretical work suggests that non-biological production of 1,3-butadiene in Titan's atmosphere is in principle possible *via* photochemical reactions, although the rate of formation of 1,3-butadiene is predicted to be negligible. Indeed, so far 1,3-butadiene has not been detected on Titan (Newby *et al.* 2007) and would likely condense onto haze given the low temperature of titan. On the other hand, photochemical processes in Titan's atmosphere lead to formation of many unsaturated hydrocarbons (e.g. $C_2H_2$,

[8] We note that on Earth a very small amount of isoprene is formed on the surface of Earth's oceans as a result of photodissociation of natural fatty acids (Ciuraru *et al.* 2015). Such production, while not a result of direct biological action is still an indirect result of biological activity and is therefore considered to be biological.

$C_2H_4$, $C_4H_2$, $C_6H_2$, $C_6H_6$ etc.), none of which contain conjugated double bonds (Maltagliati *et al.* 2015; Yung *et al.* 1984).

The lack of chemical pathways leading to the abiotic formation of isoprene in a diverse host of planetary environments was recently strengthened by a series of photochemical laboratory experiments on formation of various gases including hydrocarbons. The exposure of various mixtures of basic gases like $H_2O$, $CO_2$, $CH_4$, $NH_3$ to UV light or plasma at various temperatures in simulated atmospheric scenarios did not show isoprene formation (nor any precursor hydrocarbons to isoprene's formation, nor any other conjugated dienes) (He *et al.* 2019).

We conclude that isoprene has no known false positive sources. Consequently, its detection in an exoplanet atmosphere would be a strong indication of biological activity.

# 5. Discussion

We evaluated the candidacy of isoprene as a potential biosignature gas based on the principle that an ideal biosignature gas should satisfy all the following criteria: 1) is produced by life; 2) lacks abiotic false positives; 3) can accumulate to a detectable abundance in exoplanet atmospheres; and 4) has distinguishable spectral features. We have shown that isoprene satisfies the first two criteria (Section 2 and 4). Isoprene satisfy the 1$^{st}$ and 2$^{nd}$ criteria but can only satisfy the 3$^{rd}$ and 4$^{th}$ criteria in some scenarios. Therefore we consider isoprene as a good biosignature gas. Here we discuss several factors limiting our analysis (Section 5.1), as well as the ability to distinguish isoprene from other hydrocarbons (Section 5.2). We end the discussion by exploring isoprene's ubiquity in many life forms, and introduce the concept of isoprene as a "Biosphere Signature" (Section 5.3).

## 5.1. Limitations of Reference Data

Our assessment of isoprene as a biosignature gas is constrained by the limited availability of spectral data, our incomplete understanding of isoprene chemistry in anoxic atmospheres, and unknown but potential haze formation from isoprene in anoxic atmospheres. Here we discuss these limitations and how they may affect our assessment.

### 5.1.1 Limitations with Using Absorption Cross Sections from Lab-Measured Spectra.

Using cross sections calculated from lab-measured data "as is" without estimating uncertainties for high isoprene mixing ratios can result in atmospheric spectra that are saturated and show artificially strong, wide, and featureless absorption bands (see Figure Appendix I-4), which lead to inconclusive detection predictions. This effect is unphysical, because in some wavelength ranges, the true cross section values are likely much smaller than the instrumental noise floor. The Brauer et al. (2014) wavelength-independent uncertainty is approximately four orders of magnitude weaker than the strongest absorption peaks of isoprene. Based on Beer-lambert law, for transmission spectroscopy, the cross section uncertainty will also need to be eight orders of magnitude weaker than the strongest absorption peaks to avoid the unphysical effect. Therefore, we recommend future lab-based measurements of the isoprene cross sections to focus on measuring the strength of the weak absorption features, or location of the spectra where there are no spectral peaks, so that it is possible to differentiate between real, but weak transitions and the noise floor.

A possible work-around method is to subtract the uncertainties from the data, leaving only the strongest peaks. However, we disfavor discarding cross section values below an arbitrary value because it can remove weak, but real, absorption lines. While individually these weak lines are negligible, collectively they can drastically change the overall opacity, and subsequently conclusions on the potential for detection. We show in Figure I-4 how such an unphysical noise floor cut-off can artificially remove the weak features of isoprene and re-engineer the spectrum and associated detectability of isoprene.

In addition, we encourage measurement of isoprene spectra to cover a broad temperature/pressure range relevant for exoplanet atmospheres. A broader coverage as compared to the current standard temperature and pressure measurements would allow extrapolation of isoprene cross section to conditions more appropriate for the upper atmosphere of a variety of exoplanets. Calculating molecular line lists using *ab initio* theoretical quantum mechanical methods is computationally demanding (e.g., (Tennyson and Yurchenko 2012), often requiring years of work even for a small molecule, and is not currently possible to obtain accurate theoretical cross-sections for isoprene due to the complexity of the molecule (see (Sousa-Silva *et al.* 2019) for more in-depth discussion on alternatives). However, we emphasize that our SEAS model is sufficiently versatile to be able to quickly update its predictions on isoprene detectability whenever improved cross sections for isoprene become available.

### 5.1.2 Limitations to Isoprene Reaction Rates

In this work, we have only focused on the reaction of isoprene with UV and the dominant radical species O· and ·OH. The photochemistry of isoprene with other, potentially relevant radicals, such as H·, Cl· and $\cdot C_nH_m$ is insufficiently studied. Inclusion of these radicals may potentially increase the isoprene destruction rate. Our models may not include all chemical reactions that are relevant to anoxic, temperate terrestrial planets. Therefore more detailed studies of anoxic atmospheric chemical reaction networks are required.

Many M dwarf stars have frequent, strong flares. Flares introduce intense packets of UV radiation that are capable of destroying biosignature gases including isoprene (Segura *et al.* 2010; Tilley *et al.* 2019). Recent work by (Günther *et al.* 2019) has found 30% of late M dwarf stars and 5% of early M dwarf stars display active flaring events. Future photochemistry models, therefore, will require consideration of flaring activities and addressing how flaring activities impact habitability and accumulation of biosignature gases.

### 5.1.3 Limitations to Understanding Isoprene Induced Haze

No one has yet addressed the formation of isoprene-induced hazes in anoxic planetary atmospheres. On Earth, isoprene destruction rapidly leads to haze formation. Further studies on constraining the isoprene-haze mass ratio and mean particle size in anoxic planetary atmospheres are needed because sub-micron haze particles can heavily mute transmission spectral features at wavelengths up to 4 μm as demonstrated in Section 4.3.

In addition, high-altitude hazes in an anoxic atmosphere could act as a UV shielding mechanism for biosignature gases, therefore increasing the probability of their accumulation (Seager *et al.* 2013; Segura *et al.* 2005).

## 5.2 Isoprene Spectral Distinguishability vs Other Hydrocarbons

In addition to isoprene and methane, all hydrocarbons have spectral features at 3.1 - 3.7 μm, due to the C-H bond stretching rovibrational band and many have spectral features at 9 - 12 μm, due to the C-H bond bending rovibrational band. Figure 16 shows a comparison of isoprene with methane and other simple hydrocarbons that highlights the challenges in the unambiguous identification of isoprene. Though isoprene accounts for most of the non-methane hydrocarbons released by life on modern Earth (Sharkey *et al.* 2008), simpler hydrocarbons have more favorable abiotic formation pathways which may lead to their accumulation in the atmosphere. Therefore, we need to study isoprene distinguishability in the context of other hydrocarbons that may be readily present in the atmosphere.

The 3.1 - 3.7 μm features are the C-H bond stretching vibration features and can have three main functional groups: carbon single bonds (X-C-H) located at 3.45 μm, carbon double bonds (X=C-H) located at 3.23 μm, and carbon triple bonds (X#C-H) located at 3.03 μm (X denotes a non-hydrogen atom bonded to the carbon atom; see Section 3.1.2 for more details on these features). For molecules that have only one spectrally active functional group, distinguishing between molecules within each group requires a spectral resolution of at least R = 200. Fortunately isoprene has both the (X-C-H) bond and the (X=C-H) bond so detection of its 3.1 - 3.7 μm can be done with a spectral resolution of R = 20. We note however, that a coincidental mixture of simple hydrocarbon molecules in the atmosphere can mimic the isoprene spectral features at 3.1 - 3.7 µm. For example, and equal amount of ethane ($C_2H_6$), which only contain the (X-C-H) bond and ethylene ($C_2H_4$), which contains only the (X=C-H) bond, would be indistinguishable from isoprene if observed at low resolutions.

The isoprene spectral features at 9 - 12 µm are due to the vibrational bending of the C-H bond and contain many contain many rovibrational substructures. Distinguishing isoprene from other hydrocarbon molecules in this spectral region can be challenging. Recent work by Sousa-Silva, et al., (2019) developed a computational tool (RASCALL: Rapid Approximate Spectral Calculations for ALL) to estimate spectral data (position in the spectra and relative strength) for any molecule by combining functional group data from experimental measurements, organic chemistry, and quantum mechanics; these methods can aid in identifying which molecules struggle to be distinguished from one another, and highlight the issues surrounding the distinguishability of isoprene from other hydrocarbons.

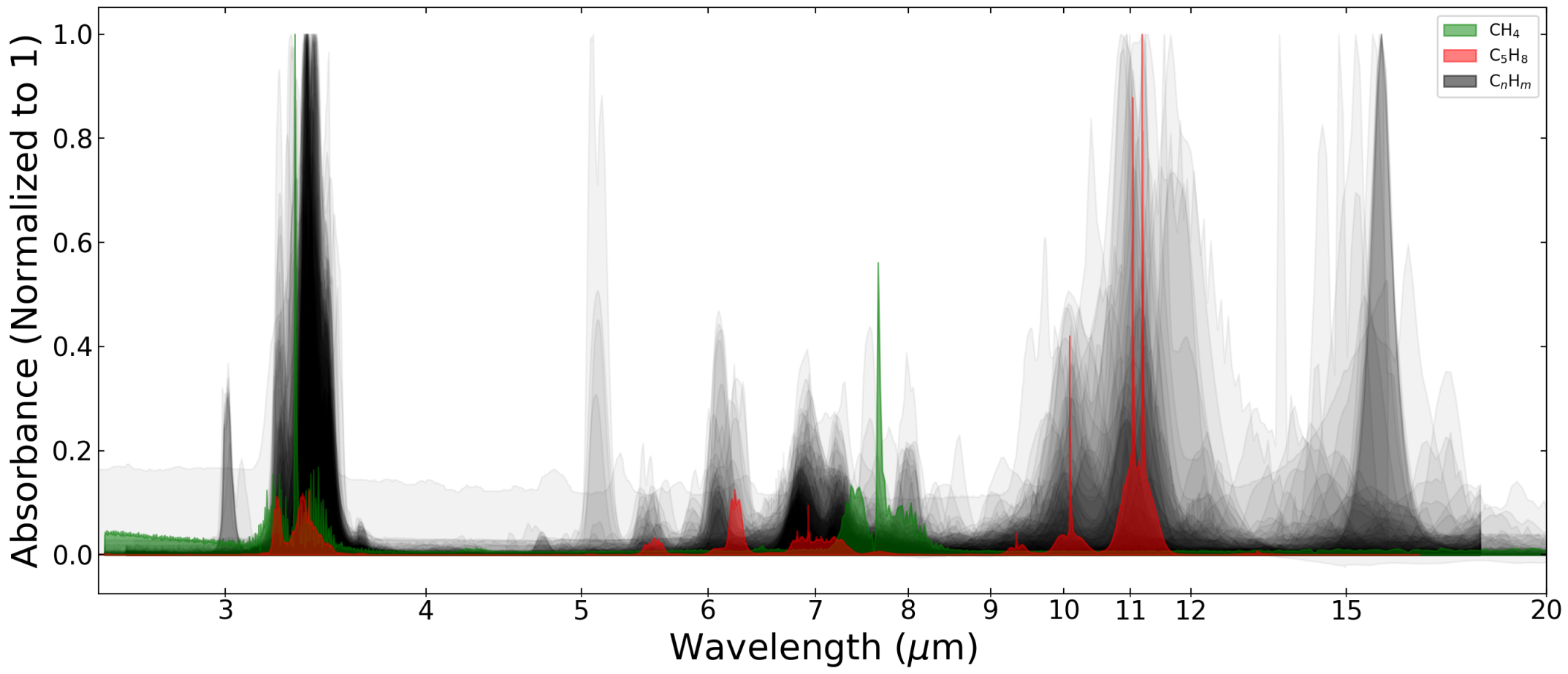

**Figure 16.** Isoprene (red) and methane (green) vs. 80 other hydrocarbon (grey) absorbances. The axes shows absorbance (normalized to 1) vs wavelength [µm]. The hydrocarbon absorbance data are taken from NIST lab-based measurements and includes all hydrocarbon with fewer than seven carbon atoms that has spectral measurements. Due to measurements being made with different instruments and experimental set-ups, we normalize the absorbances to one for a qualitative comparison. We show that although isoprene has many broad spectral features, it does not stand out when comparing with the collective sum of spectral features from other hydrocarbons. Therefore, it is possible for a coincidental combination of simple hydrocarbons to mimic isoprene spectral features.

## 5.3 Isoprene as a “Biosphere Signature”

Isoprene stands out among the studied biosignature gases not only because it is produced in high abundance by life on Earth but also due to the fact that isoprenoid biosynthetic pathways, that are responsible for synthesis of isoprene and many other terpenoids, are present in virtually every domain of life on Earth, i.e. Bacteria, Archaea and Eukarya (see Appendix IIB). While the synthesis and likely functions of isoprene on Earth are closely linked to oxygenic photosynthesis in plants, many non-photosynthesizing organisms make isoprene, and its synthesis is not dependent on an oxygenated atmosphere. It is plausible therefore to speculate that an entirely anaerobic biosphere could produce high levels of isoprene.

Even if isoprene itself is not made by every species on Earth, a wide variety of molecules that contain the isoprene-motif, or isoprenoids, are. The isoprene motif and its structural derivatives are utilized to synthesize countless chemicals in all of Earth's life. In contrast to previously proposed classical type III biosignatures such as methyl chloride, methyl sulfide and DMS, isoprenoids have a much larger coverage of the phylogenetic tree of life (Seager *et al.* 2012).

Isoprenoids are produced by many evolutionary distant organisms occupying diverse ecological niches (e.g. both marine and terrestrial surface and subsurface environments). Out of other proposed type III biosignature gases, only dimethyl sulfide is produced by a host of evolutionarily diverse organisms but nowhere near to the phylogenetic diversity of isoprenoid production. Other frequently cited type III biosignatures (like $N_2O$ and $CH_3Cl$) are produced by less diverse species. In addition all of them inhabit very similar habitats (e.g. methyl chloride is released mainly by several species of marine algae and to a lesser extent by the soil bacteria and fungi; 62% of the global $N_2O$ emissions is produced by nitrifying and denitrifying bacteria and fungi in the soil (Maeda *et al.* 2015)). Isoprene and other isoprenoids do not have these limitations. Therefore, isoprene is the most widely made molecule yet evaluated as a biosignature gas. Moreover, there are at least two distinct isoprenoid biosynthesis pathways, and isoprene synthesis likely has evolved independently multiple times so it is plausible to suggest that it might evolve on other worlds. Therefore, we postulate that isoprene and other volatile isoprenoids could be considered characteristic, unifying chemical signatures, specific and common to the whole Earth-type biosphere. In essence isoprenoids are more than biosignatures – they are "biosphere signatures" – chemicals that are a molecular fingerprint of the entire Earth-type biosphere.

## Acknowledgements

We thank the MIT BOSE Fellow program, the Change Happens Foundation, the Heising-Simons Foundation, and NASA grants 80NSSC19K0471 and NNX15AC86G for partial funding of this work. S.R. gratefully acknowledges support from the Simons Foundation, grant no. 495062.

We thank Mihkel Pajusalu, Karen Cady-Pereira, Robert Hargreaves, and Iouli Gordon, for lengthy discussions on using molecular absorption cross section calculated from lab measured spectra for exoplanet atmosphere simulations. We also thank Thomas Evans, Jason Eastman, Elisabeth Matthews and the Seager group for helping us to establish the detection metric. Finally, we thank Joanna Petkowska for contributing Figures 2 and 3).

# Appendices

## Appendix I: Atmospheric Mixing Ratios and Temperature-Pressure Profiles

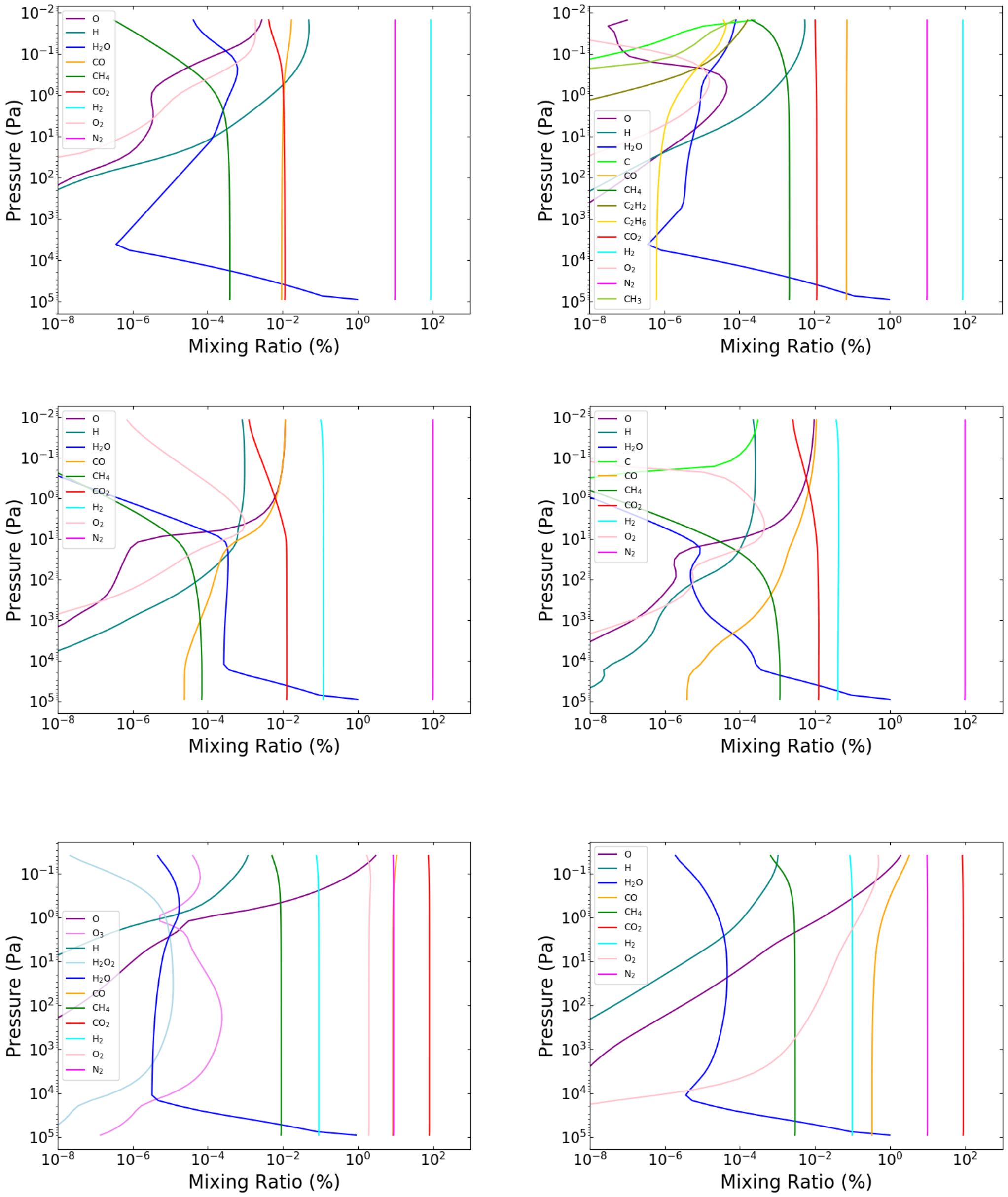

Figure I-1: Mixing ratio profiles used for simulating the atmosphere spectra of super-Earths ($M_p$ =10 $M_E$ and $R_p$ = 1.75 $R_E$); adapted from (Hu *et al.* 2012) and (Seager *et al.* 2013). Here we show volume mixing ratio [% of the total atmosphere] vs. pressure [Pa] for molecules that have a local mixing ratio of 100 ppb in any given layer (1 scale height) of the atmosphere for the six atmosphere archetypes (three major gases and two stellar types): Top left: $H_2$-dominated and M dwarf star. Top right: $H_2$-dominated and Sun-like star. Middle left: $N_2$-dominated and M dwarf star. Middle right: $N_2$-dominated and Sun-like star. Bottom left: $CO_2$-dominated and M dwarf star. Bottom right: $CO_2$-dominated and Sun-like star.

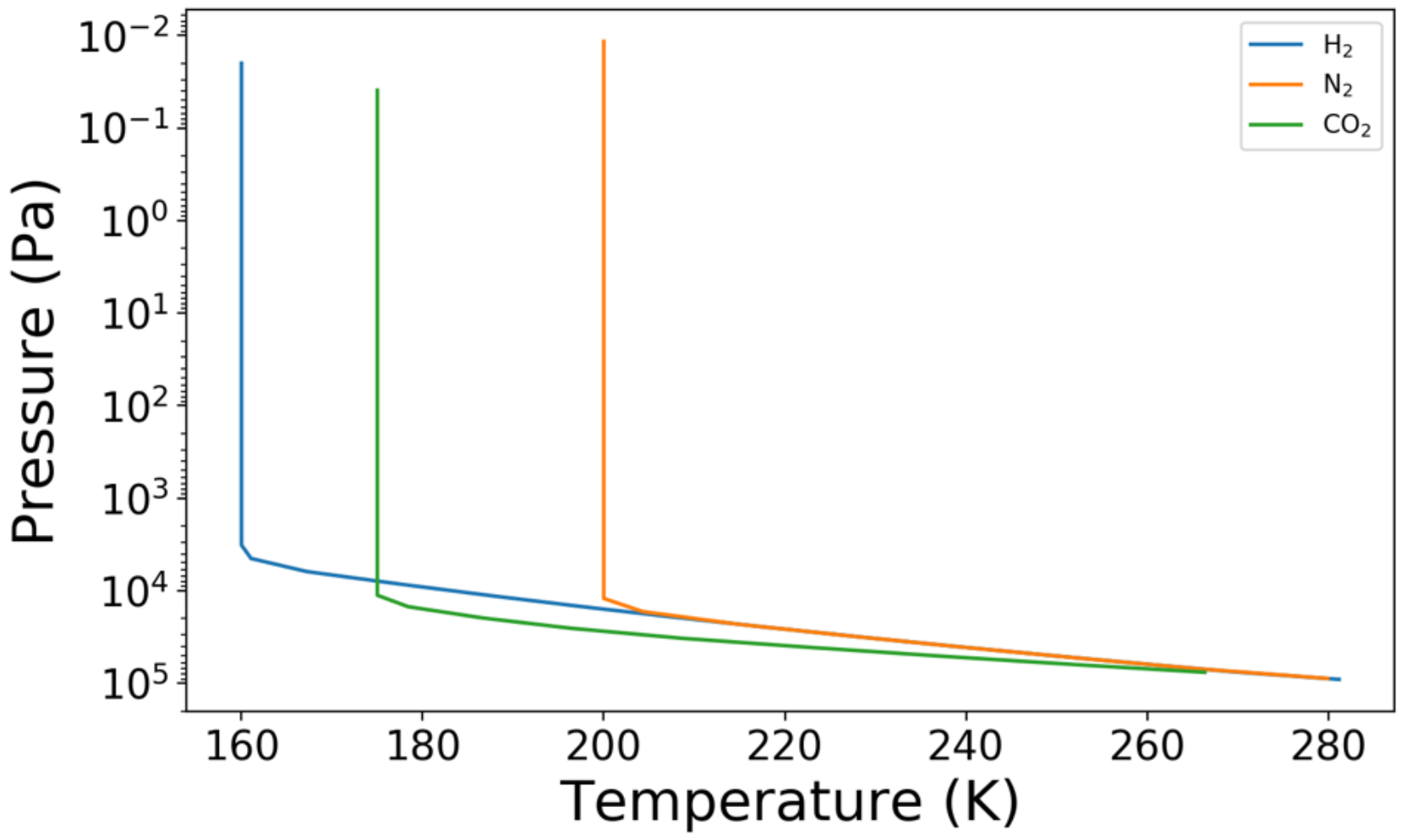

Figure I-2: Temperature-pressure profiles for $H_2$-dominated, $N_2$-dominated and $CO_2$-dominated atmosphere archetypes adapted from (Hu *et al.* 2012). Vertical axis shows pressure in units of Pa and horizontal axis shows temperature in units of K.

## B: Comparison of Cross Sections for Isoprene Other Atmospheric Gases

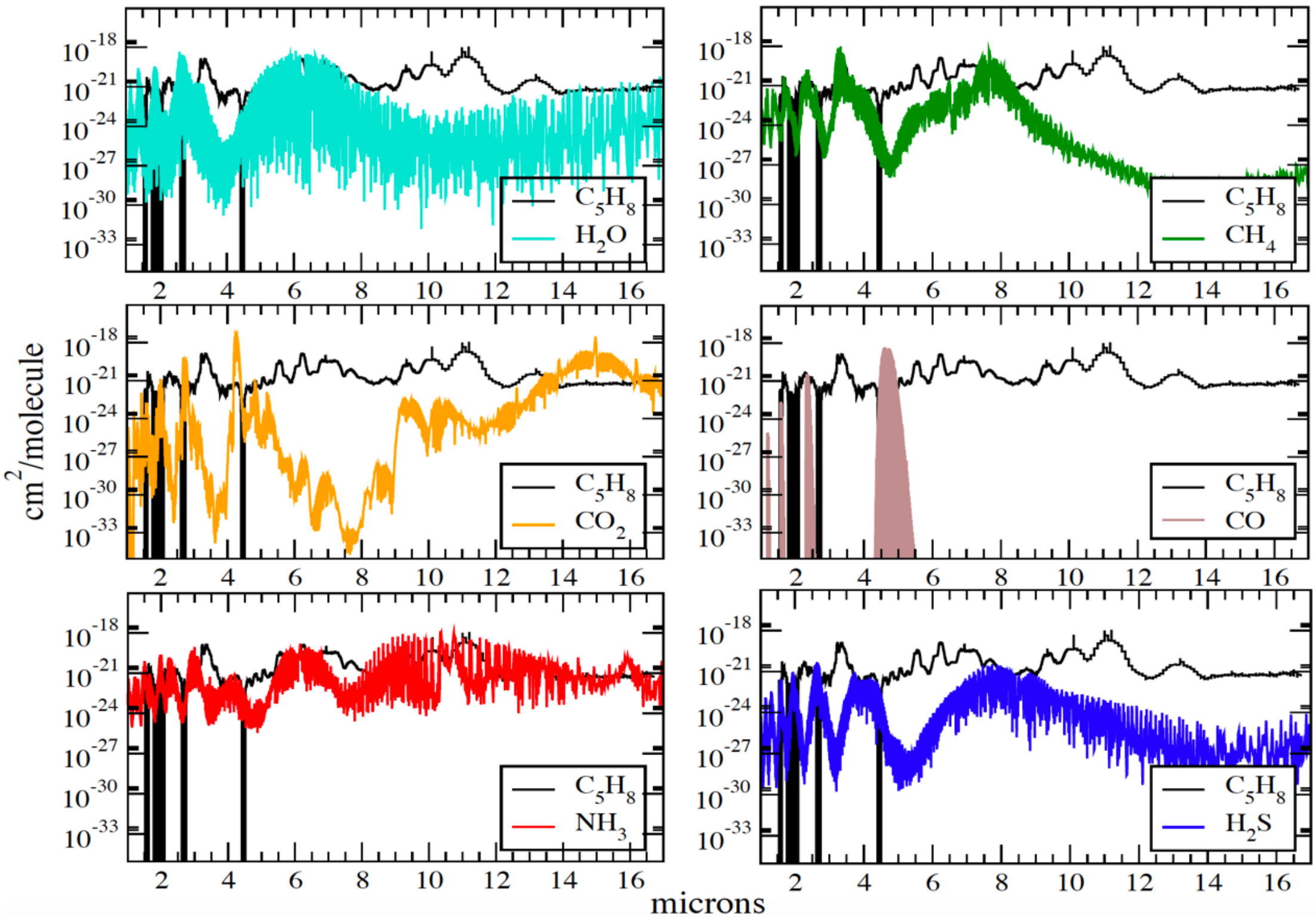

Figure I-3. Comparison of isoprene absorption cross sections to cross sections of other possible dominant gases, at room temperature, partially adapted from (Sousa-Silva *et al.* 2020). We plot the cross sections [$cm^2$ $molecule^{-1}$] vs wavelength [μm]. Cross sections for $C_5H_8$ is obtained from HITRAN (Brauer *et al.* 2014; Gordon *et al.* 2017). The $CO_2$ cross sections are from HITEMP (Rothman *et al.* 2010). All other molecular cross sections are simulated using molecular line lists from the ExoMol database (Sousa-Silva *et al.* 2015; Tennyson *et al.* 2016; Yurchenko *et al.* 2011; Yurchenko and Tennyson 2014) and are calculated at zero-pressure (i.e. Doppler-broadened lines only) using the procedure described

by (Hill *et al.* 2013). $C_5H_8$, shown in black, is distinguishable from all compared molecules due to its strong bands in the 3.1 - 3.6 μm and 9 - 12 μm regions.

## C: Consequences of an Isoprene Cross Section with an Artificial Noise Floor

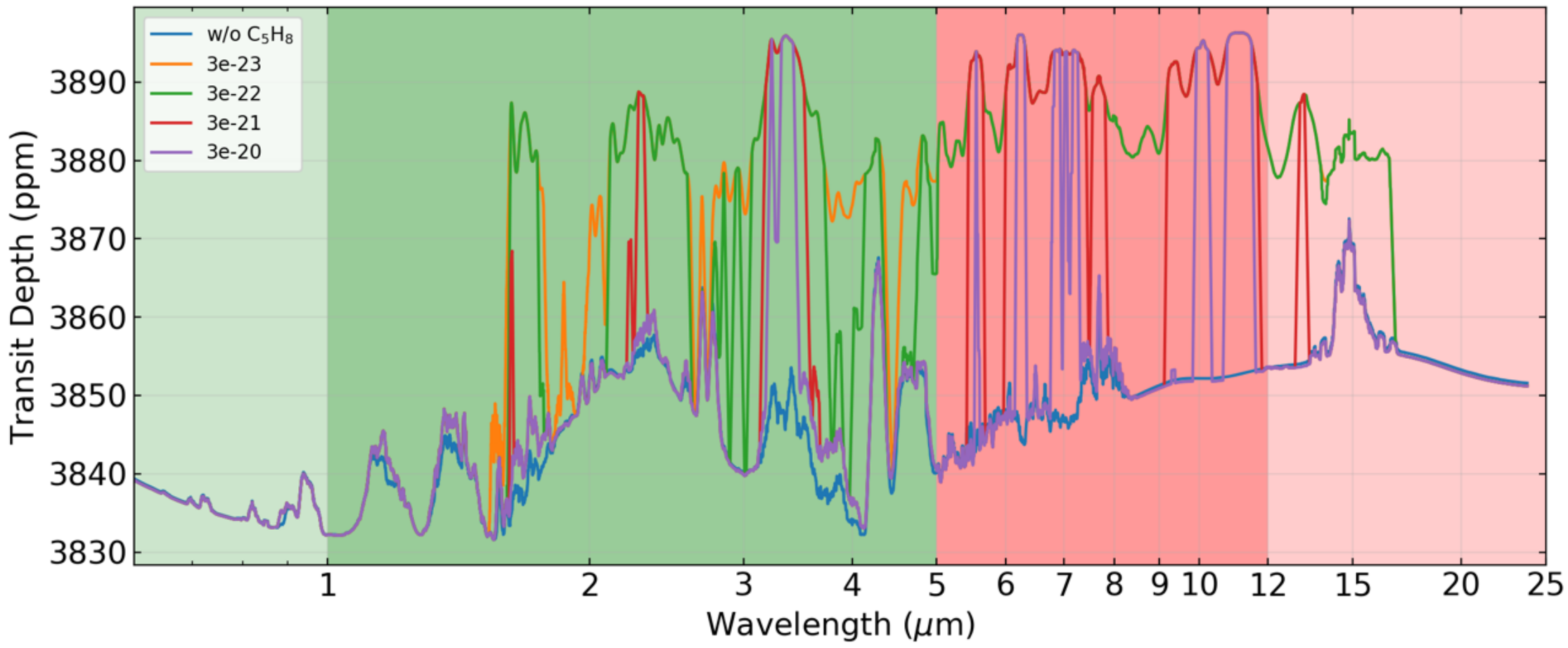

Figure I-4: Theoretical transmission spectra for an $H_2$-dominated atmosphere on a 10 $M_E$, 1.75 $R_E$ planet with a surface temperature of 288 K orbiting an active M dwarf (1 bar atmosphere composed of 90% $H_2$ and 10% $N_2$) with surface isoprene flux of $1 \times 10^{13}$ molecule $cm^{-2}$ $s^{-1}$. We show how truncating the cross sections with a uniform $3 \times 10^{-23}$ $cm^{-2}$ $molecule^{-1}$ threshold cut to $3 \times 10^{-20}$ $cm^{-2}$ $molecule^{-1}$ threshold cut will change the shape of the spectral features of isoprene in the transmission spectra and consequently the detectability assessment. Without knowing the wavelength-dependent noise floor, we cannot justify applying any threshold cut to the absorption data without “engineering” or removing spectral features.

# Appendix II: Structural Diversity and Biosynthesis of Isoprenoids

## IIA: Structural Diversity of Isoprenoids

Isoprenoids are the most numerous and structurally diverse family of natural compounds. Approximately 60% of known natural products belong to this group (Firn 2010). Isoprenoids are classified based on the number and structural organization of carbons formed by linear arrangement of isoprene subunits $(C_5H_8)_n$, according to the empirical feature known as the 'isoprene rule', and the degree of further secondary modifications to the isoprenoid polymer (e.g. cyclization and other rearrangements of the carbon skeleton). Isoprene is the "basic building block" of isoprenoids and therefore is technically the simplest representative of the isoprenoid family (Ludwiczuk *et al.* 2017).

The empirical 'isoprene rule' states that the majority of isoprenoids are synthetized from the ordered 'head-to-tail' polymerization of isoprene subunits (Ludwiczuk *et al.* 2017). While 'head-to-tail' polymerization is the most common, other, 'non-head-to-tail' condensation variants were also reported (Figure IIA1). 'Head-to-head' polymerization is common in the biosynthesis of carotenoids and triterpenoids while 'head-to-middle' polymerization and other more exotic ways to join isoprene subunits were observed in irregular monoterpenoids (Figure IIA1B). The large number of possible combinations of subunit joining allows for biosynthesis of an unimaginable number of biochemicals.

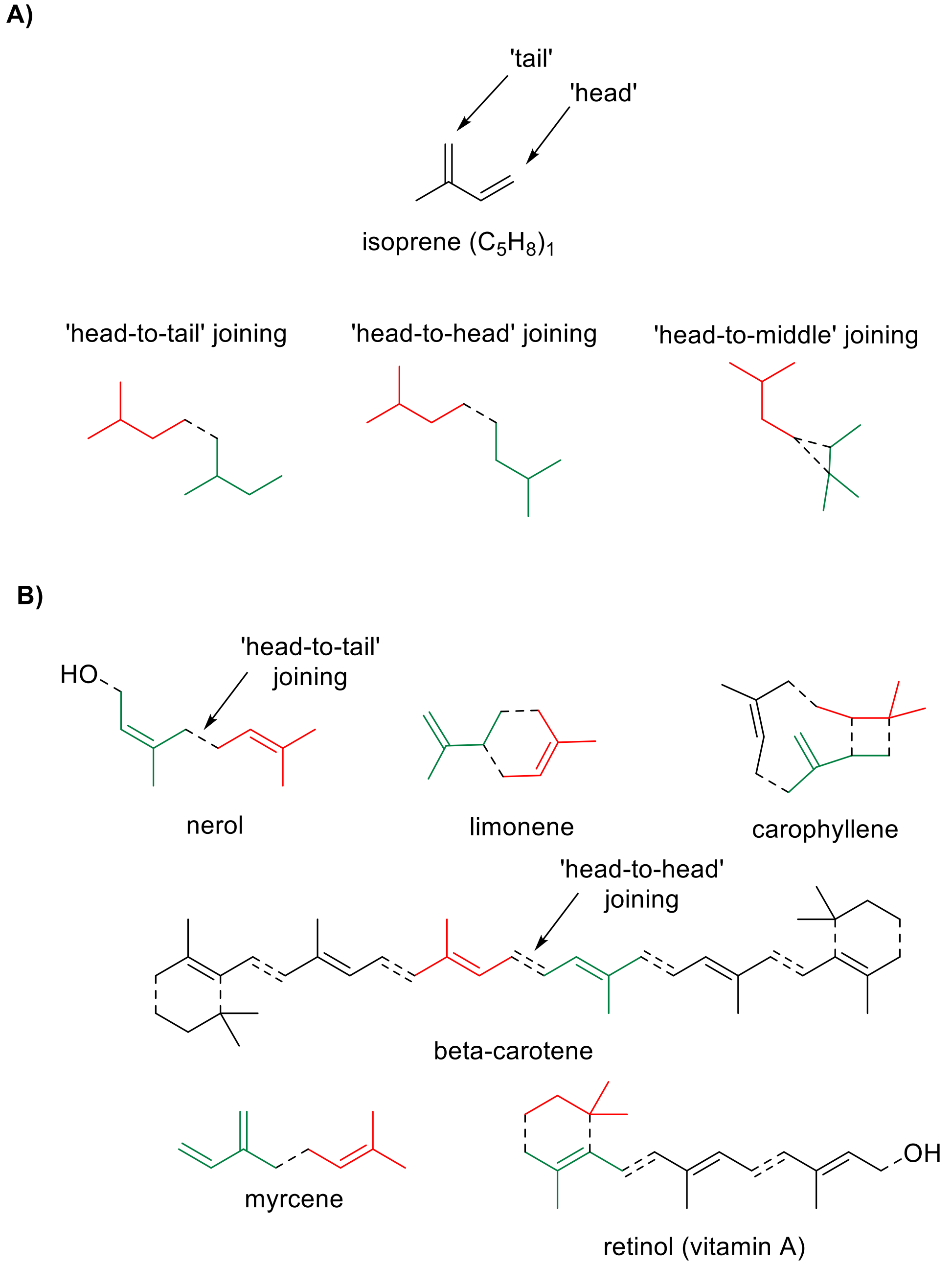

Figure IIA1. The structural diversity of isoprenoids. A) Schematic representation of examples of isoprene ($C_5H_8$)$_1$ subunit joining. The new bonds formed upon subunit joining are marked as dashed lines. The ‘head-to-tail’ polymerization of isoprene subunits is the most common mechanism in biosynthesis of isoprenoids B) Examples of biologically important isoprenoids and the type of polymerization of isoprene subunit utilized in their synthesis. The large number of possible combinations of isoprene subunit joining allows for biosynthesis of a very large number of diverse molecules.

## IIB: Biosynthesis of Isoprenoids

Isoprenoids (also known as terpenes or terpenoids), a large and diverse family of natural molecules, are polymers of isoprene or molecules that contain multiple isoprene subunits. Isoprenoids consist of two or more isoprene subunits $(C_5H_8)_n$. The simplest isoprenoids are monoterpenes, which consist of two isoprene subunits. The boiling point for monoterpenes are higher than water, and are not expected to be present in habitable planetary atmospheres. We note that isoprenoids are produced by all free-living organisms[9]. Although isoprene is not a direct biosynthetic precursor for isoprenoids, both isoprene itself as well as all isoprenoids share the same biosynthetic precursors: dimethylallyl pyrophosphate (DMAPP) and its isomer isopentenyl pyrophosphate (IPP).

The presence of isoprenoids on Earth's biology is ubiquitous. Isoprenoid biosynthetic pathways are utilized in the biosynthesis of such crucial molecules like: hopanoids and sterols, archaeal lipids, chlorophylls, carotenoids, electron carriers such as ubiquinone, not to mention a plethora of steroid hormones and signaling molecules (Rohmer 1999; Rohmer 2010). Isoprenoids are also present as post-translational modifications in proteins, providing attachment of proteins to cell membranes (e.g. geranylgeranylation or farnesylation) (Eastman *et al.* 2006; Wiemer *et al.* 2009). Isoprenoids are synthesized by polymerization of IPP and DMAPP (Figure IIB1). Depending on which of these two molecules are required for synthesis, DMAPP and IPP are interconverted by isopentenyldiphosphate isomerases (IPPI). Two evolutionary ancient biosynthetic pathways are utilized in the synthesis of IPP and DMAPP. The mevalonate (MVA) pathway is used by animals, fungi, archaea, some bacteria and in the cytosol of plants. The methylerythritol phosphate (MEP) pathway, also called the non-mevalonate or D(O)XP pathway is predominant in chloroplasts and most bacteria (Table IIB1) (Bentlage *et al.* 2015; Rohmer 1999).

The MVA and MEP biosynthetic pathways are mutually exclusive in most organisms. IPP and DMAPP are the end-products in either pathway, and are the precursors of isoprene, monoterpenoids $(C_5H_8)_2$, diterpenoids $(C_5H_8)_4$, carotenoids $(C_5H_8)_8$, chlorophylls, and plastoquinone-9 $(C_5H_8)_9$ and many other molecules.

---

[6] Many obligatorily parasitic organisms lost their ability to synthesize isoprenoids directly, however they still utilize them by stealing necessary isoprenoid biosynthetic precursors from their hosts (Imlay and Odom 2014).

| Organisms | Isoprenoid Biosynthetic Pathway | Reference |
|---|---|---|
| Bacteria (e.g. *Bacillus subtilis*) | MVA or MEP | (Kuzma *et al.* 1995; Kuzuyama and Seto 2012; Pérez-Gil and Rodríguez-Concepción 2013) |
| Archaea (e.g. *Archaeoglobus fulgidus*, *Methanocaldococcus jannaschii*) | MVA | (Grochowski and White 2008; Murakami *et al.* 2007) |
| Green Algae (e.g. *Mesostigma viride*) | MEP | (Grauvogel and Petersen 2007; Lohr *et al.* 2012) |
| Plants (e.g. *Quercus fusiformis*) | MVA and MEP | (Sharkey and Monson 2017; Sharkey *et al.* 2008) |
| Animals (e.g. *Mus musculus*, *Homo sapiens*) | MVA | (Sharkey 1996) |
| Fungi (e.g. *Meliniomyces variabilis*, *Eurotium amstelodami*) | MVA | (Bäck *et al.* 2010; Berenguer *et al.* 1991) |

Table IIB1. Isoprenoid biosynthetic pathways are ancient and universal with two distinct different starting points

A large number of enzymes utilize DMAPP and IPP to assemble different isoprenoid molecules. Many of these enzymes have a common terpenoid synthase protein fold. For example, isoprene synthase cleaves off the phosphate and releases isoprene, farnesyl diphosphate synthase (PDB entry 1fps) joins three isoprene units to create a longer linear isoprenoid chain, bornyl diphosphate synthase (PDB entry 1n20) catalyzes ring formation out of longer isoprenoid chains. Terpenoid synthases are often crucial in biosynthesis of very complicated natural products, in which the presence of the intermediate isoprene unit is not immediately apparent. An example of such enzyme is taxadiene synthase (PDB entry 3p5p) that uses isoprene units in the biosynthesis of an important anti-cancer drug taxol.

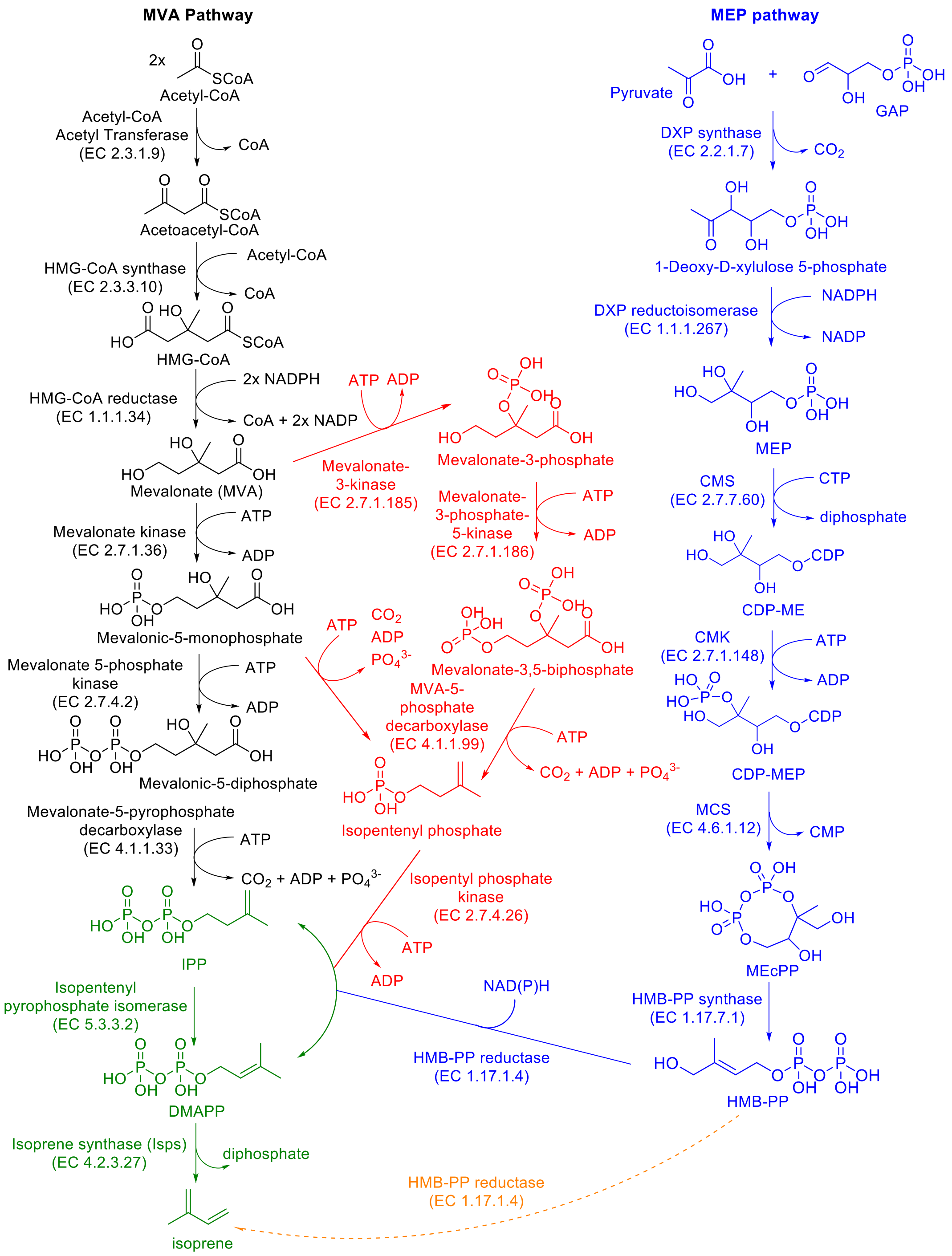

**Figure IIB1.** Mevalonic acid (MVA) and methylerythritol phosphate (MEP) biosynthetic pathways. Both the MVA pathway (left) and the MEP pathway (right) lead to the formation of isopentenyl-diphosphate (IPP) and dimethylallyl-diphosphate (DMAPP), which serve as substrates to all isoprenoids, including isoprene. The MVA pathway is present in eukaryotes, archaea, and some bacteria. The eukaryotic variant is shown in black font. Archaeal modifications to the MVA pathway are shown in red. The MEP pathway (shown in blue) is present in bacteria, green algae and in chloroplasts in plants. Metabolic steps common to both isoprenoid biosynthetic pathways are shown in green. Orange dashed line depicts a potential pathway for isoprene synthesis in bacteria *Bacillus* sp. (Ge *et al.* 2016).

# References

Abdelrahman O. A., Park D. S., Vinter K. P., Spanjers C. S., Ren L., Cho H. J., Zhang K., Fan W., Tsapatsis M., and Dauenhauer P. J. (2017) Renewable isoprene by sequential hydrogenation of itaconic acid and dehydra-decyclization of 3-methyl-tetrahydrofuran. *ACS Catalysis*, 7: 1428-1431.

Allard F., Hauschildt P. H., Alexander D. R., and Starrfield S. (1997) Model atmospheres of very low mass stars and brown dwarfs. *Annual Review of Astronomy and Astrophysics*, 35: 137-177.

Alvarez L. A., Exton D. A., Timmis K. N., Suggett D. J., and McGenity T. J. (2009) Characterization of marine isoprene-degrading communities. *Environmental microbiology*, 11: 3280-3291.

Anderson M. R. (2000) Determination of infrared optical constants for single component hydrocarbon fuels. MISSOURI UNIV-ROLLA.

Arneth A., Monson R. K., Schurgers G., Niinemets Ü., and Palmer P. I. (2008) Why are estimates of global terrestrial isoprene emissions so similar (and why is this not so for monoterpenes)? *Atmos. Chem. Phys.*, 8: 4605-4620.

Arney G., Domagal-Goldman S. D., and Meadows V. S. (2018) Organic haze as a biosignature in anoxic Earth-like atmospheres. *Astrobiology*, 18: 311-329.

Bäck J., Aaltonen H., Hellén H., Kajos M. K., Patokoski J., Taipale R., Pumpanen J., and Heinonsalo J. (2010) Variable emissions of microbial volatile organic compounds (MVOCs) from root-associated fungi isolated from Scots pine. *Atmospheric Environment*, 44: 3651-3659.

Bagnasco G., Kolm M., Ferruit P., Honnen K., Koehler J., Lemke R., Maschmann M., Melf M., Noyer G., and Rumler P. (2007) Overview of the near-infrared spectrograph (NIRSpec) instrument on-board the James Webb Space Telescope (JWST). International Society for Optics and Photonics.

Bains W., Petkowski J. J., and Seager S. (2017) Toward a List of Molecules as Potential Biosignature Gases for the Search for Life on Exoplanets: Thermodynamic Profiling Potential False Positives. In: *Astrobiology Science Conference (AbSciCon)*, Mesa, Arizona.

Batalha N. E., Mandell A., Pontoppidan K., Stevenson K. B., Lewis N. K., Kalirai J., Earl N., Greene T., Albert L., and Nielsen L. D. (2017) PandExo: a community tool for transiting exoplanet science with JWST & HST. *Publications of the Astronomical Society of the Pacific*, 129: 064501.

Beck Z. Q., Cervin M. A., Chotani G. K., Diner B. A., Fan J., Peres C. M., Sanford K. J., Scotcher M. C., Wells D. H., and Whited G. M. (2014) Recombinant Anaerobic Acetogenic Bacteria for Production of Isoprene and/or Industrial Bio-Products Using Synthesis Gas. Google Patents, pp US 2014/0234926A1-US 2014/0234926A1.

Bentlage B., Rogers T. S., Bachvaroff T. R., and Delwiche C. F. (2015) Complex Ancestries of Isoprenoid Synthesis in Dinoflagellates. *Journal of Eukaryotic Microbiology*, 63: 123-137.

Berenguer J. A., Calderon V., Herce M. D., and Sanchez J. J. (1991) Spoilage of a bakery product (sobao pasiego) by isoprene-producing molds. *Revista de Agroquimica y Tecnologia de Alimentos (Spain)*.

Berk A., Bernstein L. S., Anderson G. P., Acharya P. K., Robertson D. C., Chetwynd J. H., and Adler-Golden S. M. (1998) MODTRAN cloud and multiple scattering upgrades with application to AVIRIS. *Remote sensing of Environment*, 65: 367-375.
Bernath P. F. (2017) The Atmospheric Chemistry Experiment (ACE). *Journal of Quantitative Spectroscopy and Radiative Transfer*, 186: 3-16.
Bernath P. F., McElroy C. T., Abrams M. C., Boone C. D., Butler M., Camy-Peyret C., Carleer M., Clerbaux C., Coheur P. F., and Colin R. (2005) Atmospheric chemistry experiment (ACE): mission overview. *Geophysical Research Letters*, 32.
Brauer C. S., Blake T. A., Guenther A. B., Sharpe S. W., Sams R. L., and Johnson T. J. (2014) Quantitative infrared absorption cross sections of isoprene for atmospheric measurements.
Chu P. M., Guenther F. R., Rhoderick G. C., and Lafferty W. J. (1999) The NIST quantitative infrared database. *Journal of research of the National Institute of Standards and Technology*, 104: 59.
Ciuraru R., Fine L., Pinxteren M. v., D'Anna B., Herrmann H., and George C. (2015) Unravelling new processes at interfaces: photochemical isoprene production at the sea surface. *Environmental science & technology*, 49: 13199-13205.
Claeys M., Graham B., Vas G., Wang W., Vermeylen R., Pashynska V., Cafmeyer J., Guyon P., Andreae M. O., Artaxo P. and others. (2004) Formation of Secondary Organic Aerosols Through Photooxidation of Isoprene. *Science*, 303: 1173.
Cleveland C. C., and Yavitt J. B. (1998) Microbial Consumption of Atmospheric Isoprene in a Temperate Forest Soil. *Applied and Environmental Microbiology*, 64: 172 LP-177.
Curdt W., Landi E., and Feldman U. (2004) The SUMER spectral atlas of solar coronal features. *Astronomy & Astrophysics*, 427: 1045-1054.
Dal Corso J., Mietto P., Newton R. J., Pancost R. D., Preto N., Roghi G., and Wignall P. B. (2012) Discovery of a major negative δ13C spike in the Carnian (Late Triassic) linked to the eruption of Wrangellia flood basalts. *Geology*, 40: 79-82.
Dalzell W. H., and Sarofim A. F. (1969) Optical constants of soot and their application to heat-flux calculations. *Journal of Heat Transfer*, 91: 100-104.
Deneris E. S., Stein R. A., and Mead J. F. (1985) Acid-catalyzed formation of isoprene from a mevalonate-derived product using a rat liver cytosolic fraction. *Journal of Biological Chemistry*, 260: 1382-1385.
Dillon T. J., Dulitz K., Groß C., and Crowley J. N. (2017) Temperature-dependent rate coefficients for the reactions of the hydroxyl radical with the atmospheric biogenics isoprene, alpha-pinene and delta-3-carene. *Atmospheric Chemistry and Physics*, 17: 15137-15150.
Dlugokencky E. J., Nisbet E. G., Fisher R., and Lowry D. (2011) Global atmospheric methane: budget, changes and dangers. *Philosophical Transactions of the Royal Society A: Mathematical, Physical and Engineering Sciences*, 369: 2058-2072.
Domagal-Goldman S. D., Meadows V. S., Claire M. W., and Kasting J. F. (2011) Using biogenic sulfur gases as remotely detectable biosignatures on anoxic planets. *Astrobiology*, 11: 419-441.

Eastman R. T., Buckner F. S., Yokoyama K., Gelb M. H., and Van Voorhis W. C. (2006) Thematic review series: lipid posttranslational modifications. Fighting parasitic disease by blocking protein farnesylation. *Journal of lipid research*, 47: 233-240.

Ehrenfreund P., and Cami J. (2010) Cosmic carbon chemistry: from the interstellar medium to the early Earth. *Cold Spring Harbor perspectives in biology*, 2: a002097.

Exton D. A., Suggett D. J., McGenity T. J., and Steinke M. (2013) Chlorophyll-normalized isoprene production in laboratory cultures of marine microalgae and implications for global models. *Limnology and Oceanography*, 58: 1301-1311.

Fall R., and Copley S. D. (2000) Bacterial sources and sinks of isoprene, a reactive atmospheric hydrocarbon. *Environmental Microbiology*, 2: 123-130.

Fall R. R., Kuzma J., and Nemecek-Marshall M. (1998) Materials and methods for the bacterial production of isoprene. Google Patents.

Fan J., and Zhang R. (2004) Atmospheric oxidation mechanism of isoprene. *Environmental Chemistry*, 1: 140-149.

Fan S., Shemansky D. E., Li C., Gao P., Wan L., and Yung Y. L. (2019) Retrieval of Chemical Abundances in Titan's Upper Atmosphere from Cassini UVIS Observations with Pointing Motion. *Earth and Space Science*.

Firn R. (2010) Nature's Chemicals: The Natural Products that Shaped Our World. Oxford University Press.

France K., Froning C. S., Linsky J. L., Roberge A., Stocke J. T., Tian F., Bushinsky R., Désert J.-M., Mauas P., and Vieytes M. (2013) The ultraviolet radiation environment around M dwarf exoplanet host stars. *The Astrophysical Journal*, 763: 149.

Gardner J. P., Mather J. C., Clampin M., Doyon R., Greenhouse M. A., Hammel H. B., Hutchings J. B., Jakobsen P., Lilly S. J., Long K. S. and others. (2006) The James Webb Space Telescope. *Space Science Reviews*, 123: 485-606.

Ge D., Xue Y., and Ma Y. (2016) Two unexpected promiscuous activities of the iron–sulfur protein IspH in production of isoprene and isoamylene. *Microbial cell factories*, 15: 79.

Gelmont D., Stein R. A., and Mead J. F. (1981) Isoprene - The main hydrocarbon in human breath. *Biochemical and Biophysical Research Communications*, 99: 1456-1460.

Gershenzon J. (2008) Insects turn up their noses at sweating plants. *Proceedings of the National Academy of Sciences of the United States of America*, 105: 17211-17212.

Gordon I. E., Rothman L. S., Hill C., Kochanov R. V., Tan Y., Bernath P. F., Birk M., Boudon V., Campargue A., and Chance K. V. (2017) The HITRAN2016 molecular spectroscopic database. *Journal of Quantitative Spectroscopy and Radiative Transfer*, 203: 3-69.

Grauvogel C., and Petersen J. (2007) Isoprenoid biosynthesis authenticates the classification of the green alga Mesostigma viride as an ancient streptophyte. *Gene*, 396: 125-133.

Grenfell J. L. (2017) A review of exoplanetary biosignatures. *Physics Reports*, 713: 1-17.

Grenfell J. L. (2018) Atmospheric Biosignatures. In: *Handbook of Exoplanets*. edited by HJ Deeg and JA Belmontes, Springer International Publishing, Cham, pp 1-14.
Grochowski L. L., and White R. H. (2008) Promiscuous Anaerobes. *Annals of the New York Academy of Sciences*, 1125: 190-214.
Guenther A., Karl T., Harley P., Wiedinmyer C., Palmer P. I., and Geron C. (2006) Estimates of global terrestrial isoprene emissions using MEGAN (Model of Emissions of Gases and Aerosols from Nature). *Atmospheric Chemistry and Physics*, 6: 3181-3210.
Guenther A. B., Jiang X., Heald C. L., Sakulyanontvittaya T., Duhl T., Emmons L. K., and Wang X. (2012) The Model of Emissions of Gases and Aerosols from Nature version 2.1 (MEGAN2.1): an extended and updated framework for modeling biogenic emissions. *Geosci. Model Dev.*, 5: 1471-1492.
Günther M. N., Zhan Z., Seager S., Rimmer P. B., Ranjan S., Stassun K. G., Oelkers R. J., Daylan T., Newton E., and Gillen E. (2019) Stellar flares from the first tess data release: exploring a new sample of M-dwarfs. *arXiv preprint arXiv:1901.00443*.
Harvey C. M., and Sharkey T. D. (2016) Exogenous isoprene modulates gene expression in unstressed Arabidopsis thaliana plants. *Plant, Cell & Environment*, 39: 1251-1263.
He C., Hörst S. M., Lewis N. K., Moses J. I., Kempton E. M. R., Marley M. S., Morley C. V., Valenti J. A., and Vuitton V. (2019) Gas Phase Chemistry of Cool Exoplanet Atmospheres: Insight from Laboratory Simulations. *ACS Earth and Space Chemistry*, 3: 39-50.
He C., Hörst S. M., Lewis N. K., Yu X., Moses J. I., Kempton E. M. R., McGuiggan P., Morley C. V., Valenti J. A., and Vuitton V. (2018) Laboratory Simulations of Haze Formation in the Atmospheres of Super-Earths and Mini-Neptunes: Particle Color and Size Distribution. *The Astrophysical Journal Letters*, 856: L3.
Hess B. M., Xue J., Markillie L. M., Taylor R. C., Wiley H. S., Ahring B. K., and Linggi B. (2013) Coregulation of Terpenoid Pathway Genes and Prediction of Isoprene Production in Bacillus subtilis Using Transcriptomics. *PLOS ONE*, 8: e66104-e66104.
Hill C., Yurchenko S. N., and Tennyson J. (2013) Temperature-dependent molecular absorption cross sections for exoplanets and other atmospheres. *Icarus*, 226: 1673-1677.
Holland H. D. (2006) The oxygenation of the atmosphere and oceans. *Philosophical Transactions of the Royal Society B: Biological Sciences*, 361: 903-915.
Hörst S. M., He C., Lewis N. K., Kempton E. M. R., Marley M. S., Morley C. V., Moses J. I., Valenti J. A., and Vuitton V. (2018) Haze production rates in super-Earth and mini-Neptune atmosphere experiments. *Nature Astronomy*, 2: 303.
Hu R., Seager S., and Bains W. (2012) Photochemistry in Terrestrial Exoplanet Atmospheres. I. Photochemistry Model and Benchmark Cases. *The Astrophysical Journal*, 761: 166.
Hughes R., Bernath P., and Boone C. (2014) ACE infrared spectral atlases of the Earth′ s atmosphere. *Journal of Quantitative Spectroscopy and Radiative Transfer*, 148: 18-21.

Imlay L., and Odom A. R. (2014) Isoprenoid metabolism in apicomplexan parasites. *Current clinical microbiology reports*, 1: 37-50.

Jeans J. (1930) The universe around us.

Johns M., McCarthy P., Raybould K., Bouchez A., Farahani A., Filgueira J., Jacoby G., Shectman S., and Sheehan M. (2012) Giant Magellan Telescope: overview.SPIE Astronomical Telescopes + Instrumentation. SPIE.

Jones A. M. P., Shukla M. R., Sherif S. M., Brown P. B., and Saxena P. K. (2016) Growth regulating properties of isoprene and isoprenoid-based essential oils. *Plant Cell Reports*, 35: 91-102.

Karl T., Potosnak M., Guenther A., Clark D., Walker J., Herrick J. D., and Geron C. (2004) Exchange processes of volatile organic compounds above a tropical rain forest: Implications for modeling tropospheric chemistry above dense vegetation. *Journal of Geophysical Research: Atmospheres*, 109.

Kasting J. F. (2014) Atmospheric composition of Hadean–early Archean Earth: the importance of CO. *Geological Society of America Special Papers*, 504: 19-28.

Kasting J. F., Kopparapu R., Ramirez R. M., and Harman C. E. (2014) Remote life-detection criteria, habitable zone boundaries, and the frequency of Earth-like planets around M and late K stars. *Proceedings of the National Academy of Sciences*, 111: 12641.

Kasting J. F., Pollack J. B., and Crisp D. (1984) Effects of high CO2 levels on surface temperature and atmospheric oxidation state of the early Earth. *Journal of Atmospheric Chemistry*, 1: 403-428.

Kasting J. F., Zahnle K. J., and Walker J. C. G. (1983) Photochemistry of methane in the Earth's early atmosphere. *Precambrian Research*, 20: 121-148.

Kempton E. M. R., Lupu R., Owusu-Asare A., Slough P., and Cale B. (2017) Exo-Transmit: An open-source code for calculating transmission spectra for exoplanet atmospheres of varied composition. *Publications of the Astronomical Society of the Pacific*, 129: 044402.

Khare B. N., Sagan C., Arakawa E. T., Suits F., Callcott T. A., and Williams M. W. (1984) Optical constants of organic tholins produced in a simulated Titanian atmosphere: From soft X-ray to microwave frequencies. *Icarus*, 60: 127-137.

Khare B. N., Sagan C., Thompson W. R., Arakawa E. T., Meisse C., and Tuminello P. S. (1994) Optical properties of poly-HCN and their astronomical applications. *Canadian journal of chemistry*, 72: 678-694.

Kiang N. Y., Domagal-Goldman S., Parenteau M. N., Catling D. C., Fujii Y., Meadows V. S., Schwieterman E. W., and Walker S. I. (2018) Exoplanet Biosignatures: At the Dawn of a New Era of Planetary Observations. *Astrobiology*, 18: 619-629.

King J., Koc H., Unterkofler K., Mochalski P., Kupferthaler A., Teschl G., Teschl S., Hinterhuber H., and Amann A. (2010) Physiological modeling of isoprene dynamics in exhaled breath. *Journal of Theoretical Biology*, 267: 626-637.

Knoll A. H., Canfield D. E., and Konhauser K. O. (2012) Fundamentals of geobiology. John Wiley & Sons.

Kochanov R. V., Gordon I. E., Rothman L. S., Wcisło P., Hill C., and Wilzewski J. S. (2016) HITRAN Application Programming Interface (HAPI): A comprehensive approach to working with spectroscopic data. *Journal of Quantitative Spectroscopy and Radiative Transfer*, 177: 15-30.
Köksal M., Zimmer I., Schnitzler J.-P., and Christianson D. W. (2010) Structure of isoprene synthase illuminates the chemical mechanism of teragram atmospheric carbon emission. *Journal of molecular biology*, 402: 363-373.
Korhonen H., Carslaw K. S., Spracklen D. V., Mann G. W., and Woodhouse M. T. (2008) Influence of oceanic dimethyl sulfide emissions on cloud condensation nuclei concentrations and seasonality over the remote Southern Hemisphere oceans: A global model study. *Journal of Geophysical Research: Atmospheres*, 113.
Kuzma J., Nemecek-Marshall M., Pollock W. H., and Fall R. (1995) Bacteria produce the volatile hydrocarbon isoprene. *Current Microbiology*, 30: 97-103.
Kuzuyama T., and Seto H. (2012) Two distinct pathways for essential metabolic precursors for isoprenoid biosynthesis. *Proceedings of the Japan Academy. Series B, Physical and Biological Sciences*, 88: 41-52.
Laothawornkitkul J., Paul N. D., Vickers C. E., Possell M., Taylor J. E., Mullineaux P. M., and Hewitt C. N. (2008) Isoprene emissions influence herbivore feeding decisions. *Plant, Cell and Environment*, 31: 1410-1415.
Levy R. L., Grayson M. A., and Wolf C. J. (1973) The organic analysis of the murchison meteorite. *Geochimica et Cosmochimica Acta*, 37: 467-483.
Linstrom P. J., and Mallard W. G. (2001) The NIST Chemistry WebBook: A Chemical Data Resource on the Internet. *Journal of Chemical & Engineering Data*, 46: 1059-1063.
Loferer-Krößbacher M., Klima J., and Psenner R. (1998) Determination of bacterial cell dry mass by transmission electron microscopy and densitometric image analysis. *Applied and environmental microbiology*, 64: 688-694.
Logan B. A., Monson R. K., and Potosnak M. J. (2000) Biochemistry and physiology of foliar isoprene production. *Trends in Plant Science*, 5: 477-481.
Lohr M., Schwender J., and Polle J. E. W. (2012) Isoprenoid biosynthesis in eukaryotic phototrophs: A spotlight on algae. *Plant Science*, 185-186: 9-22.
Ludwiczuk A., Skalicka-Woźniak K., and Georgiev M. I. (2017) Terpenoids. *Pharmacognosy*: 233-266.
Maeda K., Spor A., Edel-Hermann V., Heraud C., Breuil M.-C., Bizouard F., Toyoda S., Yoshida N., Steinberg C., and Philippot L. (2015) N(2)O production, a widespread trait in fungi. *Scientific Reports*, 5: 9697-9697.
Maltagliati L., Bézard B., Vinatier S., Hedman M. M., Lellouch E., Nicholson P. D., Sotin C., de Kok R. J., and Sicardy B. (2015) Titan's atmosphere as observed by Cassini/VIMS solar occultations: CH4, CO and evidence for C2H6 absorption. *Icarus*, 248: 1-24.
McBride E. J., Millar T. J., and Kohanoff J. J. (2013) Organic synthesis in the interstellar medium by low-energy carbon irradiation. *The Journal of Physical Chemistry A*, 117: 9666-9672.
McElroy D., Walsh C., Markwick A. J., Cordiner M. A., Smith K., and Millar T. J. (2013) The UMIST database for astrochemistry 2012. *Astronomy & Astrophysics*, 550: A36.

McGenity T. J., Crombie A. T., and Murrell J. C. (2018) Microbial cycling of isoprene, the most abundantly produced biological volatile organic compound on Earth. *ISME Journal*, 12: 931-941.
Michelozzi M., Raschi A., Tognetti R., and Tosi L. (1997) Eco-ethological analysis of the interaction between isoprene and the behaviour of Collembola. *Pedobiologia*, 41: 210-214.
Miller-Ricci E., Meyer M. R., Seager S., and Elkins-Tanton L. (2009) On the emergent spectra of hot protoplanet collision afterglows. *The Astrophysical Journal*, 704: 770.
Moore R. M., Oram D. E., and Penkett S. A. (1994) Production of isoprene by marine phytoplankton cultures. *Geophysical Research Letters*, 21: 2507-2510.
Murakami M., Shibuya K., Nakayama T., Nishino T., Yoshimura T., and Hemmi H. (2007) Geranylgeranyl reductase involved in the biosynthesis of archaeal membrane lipids in the hyperthermophilic archaeon Archaeoglobus fulgidus. *The FEBS Journal*, 274: 805-814.
Murphy N., Roswitha B., Weber K. A., Aldridge J. T., and Carr S. R. (2017) Production Of Isoprene By Methane-producing Archaea. Google Patents.
Newby J. J., Stearns J. A., Liu C.-P., and Zwier T. S. (2007) Photochemical and Discharge-Driven Pathways to Aromatic Products from 1,3-Butadiene.
Panchenko Y. N., and De Maré G. R. (2008) Vibrational analysis of buta-1, 3-diene and its deutero and 13 C derivatives and some of their rotational isomers. *Journal of structural chemistry*, 49: 235.
Pascale E., Bezawada N., Barstow J., Beaulieu J.-P., Bowles N., du Foresto V. C., Coustenis A., Decin L., Drossart P., and Eccleston P. (2018) The ARIEL space mission.Space Telescopes and Instrumentation 2018: Optical, Infrared, and Millimeter Wave. International Society for Optics and Photonics.
Paulson S. E., Flagan R. C., and Seinfeld J. H. (1992) Atmospheric photooxidation of isoprene part I: The hydroxyl radical and ground state atomic oxygen reactions. *International Journal of Chemical Kinetics*, 24: 79-101.
Peñuelas J., Llusia J., Asensio D., and Munné-Bosch S. (2005) Linking isoprene with plant thermotolerance, antioxidants and monoterpene emissions. *Plant, Cell & Environment*, 28: 278-286.
Pérez-Gil J., and Rodríguez-Concepción M. (2013) Metabolic plasticity for isoprenoid biosynthesis in bacteria. *Biochemical Journal*, 452: 19 LP-25.
Pilcher C. B. (2003) Biosignatures of Early Earths. *Astrobiology*, 3: 471-486.
Pizzarello S., and Shock E. (2010) The organic composition of carbonaceous meteorites: the evolutionary story ahead of biochemistry. *Cold Spring Harbor perspectives in biology*, 2: a002105.
Rackham B. V., Apai D., and Giampapa M. S. (2018) The transit light source effect: false spectral features and incorrect densities for M-dwarf transiting planets. *The Astrophysical Journal*, 853: 122.
Rohmer M. (1999) The discovery of a mevalonate-independent pathway for isoprenoid biosynthesis in bacteria, algae and higher plants†. *Natural Product Reports*, 16: 565-574.
Rohmer M. (2010) Methylerythritol Phosphate Pathway. *Comprehensive Natural Products II*: 517-555.

Rothman L. S., Gordon I. E., Barber R. J., Dothe H., Gamache R. R., Goldman A., Perevalov V. I., Tashkun S. A., and Tennyson J. (2010) HITEMP, the high-temperature molecular spectroscopic database. *Journal of Quantitative Spectroscopy and Radiative Transfer*, 111: 2139-2150.

Royer D. L., Berner R. A., Montañez I. P., Tabor N. J., and Beerling D. J. (2004) Co~2 as a primary driver of phanerozoic climate. *GSA today*, 14: 4-10.

Schöller C., Molin S., and Wilkins K. (1997) Volatile metabolites from some gram-negative bacteria. *Chemosphere*, 35: 1487-1495.

Schöller C. E. G., Gürtler H., Pedersen R., Molin S., and Wilkins K. (2002) Volatile metabolites from actinomycetes. *Journal of Agricultural and Food Chemistry*, 50: 2615-2621.

Schreier F., Städt S., Hedelt P., and Godolt M. (2018) Transmission spectroscopy with the ACE-FTS infrared spectral atlas of Earth: A model validation and feasibility study. *Molecular Astrophysics*, 11: 1-22.

Schwieterman E. W., Kiang N. Y., Parenteau M. N., Harman C. E., DasSarma S., Fisher T. M., Arney G. N., Hartnett H. E., Reinhard C. T., Olson S. L. and others. (2018) Exoplanet Biosignatures: A Review of Remotely Detectable Signs of Life. *Astrobiology*, 18: 663-708.

Seager S. (2010) Exoplanet Atmospheres: Physical Processes (Princeton, NJ). Princeton Univ. Press.

Seager S., Bains W., and Hu R. (2013) Biosignature gases in H2-dominated atmospheres on rocky exoplanets. *The Astrophysical Journal*, 777: 95.

Seager S., Bains W., and Petkowski J. J. (2016) Toward a List of Molecules as Potential Biosignature Gases for the Search for Life on Exoplanets and Applications to Terrestrial Biochemistry. *Astrobiology*, 16: 465-485.

Seager S., Huang J., Petkowski J. J., and Pajusalu M. (2020) Laboratory studies on the viability of life in H2-dominated exoplanet atmospheres. *Nature Astronomy*, 4: 802-806.

Seager S., Schrenk M., and Bains W. (2012) An astrophysical view of Earth-based metabolic biosignature gases. *Astrobiology*, 12: 61-82.

Segura A., Kasting J. F., Meadows V., Cohen M., Scalo J., Crisp D., Butler R. A., and Tinetti G. (2005) Biosignatures from Earth-like planets around M dwarfs. *Astrobiology*, 5: 706-725.

Segura A., Walkowicz L. M., Meadows V., Kasting J., and Hawley S. (2010) The effect of a strong stellar flare on the atmospheric chemistry of an Earth-like planet orbiting an M dwarf. *Astrobiology*, 10: 751-771.

Seinfeld J. H., and Pandis S. N. (2016) Atmospheric chemistry and physics: from air pollution to climate change. John Wiley & Sons.

Sephton M. A. (2004) Organic matter in ancient meteorites. *Astronomy & Geophysics*, 45: 2.08-2.14.

Sharkey T. D. (1996) Isoprene synthesis by plants and animals. *Endeavour*, 20: 74-78.

Sharkey T. D., and Loreto F. (1993) Water stress, temperature, and light effects on the capacity for isoprene emission and photosynthesis of kudzu leaves. *Oecologia*, 95: 328-333.

Sharkey T. D., and Monson R. K. (2017) Isoprene research – 60 years later, the biology is still enigmatic. *Plant Cell and Environment*, 40: 1671-1678.

Sharkey T. D., Wiberley A. E., and Donohue A. R. (2008) Isoprene emission from plants: Why and how. *Annals of Botany*, 101: 5-18.

Shennan J. L. (2005) Utilisation of C2–C4 gaseous hydrocarbons and isoprene by microorganisms. *Journal of Chemical Technology & Biotechnology*, 81: 237-256.
Sivy T. L., Shirk M. C., and Fall R. (2002) Isoprene synthase activity parallels fluctuations of isoprene release during growth of Bacillus subtilis. *Biochemical and Biophysical Research Communications*, 294: 71-75.
Skidmore W., Teams T. M. T. I. S. D., and Committee T. M. T. S. A. (2015) Thirty Meter Telescope Detailed Science Case: 2015. *Research in Astronomy and Astrophysics*, 15: 1945.
Sousa-Silva C., Al-Refaie A. F., Tennyson J., and Yurchenko S. N. (2015) ExoMol line lists – VII. The rotation–vibration spectrum of phosphine up to 1500 K. *Monthly Notices of the Royal Astronomical Society*, 446: 2337-2347.
Sousa-Silva C., Petkowski J. J., and Seager S. (2019) Molecular Simulations for the Spectroscopic Detection of Atmospheric Gases. *Physical Chemistry Chemical Physics*, 21: 18970-18987.
Sousa-Silva C., Seager S., Ranjan S., Petkowski J. J., Zhan Z., Hu R., and Bains W. (2020) Phosphine as a Biosignature Gas in Exoplanet Atmospheres. *Astrobiology*, 20.
Takagi S., Mahieux A., Wilquet V., Robert S., Vandaele A. C., and Iwagami N. (2019) An uppermost haze layer above 100 km found over Venus by the SOIR instrument onboard Venus Express. *Earth, Planets and Space*, 71: 124.
Tamai R., and Spyromilio J. (2014) European Extremely Large Telescope: progress report.SPIE Astronomical Telescopes + Instrumentation. SPIE.
Taylor T. C., Smith M. N., Slot M., and Feeley K. J. (2019) The capacity to emit isoprene differentiates the photosynthetic temperature responses of tropical plant species. *Plant, Cell & Environment*, 0.
Teng A. P., Crounse J. D., and Wennberg P. O. (2017) Isoprene Peroxy Radical Dynamics. *Journal of the American Chemical Society*, 139: 5367-5377.
Tennyson J., and Yurchenko S. N. (2012) ExoMol: molecular line lists for exoplanet and other atmospheres. *Monthly Notices of the Royal Astronomical Society*, 425: 21-33.
Tennyson J., Yurchenko S. N., Al-Refaie A. F., Barton E. J., Chubb K. L., Coles P. A., Diamantopoulou S., Gorman M. N., Hill C., Lam A. Z. and others. (2016) The ExoMol database: Molecular line lists for exoplanet and other hot atmospheres. *Journal of Molecular Spectroscopy*, 327: 73-94.
Tessenyi M., Tinetti G., Savini G., and Pascale E. (2013) Molecular detectability in exoplanetary emission spectra. *Icarus*, 226: 1654-1672.
Tian H., Chen G., Lu C., Xu X., Ren W., Zhang B., Banger K., Tao B., Pan S., and Liu M. (2015) Global methane and nitrous oxide emissions from terrestrial ecosystems due to multiple environmental changes. *Ecosystem Health and Sustainability*, 1: 1-20.
Tilley M. A., Segura A., Meadows V., Hawley S., and Davenport J. (2019) Modeling Repeated M Dwarf Flaring at an Earth-like Planet in the Habitable Zone: Atmospheric Effects for an Unmagnetized Planet. *Astrobiology*, 19: 64-86.
Velikova V., Sharkey T. D., and Loreto F. (2012) Stabilization of thylakoid membranes in isoprene-emitting plants reduces formation of reactive oxygen species. *Plant Signaling & Behavior*, 7: 139-141.

Vickers C. E., Gershenzon J., Vickers C. E., Gershenzon J., Lerdau M. T., and Loreto F. (2009) A unified mechanism of action for volatile isoprenoids in plant abiotic stress PERSPECTIVE A unified mechanism of action for volatile isoprenoids in plant abiotic stress. *Nature Chemical Biology*, 5: 283-291.

Wagner W. P., Nemecek-Marshall M., and Fall R. (1999) Three Distinct Phases of Isoprene Formation during Growth and Sporulation of <em>Bacillus subtilis</em>. *Journal of Bacteriology*, 181: 4700 LP-4703.

Wiemer A. J., Wiemer R. J. H., and David F. (2009) The Intermediate Enzymes of Isoprenoid Metabolism as Anticancer Targets. *Anti-Cancer Agents in Medicinal Chemistry*, 9: 526-542.

Wiscombe W. J. (1979) Mie scattering calculations: Advances in technique and fast, vector-speed computer codes. National Technical Information Service, US Department of Commerce.

Wright G. S., Rieke G., Boeker T., Colina L., Van Dishoeck E., Driggers P., Friedman S., Glasse A., Goodson G., and Greene T. (2010) Progress with the design and development of MIRI, the mid-IR instrument for JWST. International Society for Optics and Photonics.

Xue J., and Ahring B. K. (2011) Enhancing Isoprene Production by Genetic Modification of the 1-Deoxy-d-Xylulose-5-Phosphate Pathway in Bacillus subtilis. *Applied and Environmental Microbiology*, 77: 2399-2405.

Yokouchi Y., Takenaka A., Miyazaki Y., Kawamura K., and Hiura T. (2015) Emission of methyl chloride from a fern growing in subtropical, temperate, and cool-temperate climate zones. *Journal of Geophysical Research: Biogeosciences*, 120: 1142-1149.

Yung Y. L., Allen M., and Pinto J. P. (1984) Photochemistry of the atmosphere of Titan: Comparison between model and observations. *Astrophysical Journal Supplement Series*, 55: 465-506.

Yurchenko S. N., Al-Refaie A. F., and Tennyson J. (2018) ExoCross: a general program for generating spectra from molecular line lists. *Astronomy & Astrophysics*, 614: A131.

Yurchenko S. N., Barber R. J., and Tennyson J. (2011) A variationally computed line list for hot NH3. *Monthly Notices of the Royal Astronomical Society*, 413: 1828-1834.

Yurchenko S. N., and Tennyson J. (2014) ExoMol line lists – IV. The rotation–vibration spectrum of methane up to 1500 K. *Monthly Notices of the Royal Astronomical Society*, 440: 1649-1661.

Zahnle K. J. (1986) Photochemistry of methane and the formation of hydrocyanic acid (HCN) in the Earth's early atmosphere. *Journal of Geophysical Research: Atmospheres*, 91: 2819-2834.

Zhang R., Suh I., Lei W., Clinkenbeard A. D., and North S. W. (2000) Kinetic studies of OH-initiated reactions of isoprene. *Journal of Geophysical Research: Atmospheres*, 105: 24627-24635.

Zhang X., Strobel D. F., and Imanaka H. (2017) Haze heats Pluto's atmosphere yet explains its cold temperature. *Nature*, 551: 352-355.

Zuo Z., Weraduwage S. M., Lantz A. T., Sanchez L. M., Weise S. E., Wang J., Childs K. L., and Sharkey T. D. (2019) Isoprene acts as a signaling molecule in gene networks important for stress responses and plant growth. *Plant physiology*, 180: 124-152.